\documentclass[structabstract]{aa}
\usepackage{txfonts}
\usepackage{graphicx}
\usepackage{color}
\usepackage{longtable}
\bibpunct{(}{)}{;}{a}{}{,}             
\usepackage{natbib,twoopt}
\usepackage[colorlinks]{hyperref}
\usepackage{etoolbox}
\usepackage{lscape}
\usepackage{threeparttable}
\usepackage{soul}
\usepackage{xcolor} 
\usepackage{dcolumn,booktabs}
\usepackage{tablefootnote}
\usepackage{siunitx}
\usepackage{dcolumn}
\newcolumntype{d}[1]{D{.}{.}{#1}}
\newcommand{\Teff}[1]{$T_{\rm{eff}}$}{}

\newcommand{\eps}[1]{\log\varepsilon_{\rm #1}}
\newcommand{\kH}{$S_{\!\rm H}$}    
\newcommand{\dnlte}{$\Delta_{\rm NLTE}$}
\newcommand{\Eexc}{$E_{\rm exc}$}
\newcommand{\Ms}{\ensuremath{\mathrm{M}_\odot}}
\begin{document}

%
%
\title{Mono-enriched stars and Galactic chemical evolution -- Possible biases in observations and theory\thanks{Based on data acquired with PEPSI using the Large Binocular Telescope (LBT). The LBT is an international collaboration among institutions in the United States, Italy, and Germany.},\thanks{The line list will appear online on CDS.}}

\author{Hansen, C. J.\inst{1}, Koch, A.\inst{2}, Mashonkina, L.\inst{3}, Magg, M.\inst{4,5}, Bergemann, M.\inst{1}, Sitnova, T.\inst{3}, Gallagher, A. J.\inst{1}, Ilyin, I.\inst{6}, Caffau, E.\inst{7},  Zhang, H.W.\inst{8,9},  Strassmeier, K. G.\inst{6}, Klessen, R. S.\inst{4,10}}

\titlerunning{PEPSI}
\authorrunning{C.~J. Hansen et al.}
\offprints{C.~J. Hansen, \email{hansen@mpia.de}}
\institute{Max Planck Institute for Astronomy, K\"onigstuhl 17, 69117 Heidelberg, Germany
\and
Zentrum f\"ur Astronomie der Universit\"at Heidelberg, Astronomisches Rechen-Institut, M\"onchhofstr. 12, 69120 Heidelberg, Germany 
\and
Institute of Astronomy, Russian Academy of Sciences, Pyatnitskaya 48, 119017, Moscow, Russia
\and Universit\"at Heidelberg, Zentrum f\"ur Astronomie, Institut f\"ur Theoretische Astrophysik, 69120 Heidelberg, Germany
\and International Max Planck Research School for Astronomy and Cosmic Physics at the University of Heidelberg (IMPRS-HD)
\and
Leibniz-Institut für Astrophysik Potsdam (AIP), An der Sternwarte 16, 14482 Potsdam, Germany --
e-mail: ilyin@aip.de, kstrassmeier@aip.de
[orcid: 0000-0002-6192-6494]
\and
GEPI, Observatoire de Paris, Universit\'{e} PSL, CNRS,  5 Place Jules Janssen, 92190 Meudon, France
\and
Department of Astronomy, School of Physics, Peking University, Beijing 100871, P.R. China
\and
Kavli Institute for Astronomy and Astrophysics, Peking University, Beijing 100871, P.R. China
\and
Universit{\"a}t Heidelberg, Interdisziplin{\"a}res Zentrum f{\"u}r Wissenschaftliches Rechnen, Im Neuenheimer Feld 205,  69120 Heidelberg, Germany
}
\date{Received July 1 2020; accepted September 15, 2020}
\abstract{A long sought after goal using chemical abundance patterns derived from metal-poor stars is to understand the chemical evolution of the Galaxy and to pin down the nature of the first stars (Pop III). Metal-poor, old, unevolved stars are excellent tracers as they preserve the abundance pattern of the gas from which they were born, and hence they are frequently targeted in chemical tagging studies. Here, we use a sample of 14 metal-poor stars observed with the high-resolution spectrograph called the Potsdam Echelle Polarimetric and
Spectroscopic Instrument (PEPSI) at the Large Binocular Telescope (LBT) to derive abundances of 32 elements (34 including upper limits). We present well-sampled abundance patterns for all stars obtained using local thermodynamic equilibrium (LTE) radiative transfer codes and one-dimensional (1D) hydrostatic model atmospheres. However, it is currently well-known that the assumptions of 1D and LTE may hide several issues, thereby introducing biases in our interpretation as to the nature of the first stars and the chemical evolution of the Galaxy. Hence, we use non-LTE (NLTE) and correct the abundances using three-dimensional (3D) model atmospheres to present a physically more reliable pattern. In order to infer the nature of the first stars, we compare unevolved, cool stars, which have been enriched by a single event (`mono-enriched'), with a set of yield predictions to pin down the mass and energy of the Pop~III progenitor. To date, only few bona fide second generation stars that are mono-enriched are known. A simple $\chi^2$-fit may bias our inferred mass and energy just as much as the simple 1D LTE abundance pattern, and we therefore carried out our study with an improved fitting technique considering dilution and mixing. Our sample presents Carbon Enhanced Metal-Poor (CEMP) stars, some of which are promising bona fide second generation (mono-enriched) stars. The unevolved, dwarf BD+09\_2190 shows a mono-enriched signature which, combined with kinematical data, indicates that it moves in the outer halo and likely has been accreted onto the Milky Way early on. 
The Pop~III progenitor was likely of 25.5\,M$_{\odot}$ and 0.6\,foe (0.6\,$10^{51}$\,erg) in LTE and 19.2\,M$_{\odot}$ and 1.5\,foe in NLTE, respectively. Finally, we explore the predominant donor and formation site of the rapid and slow neutron-capture elements. In BD-10\_3742, we find an almost clean r-process trace, as is represented in the star HD20, which is a `metal-poor Sun benchmark' for the r-process, while TYC5481-00786-1 is a promising CEMP-r/-s candidate that may be enriched by an asymptotic giant branch star of an intermediate mass and metallicity.} 
\maketitle


\section{Introduction}
Understanding the chemical evolution of the Milky Way (MW) has been a long-standing quest that crucially affects several branches of astrophysics. Key aspects in such analyses are stellar abundances, yield predictions, and the modelling of how gas flows and condensates into later generations of stars. These topics are central to our following analysis, where we target the earliest  enrichment from the first, so-called Population III (Pop III) stars, up to later enrichment events from asymptotic giant branch (AGB) stars.
The vast majority of these studies focus on one-dimensional (1D) local thermodynamic equilibrium (LTE) abundances and explore them in the grand scheme of Galactic chemical evolution (GCE), as was observationally done in \citet{Edvardsson1993} and \citet{Cayrel2004}, for example. However, recent improvements in atomic physics and more computational power allow for higher dimensional model atmosphere calculations and a more sophisticated physical treatment of atomic transitions, including improved radiative and collisional rates. When combined, these allow for more precise and accurate abundance determinations and they are an improvement from the simpler 1D LTE to three-dimensional (3D) and non-LTE (NLTE) results.

The chemo-dynamic behaviour of the major Galactic components, the halo as well as the thick and thin disc, are known to differ in several ways. The MW halo is typically more metal-poor (peaking around [Fe/H] = $-1.6$; \citealt{Schoerck2009,Youakim2020}), exhibiting fast moving stars on elliptical orbits reaching apocentres several kiloparsecs above the plane and beyond. For comparison,  disc stars are on average more metal-rich, and they move on more circular orbits close to and/or within the plane \citep[e.g.][]{Bensby2003A}.
By combining the orbital properties with the chemical composition of stars, we can trace the spatial and chemical origin of the stars with a higher level of confidence and, in turn, label the evolution and chemistry of the Galaxy. Earlier studies, such as
\citet{Nissen2010}, showed how the $\alpha-$element abundances combined with the 3D phase space velocities could help determine whether stars were formed in situ in the MW and likely enriched by more massive supernovae (SN), or whether they have been accreted, predominantly 
move on fast, retrograde orbits in the outer halo, and exhibit a lower [$\alpha$/Fe] 
\citep[see also][]{Cooper2013,Pillepich2015,Myeong2019}. The underlying scenario is likely the accretion of dwarf galaxies onto the MW, which would explain the counter-rotating stellar orbits. Moreover, the low-mass dwarf galaxies typically retain a small gas reservoir and may, as a result, predominantly form lower-mass supernovae, which in turn yield smaller amounts of $\alpha-$elements \citep{Tinsley1979}.

The GCE scenarios are best studied in large, statistically significant samples. However, the precise determination of stellar patterns offers an intriguing counterpart to help learn a great level of detail about the galacto-chemical origin of stars in smaller samples. As an example, the level of $\alpha$-content can provide information on the mass of the previous generation of SN progenitors, while the Fe-peak elements or the odd-even element abundances can reveal information on the explosion energy \citep[e.g.][]{Kobayashi2006}. The heavy elements ($Z>30$) are formed via neutron-captures in most cases, and they split into two categories depending on the rates of captures with respect to the following $\beta-$decay. Here, the slow neutron-capture ($s$-)process forms elements, such as Ba and Pb, while the rapid ($r$-)process forms, Eu, Th, and U, for example. The $s$-process is typically associated with AGB stars or massive rotating stars \citep{Busso1999, Kappeler2011,Meynet2006}, while the $r$-process site is still debated \citep{Horowitz2019, Cote2019}. However, recent discoveries have confirmed that neutron star mergers can produce $r$-process material in the event \citep{Watson2019}, while magneto-rotational SN provide a promising extra site, both at higher and lower metallicities \citep{Winteler2012,Cote2019}.

When trying to trace the individual sources of enrichment, be it a rare or common SN, the best  way to explore their nature in great detail is by studying the chemical abundances in the following generation of old, metal-poor stars and indirectly inferring the 
nature of the SN that provided the enriched material \citep[e.g.][]{Cayrel2004, Hansen2011a}. The surfaces of old, cool, unevolved stars preserve the gas composition from which they were born and hence offer a perfect laboratory for understanding how and where the elements formed. This is because we can derive stellar abundances more accurately in the photospheres of cool stars, especially because calculations of radiative transfer for cool stars can now include 3D hydrodynamics and detailed NLTE, yielding more robust results compared to what is currently possible in fast expanding SN ejecta.

At the lowest metallicities, we are most likely to find a generation of stars that are mono-enriched, that is to say enriched by only one supernova. This leaves the cleanest fingerprint of the nature of that SN, which is why metal-poor stars are a much sought after diagnostic. At the lowest metallicities, an increasing number of stars with high C-abundances have been discovered in the past decades \citep{Frebel2006, Lee2013,Placco2014a}. These stars are commonly referred to as CEMP stars (Carbon Enhanced Metal-Poor stars), and they tend to have a factor of 5--10 times higher [C/Fe] than the Sun \citep{Beers2005, Aoki2007,Hansen2016}. These stars come in different flavours depending on their heavy element composition, where a rich $s$-process composition likely indicates that the star is or was in a binary system with an AGB star transferring mass to the companion, while a lack of heavy elements at the lowest metallicities seems to point towards the star being a bona fide second generation star \citep{Ito2009,Bonifacio2015,Placco2014b,Hansen2019}. This indicates that the first stars (Pop III stars) likely produced large amounts of C very early on \citep[e.g.][]{Meynet2006}. 

Model predictions have shown that a trace of the Pop III stars could be inferred by looking at the [Mg/C] abundances in the most metal-poor stars where a low value could indicate mono-enriched Population II stars \citep{Hartwig2018}. 
A single element ratio is, however, no guarantee that the star is truly mono-enriched, which is why the stellar abundance pattern is often compared to a set of SN yield predictions with the goal of finding the best fitting model and inferring the mass and energy of the SN progenitor. This way, the initial mass function (IMF) in GCE modelling can be retrieved.  Several studies have attempted to place constraints on the nature of the first stars using 1D, LTE abundance patterns of old, Fe-poor ([Fe/H]~$\lesssim -3$) stars \citep[see e.g.][where the latter also includes the NLTE for six elements]{Cayrel2004, Hansen2011b, Placco2016,Aguado2018, Frebel2019}.

The yield predictions required for this process are obtained from simulations of the nucleosynthesis in Pop~III stars and their SNe. In most cases, the SNe are modelled as one dimensional \citep{HegerWoosley2002, Heger2010, Kobayashi2011, Takahashi2014, Ishigaki2018}. Simulations of Pop~III SNe in two \citep{Tominaga2009, Chen2017, Choplin2020} and three \citep{Chan2020} dimensions have been conducted, but so far it is not feasible to perform these for large ranges of stellar masses and explosion energies. Therefore, usually the yields for abundance fitting are taken from one-dimensional simulations. The inherently multidimensional mixing processes within SNe as well as potential anisotropies are usually treated by the introduction of an additional mixing model with one or several free parameters. For example, \citet{Heger2010} smoothed the abundance over a characteristic mass scale before the part of the SN fell back on the compact remnant. \citet{Ishigaki2018} used a three-parametric model that separates the SN into several phases, one of which is fully ejected, one is mixed and partially ejected, and one falls back onto the central object. This model is designed to reproduce the combined yields obtained from two-dimensional simulations of jet-SNe \citep{Tominaga2009}.

However, recent theoretical works have shown that this process may be as flawed as using the simple 1D LTE abundance pattern in the model comparison \citep{Magg2020}. Examples of tracing the mass of the progenitor supernova, which enriched one of the most Fe-poor stars, have been conducted in several studies \citep[1D LTE versus 3D NLTE,][]{Caffau2011,Keller2014, Nordlander2017}, where, for example, very different masses were inferred for the SN progenitor yielding the birth gas in the Fe-poor star when using 1D, LTE versus 3D, NLTE abundance patterns. From a theoretical perspective, it is extremely important to consider whether or not the mixing and dilution of the SN ejecta into the ISM physically make sense. If the dilution of metals in the ISM is left as a free parameter, in effect, only abundance ratios (patterns) are considered. However, low-yield faint SNe may reproduce the abundance ratios (pattern), but they are still unable to produce sufficient amounts of metals to explain the observed (absolute) abundances. Hence, it is equally important to carefully treat the SN model predictions, their dilution, and mixing, as well as stellar abundances, since simplified views can easily bias the results (e.g. mass and energy) by tens of solar masses and explosion energies ($10^{51}$\,erg - `foe') \citep[see e.g.][]{Ezzeddine2019}. 

Here, we shed light on how these types of biases easily occur and how they can bias the conclusions when drawn from the simple $\chi^2$ fitting of SN yields to 1D LTE abundances, as well as further affecting the first generation of stars and the IMF in GCE modelling. We also attempt to trace the origin of the gas composition in more metal-rich stars.

In Sect.~\ref{sec:obs} we describe the observations, sample, and data reduction. Section~\ref{sec:par} outlines the stellar parameter determination and related uncertainties, followed by Sect.~\ref{sec:analysis}, which details the 1D LTE analysis, the NLTE corrections, and finally the 3D NLTE corrected abundances for a subset of elements. The results are laid out in Sect. \ref{sec:results} and they are elaborated on in the discussion in Sect.~\ref{sec:discussion}, where yields, kinematics, and improved fitting techniques are described. Finally, our conclusions are presented in Sect.~\ref{sec:concl}.

\section{Sample, observations, and data reduction}\label{sec:obs}

\subsection{The observations}

The spectra were obtained with the Potsdam Echelle Polarimetric and
Spectroscopic Instrument \citep[PEPSI,][]{Strassmeier2015} at the $2\times
8.4$\,m Large Binocular Telescope (LBT) on Mt.\ Graham, Arizona, USA.
The white-pupil fibre-fed spectrograph was used with the resolving power mode of
120\,000 with $200\,\mu$m fibre ($1.5''$ on sky) and the five-slice image slicer. PEPSI
has blue and red arms with three cross-dispersers (CD) in every arm. The
\'echelle image recorded, in each arm, on a $10.3\times 10.3$\,K STA1600LN CCD
with $9\,\mu$m pixel size and 16 amplifiers.

The spectra were observed in December 2017 and January 2018 with the
pre-selected number of blue and red CDs with the two LBT mirrors. Depending on
the stellar brightness, the signal-to-noise ratio (S/N) per CCD pixel averaged over each CD wavelength range is about 360 in the red CD6 (8200\,\AA)
and 190 in the blue CD2 (4500\,\AA) attained in one hour integration time.
The spectrograph is located in a pressure-controlled chamber at a
constant temperature and humidity to ensure that the refractive index of the air
inside stays constant over a long-term period and providing the radial velocity
stability at about 5\,m\,s$^{-1}$.

\subsection{The data reduction}\label{sec:datared}

The data reduction is done via the Spectroscopic Data Systems for PEPSI (SDS4PEPSI) with its
pipeline adapted to the PEPSI data calibration flow and image specific
content. It is designed based on \citet{Ilyin2000} and its recent
description is given in \citet{Strassmeier2018}.

The specific steps of image processing include bias subtraction and variance
estimation of the source images, super-master flat field correction for the CCD
spatial noise, a definition of \'echelle orders from the tracing flats, and scattered
light subtraction. Then what follows is the wavelength solution for the ThAr images, the optimal extraction
of image slicers and cosmic spikes elimination of the target image, wavelength
calibration, and merging slices in each order. A normalisation to the master flat
field spectrum is carried out to remove CCD fringes and blaze function. A global 2D fit to the
continuum of the normalised image and a rectification of all spectral orders in
the image to a 1D spectrum for a given cross-disperser is conducted as well.

The spectra from two sides of the telescope are averaged with weights into one
spectrum and corrected for the barycentric velocity of the Solar System. The
wavelength scale is preserved for each pixel, as given by the wavelength solution
without rebinning.  The wavelength solution uses about 3000 ThAr lines and has
an error on the fit at the image centre of 4\,m\,s$^{-1}$.

\subsection{The sample}\label{sec:sample}
\begin{figure}
    \centering
    \includegraphics[scale=0.7]{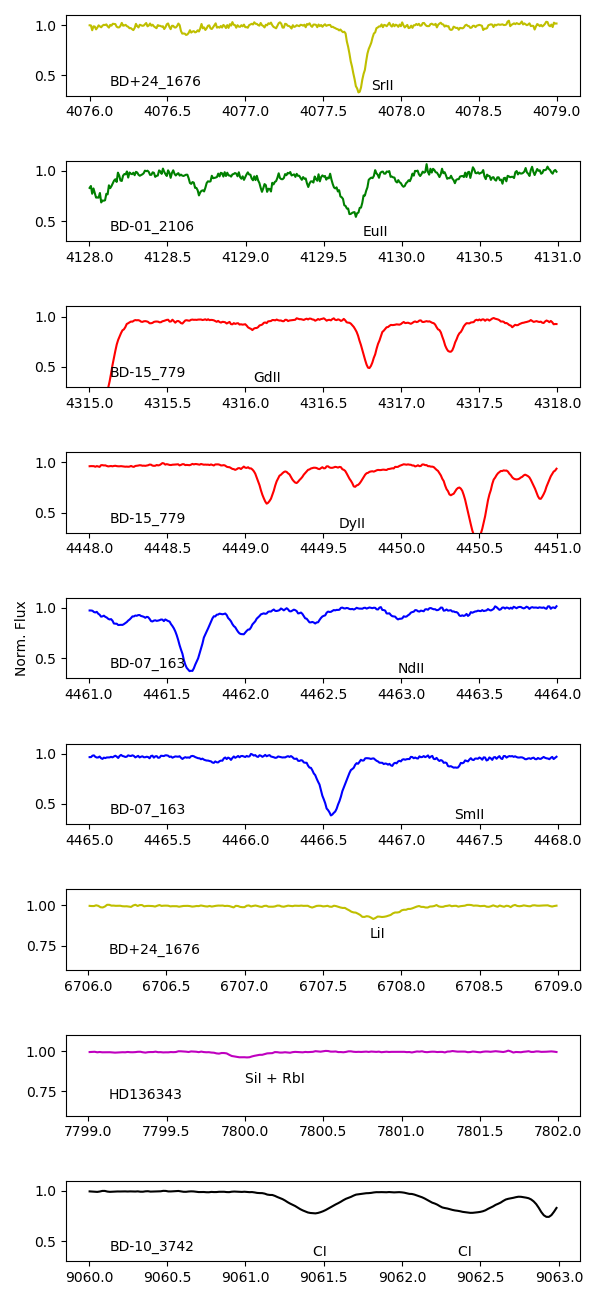}
    \caption{Various lines from the PEPSI spectra of some of our sample stars showing normalised flux versus wavelength $[\AA]$.}
    \label{fig:spec}
\end{figure}

The sample was selected based on a literature compilation covering the Galactic discs and halo. Hence, we selected relatively bright stars ($9.5<$~V~$<12.5$), covering a broad range of stellar parameters (e.g. $\sim -2.8 <$~[Fe/H]~$<-0.9$), to allow us to explore the Galactic chemical evolution using very high-resolution PEPSI spectra. The goal was to derive stellar abundance patterns in LTE and NLTE in moderate to very metal-poor stars. This resulted in a sample of 14 stars (ten giants and four dwarfs), based on \citet{Hollek2015},  \citet{Hansen2012}, and  \citet{Ruchti2013} (see Fig.~\ref{fig:spec}).

Here we target heavy elements, and their strongest transitions are typically located in the blue wavelength range \citep{Hansen2015}. However, as a trade-off for efficiency and exposure time, we typically chose setting 2 (CD2) in the PEPSI setup to get a blue setting but with higher efficiency than setting 1 (CD1). Details on the sample, observations, and adopted instrument setting can be seen in Table~\ref{tab:obslog}.
\begin{landscape}
\begin{table}
\centering
\caption{Observation log for the PEPSI spectra.\label{tab:obslog}}
\begin{threeparttable}
\begin{tabular}{lrrrrcccccc}
\hline
Star ID (Alternative/RAVE ID) &    &  RA (J2000.0) &   Dec  (J2000.0) & Vmag &Date       &   Time  & Exp. time &               Setting&  Wavelength [\AA] &  Median S/N$^*$\\
\hline
\object{BD-01 2439} &              & 10 46 10.97 & $-$02 38 06.70 & 10.22 & 31/01/2018 & 10:42:32.7& 01:00:00.000 & 2:& 4265--4800 &   156\\
(RAVE J104610.9-023806) &                 & &  &         & 31/01/2018 & 10:42:32.6 &01:00:00.000    & 6:& 7419--9067 &  393\\
\object{BD-07 163}       &                     & 01 02 06.19 & $-$06 24 36.50 &       10.50 & 13/12/2017 & 03:29:37.7 &01:30:00.000 & 1:& 3837--4265 &  29\\
                   &                   &  &  &   & 13/12/2017 & 01:56:01.2 &01:30:00.000 & 2:& 4265--4800 &   93\\
                   &                   & &  &   & 13/12/2017 & 03:29:35.9 &01:30:00.000 & 5:& 6278--7419 &   360\\
                   &                   &  &  &   & 13/12/2017 & 01:55:59.4 &01:30:00.000 & 6:& 7419--9067 &   290\\
\object{BD-08 619}&                    & 03 17 29.30 & $-$07 38 07.08 &   9.68  & 14/12/2017 & 06:23:21.4 &01:00:00.000 & 2:& 4265 - 4800 &   161\\
                   &                   &  &  &   & 14/12/2017 & 06:23:21.4 &01:00:40.000 & 6:& 7419--9067 &   316\\
\object{BD+09 2190}     &                      & 09 29 15.56 & +08 38 00.46 &       11.02     & 13/12/2017 & 22:42:27.2 &01:39:35.600 & 2:& 4265 - 4800 &   86\\
                        &                  &  &  &          & 13/12/2017 & 22:42:27.6 &01:39:35.600 & 6:& 7419--9067 &   181\\
\object{BD-10 3742} &    & 13 43 40.69 & $-$11 26 20.55 & 10.34     &31/01/2018 & 11:52:27.7 &01:00:00.000 & 2:& 4265 - 4800 &  131\\
(RAVE J134340.7-112620)                 &              &  &  &    & 31/01/2018 & 11:52:27.7 &01:00:00.000 & 6:& 7419--9067 &   427\\
\object{BD-12 106} &                           & 00 37 43.27 & $-$12 04 39.17 &  11.15  & 14/12/2017 & 02:28:27.6 &01:30:00.000 & 1:& 3837 - 4265 &  18\\
        &                      &  & &    & 14/12/2017 & 00:56:20.7 &01:30:00.000         & 2:& 4265--4800 &   78\\
                &                      &  &  &  & 14/12/2017 & 02:28:27.6 &01:30:10.000 & 5:& 6278--7419 &   265\\
                &                      &  &  &  & 14/12/2017 & 00:56:20.7 &01:30:10.000 & 6:& 7419--9067 &   256\\
\object{BD-15 779}&                            & 04 24 45.64 & $-$15 01 50.69  &      10.07 & 15/12/2017 & 07:35:53.3 &01:02:00.000 &  1:& 3837 - 4265 &  8 \\
                &                      &  &  &   & 13/12/2017 & 06:49:07.6 &01:24:35.200  & 2:& 4265--4800 &  131\\
                &                      & &   &   & 15/12/2017 & 07:35:53.3 &01:02:00.000  & 5:& 6278--7419 &  230\\
                &                      &  &   &  & 13/12/2017 & 06:49:05.9 &01:24:35.200  & 6:& 7419--9067 &  473\\
\object{BD-18 477}&            & 02 47 50.67 & $-$18 16 36.83 &   9.86  & 14/12/2017 & 05:16:23.8 &01:00:00.000 &  2:& 4265 - 4800 &  185\\
(HD 17492)              &                      &  &  &  & 14/12/2017 & 05:16:350 & 01:01:00.000 & 6: & 7419--9067 &  350\\
\object{BD+24 1676}     &                      & 07 30 41.26 & +24 05 10.25 &       10.65 & 14/12/2017 & 08:03:17.1 &01:30:00.000 & 1:& 3837 - 4265 &  36\\
                &                      &  &  &   & 13/12/2017 & 17:17:18.3 &01:33:05.200 & 2:& 4265--4800 &  105\\
                &                      &  &  &  & 14/12/2017 & 08:03:17.1 &01:30:50.000 & 5:& 6278--7419 &  237\\
                &                      &  &  &   & 13/12/2017 & 17:24:18.8 &01:33:55.200 & 6:& 7419--9067 &  216\\
\object{HD136343} &            & 15 20 33.72 & $-$09 43 04.62 & 9.85      & 31/01/2018 & 12:56:25.3 &00:45:00.000  & 2:& 4265 - 4800 &  147\\
(RAVE J152033.7-094304) &                      &  &  &       & 31/01/2018 & 12:56:25.3 &00:45:00.000       & 6:& 7419--9067 &  428\\
\object{HE0420+0123a}&               & 04 23 14.54 & +01 30 48.34 &     11.48  & 17/12/2017 & 04:51:15.3 &00:23:11.635 &  2:& 4265 - 4800 &  10 \\
                &                      &  &  &    & 17/12/2017 & 04:51:15.2 &00:23:11.653 & 6:& 7419--9067 &   34 \\
\object{TYC 5329-1927-1} &           & 05 06 20.12 & $-$13 44 21.98 &     11.71  & 15/12/2017 & 08:41:12.3 &00:46:00.000 &  3:& 4800 - 5441 &   21 \\
                &                      &  &  &   & 15/12/2017 & 08:41:12.3 &00:46:00.000  & 6:& 7419--9067 &   61 \\
\object{TYC 5481-00786-1}&                     & 10 08 00.65 & -10 08 52.25 &        10.73    & 31/01/2018 & 09:31:56.8 &01:00:00.000 &  2:& 4265 - 4800 &   106 \\
                &                      & &  &       & 31/01/2018 & 09:31:56.8 &01:00:00.000 & 6:& 7419--9067 &  336 \\
\object{2MASS J00233067-1631428}&  & 00 23 30.68 & $-$16 31 43.12 & 12.28   & 16/12/2017 & 01:04:37.9 &00:45:00.000 &  2:& 4265 - 4800 &   16 \\
                        &              &  &  &   & 16/12/2017 & 01:04:37.9 &00:45:00.000 &  6:& 7419--9067 &   64 \\
\hline
\end{tabular}
\begin{tablenotes}
\item[]$^*$The S/N is given per pixel and the observing time is in the format hh:mm:ss.
\end{tablenotes}
\end{threeparttable}
\end{table}
\end{landscape}

\section{Stellar parameters}\label{sec:par}
As the targets have been previously analysed in the literature, we adopt the photometric temperatures from \citet{Hansen2012} and  \citet{Ruchti2013},
which have been calculated via the infrared flux method.
We then checked for excitation balance and slightly altered the input literature values to achieve this. In most cases (12 stars), the adopted photometric temperatures almost directly led to a balanced excitation potential\footnote{This thereby ensures that all Fe lines yield the same Fe abundance, regardless of excitation potential.} (labelled 'Teb' in Table~\ref{tab:stelpar}). In two of these stars (BD+09\_2190 and BD-01\_2439), we had to alter the temperature by 15\,K and 100\,K, respectively, to obtain balance. 
However, for two stars, we either had few lines or lower quality spectra, so a balance could not be reasonably achieved, and we kept the literature values.

Gravities were calculated using parallaxes from Gaia data release 2 \citep[Gaia DR2,][]{GaiaDR2}. Together with temperature, V magnitudes and dereddening (from \citealt{Schlafly2011} via the IRSA\footnote{https://irsa.ipac.caltech.edu/applications/DUST/} interface), bolometric corrections (BC), and assuming masses of 1 M$_{\odot}$, we derived log\,$g$. Here, the bolometric correction was calculated using our initial temperature and metallicity ([Fe/H]). 
This approach is similar to 'method 2' in \citet{Ruchti2013}, and our values are generally in agreement within 0.1--0.2\,dex. Only in a few cases does the difference reach 0.4. The stars tagged with 'gib' in Table~\ref{tab:stelpar} also fulfil ionisation balance where Fe I$_{\rm LTE}$ and Fe II$_{\rm LTE}$ agree to within 0.1dex (typically even within 0.05 dex). If the label '(gib)' is used, the balance is just above 0.1 dex (in LTE).

The metallicity, [Fe/H], was based on an average of Fe I and Fe II lines. However, due to the spectral ranges we chose in order to measure heavy element lines, we missed out on several Fe II lines. As a result, we measured equivalent widths (EW) of 1--3 Fe II and 8--30 Fe I lines. Hence, our [Fe/H] is mainly driven by Fe I lines, and it may be biased by deviations from LTE, which, however, we corrected for (see Sect.~\ref{sec:NLTE}). The uncertainties listed in Table~\ref{tab:stelpar} are the errors on the mean.

The microturbulence ($\xi$) was fixed by requiring that all Fe lines yield the same abundance, regardless of EW. In three stars, we only measured a few Fe lines that were distributed, such that optimising a linear trend was hard to achieve, so we adopted the empirical scaling from \citet{Mashonkina2017} to obtain the microturbulence in these three cases (labeled with a 'v' in Table~\ref{tab:stelpar}).
\begin{table*}
\centering
\caption{Stellar parameters of the sample. Comments: Here 'Teb' indicates the excitation equilibrium, 'gib' is the ionisation equilibrium, and the uncertainty on A(Fe) is the error on the mean absolute Fe abundance (st.dev/$\sqrt{N_{lines}}$). R13 refers to \citet{Ruchti2013} and H12 to \citet{Hansen2012}. The '*' indicates that gravity is more uncertain due to the larger distance to the star. \label{tab:stelpar}}
\vspace{-0.5cm}
\begin{tabular}{llccccl}\\
\hline
\hline
ID & T$_{\rm{eff}}$ [K]& log$g$ [dex] & A(Fe)$_{\rm LTE}$ & [Fe/H]$_{\rm LTE}$ & $\xi$ [km/s] &comment \\
\hline
BD-01\_2439 &  $  5288\pm100$ &  $ 2.47\pm0.09$ &  $ 6.41\pm0.07$ &  $ -1.09$  &  $ 1.8\pm0.1$  & R13, Teb, gib \\ 
BD-07\_163  &  $  5564\pm50 $ &  $ 2.50\pm0.23$ &  $ 6.06\pm0.03$ &  $ -1.44$  &  $ 2.3\pm0.1$  & R13, Teb  \\
BD-08\_619  &  $  5993\pm100$ &  $ 4.27\pm0.05$ &  $ 6.42\pm0.05$ &  $ -1.10$  &  $ 1.6\pm0.2$  & R13, Teb, (gib) \\
BD+09\_2190 &  $  6465\pm50 $ &  $ 4.32\pm0.05$ &  $ 4.75\pm0.02$ &  $ -2.75$  &  $ 1.4\pm0.1$  & H12, Teb \\
BD-10\_3742 &  $  4678\pm120$ &  $ 1.38\pm0.04$ &  $ 5.53\pm0.07$ &  $ -1.96$  &  $ 1.9\pm0.1$  & R13, Teb, (gib) \\
BD-12\_106  &  $  4889\pm50 $ &  $ 2.03\pm0.05$ &  $ 5.39\pm0.04$ &  $ -2.11$  &  $ 1.5\pm0.2$  & R13, Teb \\
BD-15\_779  &  $  4805\pm100$ &  $ 1.89\pm0.14$ &  $ 6.02\pm0.03$ &  $ -1.48$  &  $ 1.4\pm0.1$  & R13, Teb, (gib) \\
BD+24\_1676 &  $  6327\pm50 $ &  $ 4.17\pm0.14$ &  $ 4.97\pm0.02$ &  $ -2.53$  &  $ 1.4\pm0.1$  & H12, Teb, gib\\
HD136343 &  $  5082\pm50 $ &  $ 2.35\pm0.08$ &  $ 6.61\pm0.10$ &  $ -0.89$  &  $ 1.6\pm0.1$  & R13, Teb, vt\\
HD17492  &  $  6036\pm50 $ &  $ 4.34\pm0.02$ &  $ 6.55\pm0.05$ &  $ -0.95$  &  $ 1.4\pm0.1$  & R13, Teb \\
HE0420+0123a   &  $  5055\pm50 $ &  $ 2.52\pm0.05$ &  $ 5.05\pm0.05$ &  $ -2.45$  &  $ 1.6\pm0.2$  & R13, Teb, gib, vt \\
TYC5329-1927-1*  &  $  4669\pm50 $ &  $ 1.52\pm0.12$ &  $ 5.36\pm0.04$ &  $ -2.14$  &  $ 1.9\pm0.1$  & R13, Teb, vt \\
TYC5481-00786-1  &  $  4864\pm90 $ &  $ 1.77\pm0.11$ &  $ 6.02\pm0.10$ &  $ -1.48$  &  $ 1.7\pm0.1$  & R13, gib \\
2MASS J00233067-1631428   &  $  5443\pm100 $ &  $ 3.60\pm0.09$ &  $ 5.10\pm0.05$ &  $ -2.40$  &  $ 1.6\pm0.3 $  & R13 \\
\hline\\
\end{tabular}
\end{table*}

\subsection*{Uncertainties}
The uncertainties on the temperatures were estimated based on the residual slopes when attempting to obtain a perfect excitation potential balance with zero slope. For the two stars, where the excitation potential balance was not achieved, we adopted an uncertainty of $\sim$100\,K, which is in agreement with \citet{Ruchti2013}.

The main source of error in our gravities originates in the parallaxes. We computed the total error by varying the parallax, the initial temperature, and the metallicity by their respective errors and we computed new gravities. This change was adopted as the error on the gravity (see Table~\ref{tab:stelpar}). For the most distant star (TYC 5329), which is  just beyond 3\,kpc, we have listed a slightly larger uncertainty (as indicated by the '*' in Table~\ref{tab:stelpar}) in order to accommodate uncertainties on parallaxes and distances and to not simply treat the latter as a `1/parallax'. Using the probabilities from \citet{BailerJones2018} and their Bayesian distance computation of this star results in a slightly higher gravity (by 0.12\,dex), which we have taken into account in the listed uncertainty. Most of the stars are within $\sim2.2$\,kpc and the parallax error is normally less than $\sim10\%$\footnote{The parallax error is between 0.5 and 12.8\% with the majority below 10\%.}, hence possible distance discrepancies are well accounted for by the associated error, which we used to estimate the uncertainty on the gravity.

For the metallicity, 
we adopted the line-to-line abundance scatter as the statistical error on [Fe/H]. In cases where an ionisation balance is not achieved, the Fe II lines increase this error slightly.

In the case of the microturbulence, $\xi$, the uncertainty in the slope was adopted as the uncertainty on the value (i.e. the deviation from a perfect zero slope).\ Additionally, for three stars where the empirical scaling relation was used, we varied the input parameters (\Teff\ , log\,$g$, and [Fe/H]) by their respective uncertainties and errors.

\begin{figure*}[!hbtp]
    \centering
     \includegraphics[scale=0.52]{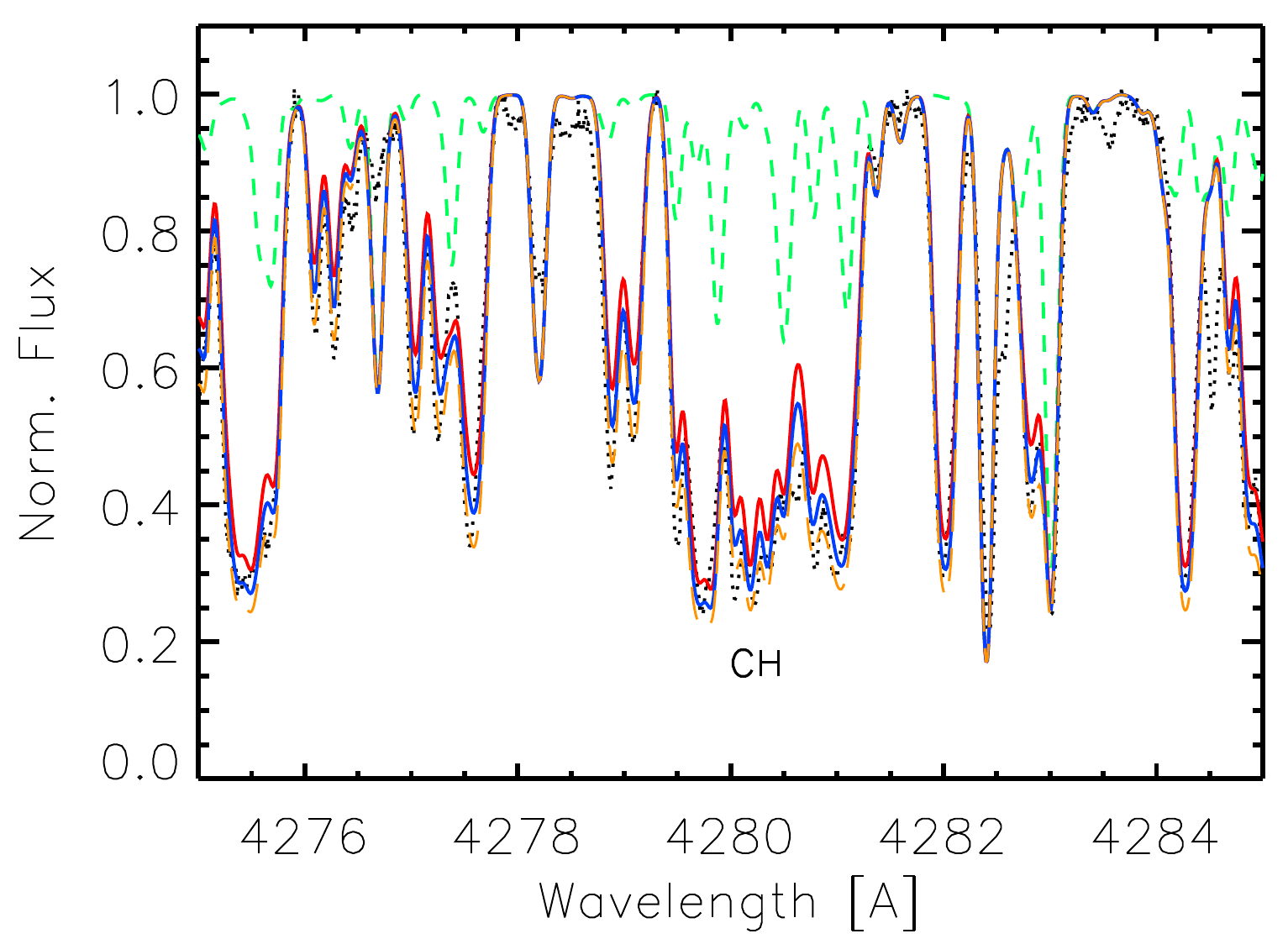}
    \includegraphics[scale=0.52]{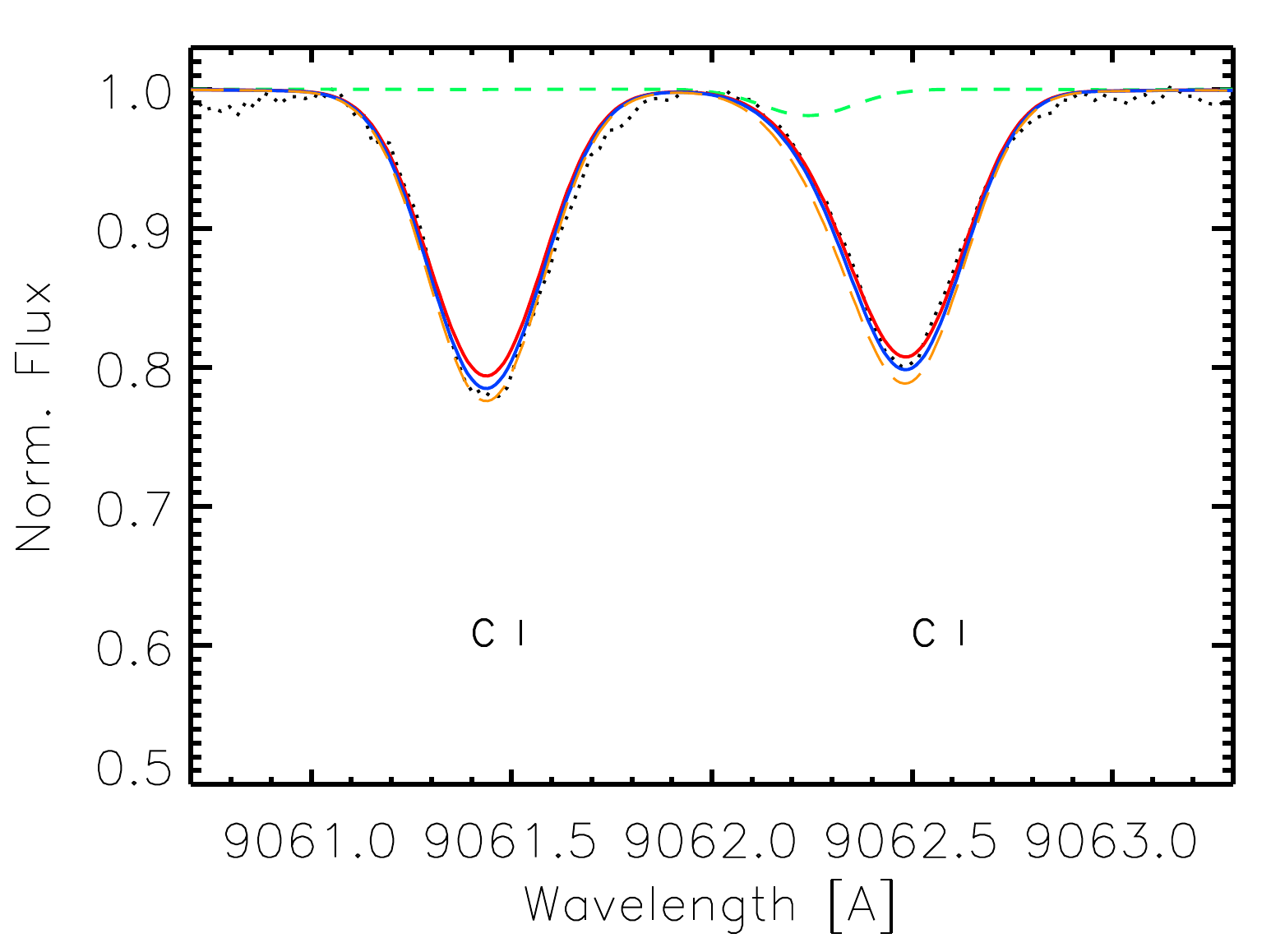}
    \includegraphics[scale=0.52]{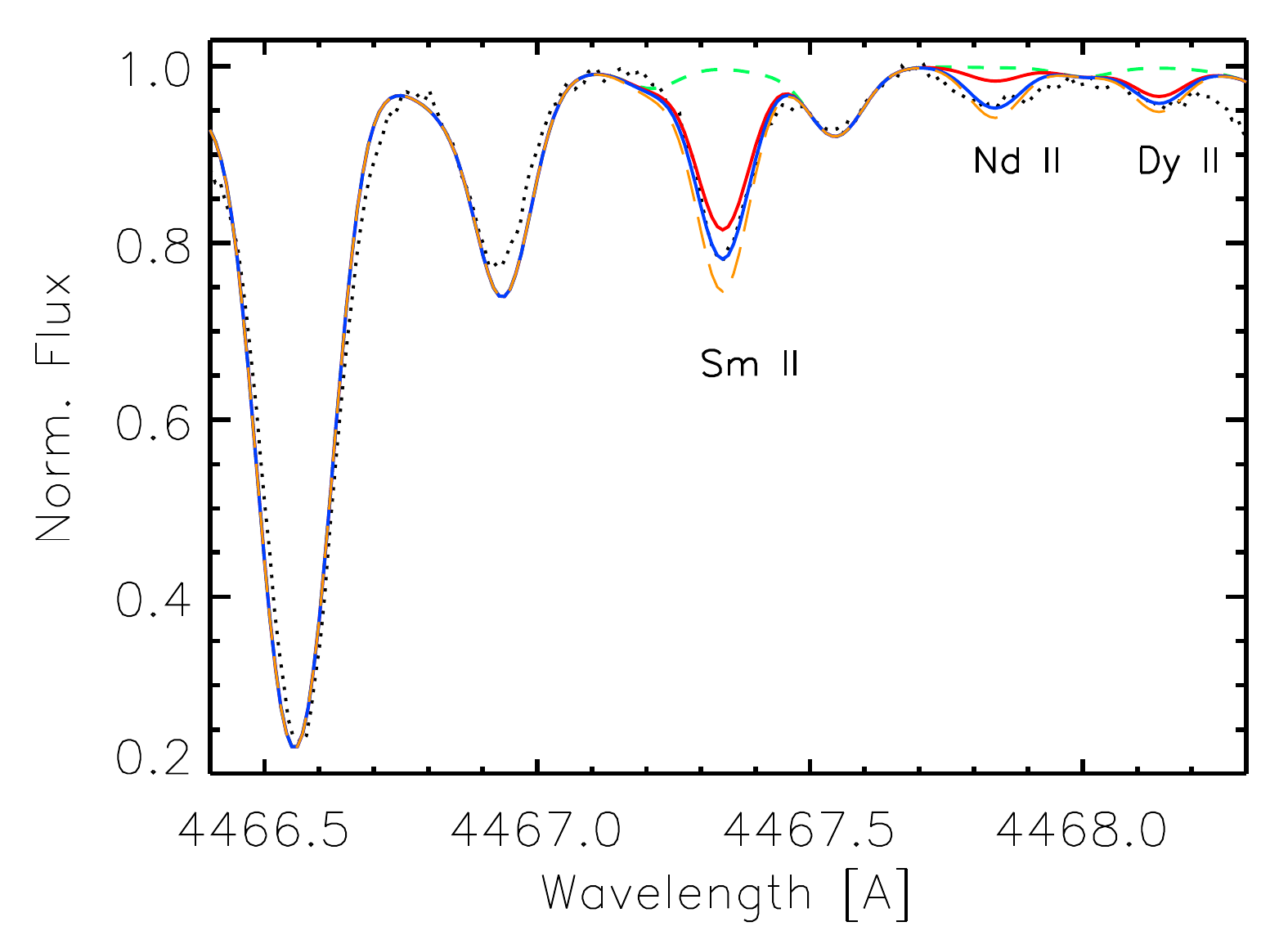}
    \includegraphics[scale=0.52]{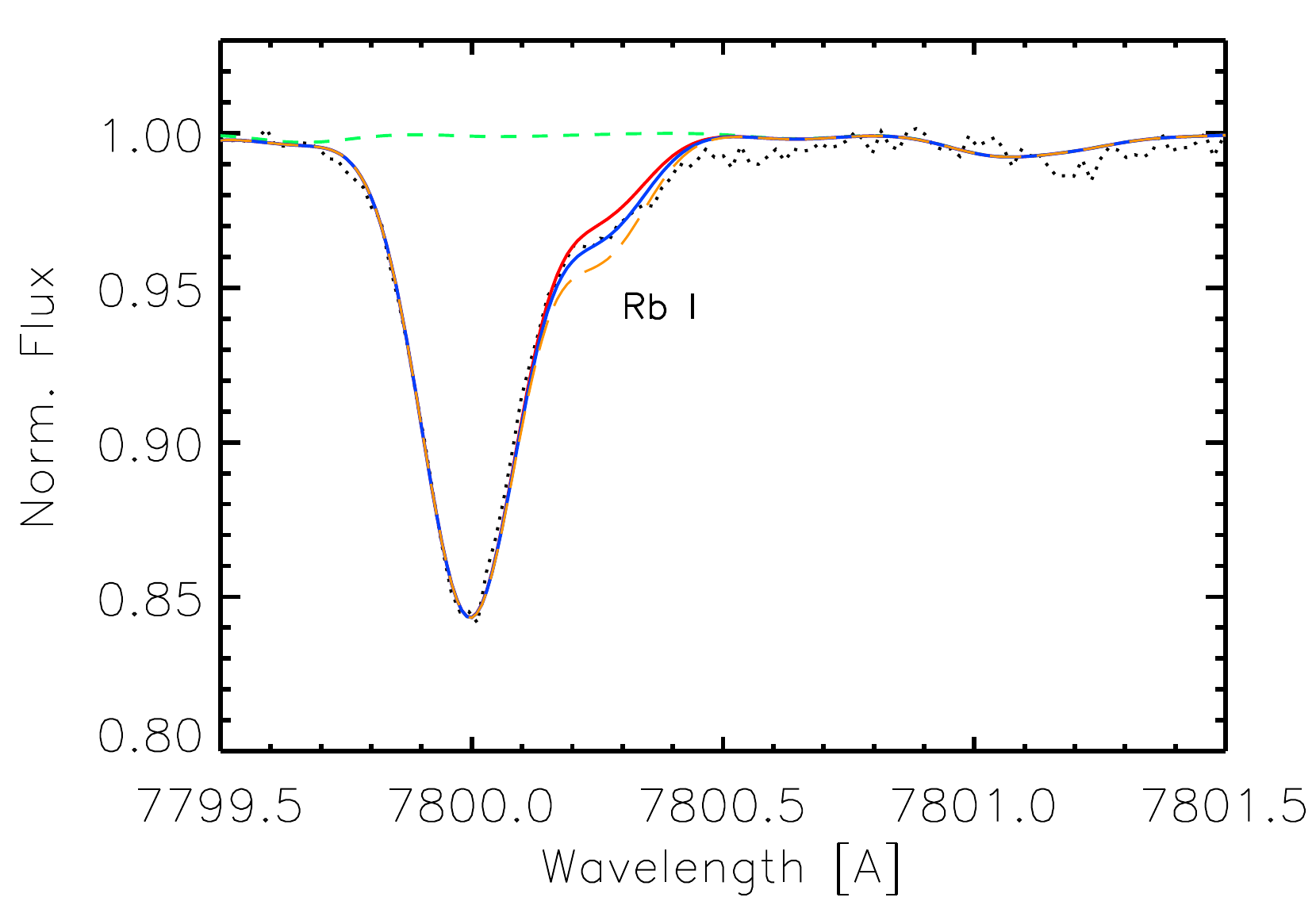}
    \caption{Spectrum synthesis of C, Sm, (Nd), Dy, and Rb in various sample stars. Specifically, molecular and atomic C in TYC5481. We note that [\ion{C}{i}/Fe]~=~0.6$\pm0.1$; [CH/Fe]~$=0.1\pm0.1$; in BD-0.1\_2439, [Sm/Fe]~$=0.3\pm0.1$ and the shown Nd line was not used in our average Nd value due to blends and a poor fit -- synthesis shows [Nd/Fe]~=~0.3, 0.8, 0.9; [Dy/Fe]~$=0.5\pm0.1$; and in HD136343, [Rb/Fe]~$=0.4\pm0.1$. In all cases, the green dashed line indicates [X/Fe]~$=-5$.}
    \label{fig:synt}
\end{figure*}

\section{Abundance analysis} \label{sec:analysis}
Our present analysis consists of both EW measurements and spectrum synthesis. We present the abundances derived under the assumptions of 1D LTE, as well as NLTE and 3D corrected abundances. The EWs also serve as an extra check when computing the NLTE corrections. First we describe the LTE analysis (Sect.~\ref{sec:LTE}) and in Sect.~\ref{sec:NLTE} we outline the details of the NLTE corrections. In addition, we present 3D corrections for C, O, and Fe in Sect.~\ref{sec:3DNLTE}.

\subsection{1D LTE abundances}\label{sec:LTE}
The 1D LTE abundances were derived using MOOG \citep[][version 2014]{Sneden1973} and MARCS atmosphere models \citep[][adopting interpolated plane parallel and spherical models for dwarfs and giants, respectively]{Gustafsson2008}. The line list is provided online on CDS\footnote{See CDS}. Relative abundances were calculated using the solar values from \citet{Asplund2009}.

As a first attempt, we measured all EWs in IRAF by fitting Gaussian or Voigt profiles to the absorption features. However, due to the broad range in stellar metallicity and signal-to-noise ratios of the PEPSI spectra ($8<S/N<428$), some lines are blended, and despite deblending attempts some EWs might still be misleading. Hence, we introduced cuts to further reduce blended or saturated lines, and we typically removed EW measurements, if one or more of the following criteria were met:
\begin{itemize}
    \item $0.1>FWHM_{Gaussian}$ or $FWHM_{Gaussian}>0.3$ (noise/blends),
    \item $|\lambda_{measured~line}-\lambda_{line~centre}|> 0.05$\,\AA,
    \item EW $>$ 250\,m\AA\  (saturation), and
    \item $log \varepsilon_{line} - \langle log \varepsilon\rangle > \pm 0.5$\,dex. 
\end{itemize}
The range of the first item depends on the setting and wavelength range in which the lines fall. In general, the abundance was synthesised if the lines are located in the blue ($\lambda<4200$\AA). We also gave preference to Gaussian fit lines, and if the line was very strong and required a Voigt profile to be fit, we synthesised the line and generally used the derived value as a limit to avoid saturated or insensitive lines.

We derived abundances of 32 (34) elements (including limits), namely of \ion{Li}{I}, \ion{C}{I} (CH), \ion{O}{I}, \ion{Na}{I}, \ion{Mg}{I}, \ion{Si}{I+II}, \ion{S}{I}, \ion{K}{I}, \ion{Ca}{I}, \ion{Sc}{II}, \ion{Ti}{I+II}, \ion{V}{I+II},  \ion{Cr}{I}, \ion{Mn}{I}, \ion{Fe}{I+II}, \ion{Co}{I}, \ion{Ni}{I}, \ion{Cu}{I}, \ion{Zn}{I}, \ion{Rb}{I}, \ion{Sr}{II}, \ion{Y}{II}, \ion{Zr}{II}, \ion{Ba}{II}, \ion{La}{II}, \ion{Ce}{II}, \ion{Pr}{II}, \ion{Nd}{II}, \ion{Sm}{II}, \ion{Eu}{II}, \ion{Gd}{II}, and \ion{Dy}{II} (\ion{Pb}{II} and \ion{Th}{II}).
The abundances of each element (neutral or ionised species) can be found in the online Table and the lines that were synthesised or if the Gaussian fit EWs were used to derive the final abundances are flagged. A few examples of spectrum line syntheses are shown in Fig.~\ref{fig:synt}.
For consistency with the NLTE analysis, we only used lines which can be NLTE corrected as well in order to ensure a better and more equal foundation for comparing the LTE versus NLTE abundance behaviour. We only deviated from this criterion for three abundances because the average remains the same whether we used all or the reduced number of lines. 

\subsection{1D NLTE abundances}\label{sec:NLTE}
The  present investigation is based on the NLTE methods developed in our
earlier studies and documented in a number of papers (see Table~\ref{Tab:nlte} for the references) in which the atomic data and the
questions on line formation have been considered in detail. For a number of chemical species, their model atoms were updated by employing quantum-mechanical rate coefficients for inelastic processes in collisions with neutral hydrogen atoms. 

\begin{figure}  
\begin{center}
 \hspace{-6mm}  
 \includegraphics[scale=0.6]{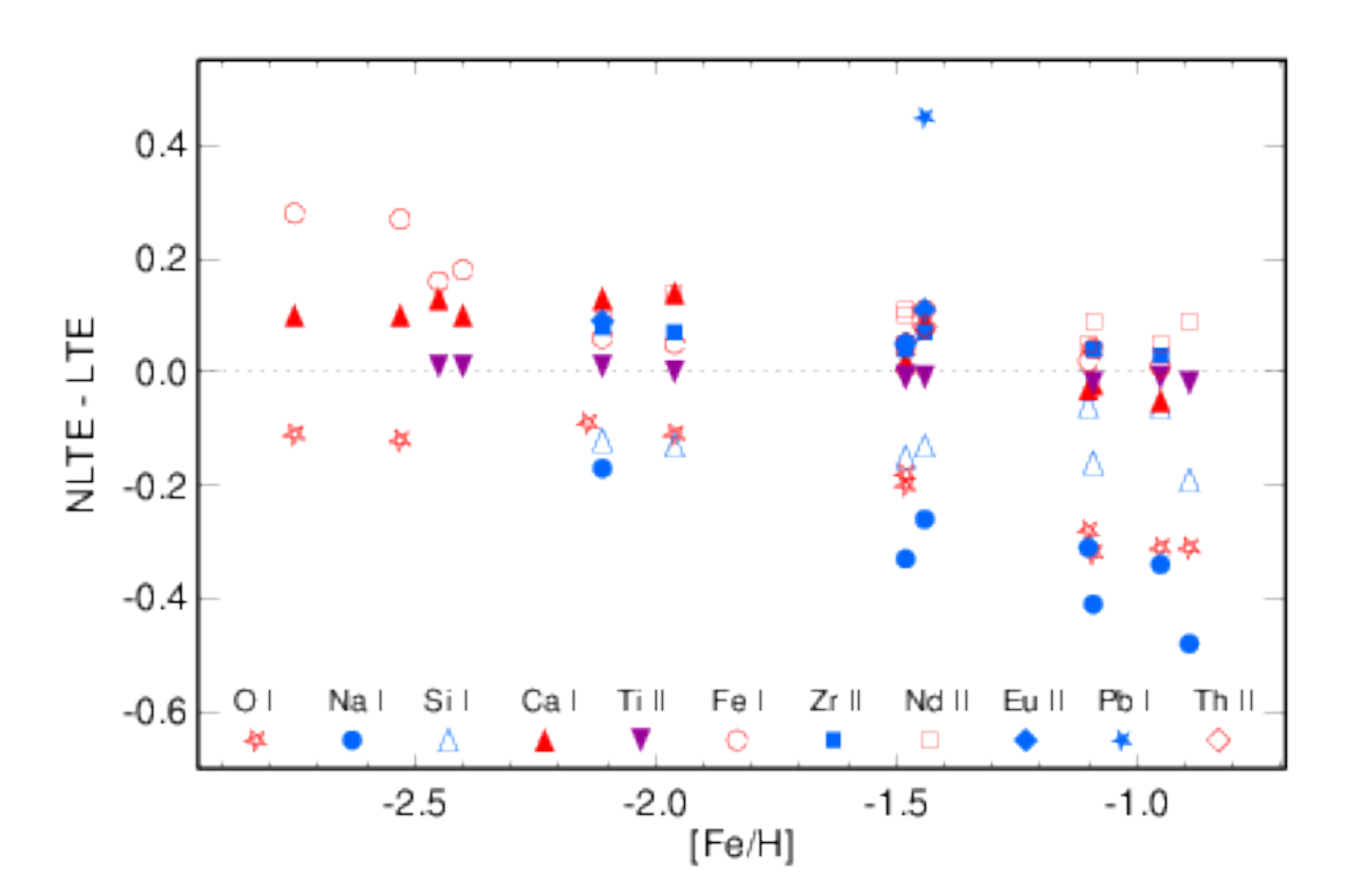}
  \caption{\label{fig:1dnlte} NLTE abundance corrections (dex) for representative lines of the NLTE species in the sample stars.  }
\end{center}
\end{figure}

We briefly describe the departures from LTE for the investigated lines. Figure~\ref{fig:1dnlte} displays the NLTE abundance corrections, $\Delta_{\rm NLTE} = \eps{NLTE} - \eps{LTE}$, for some representative lines, namely, \ion{O}{i} 7771~\AA, \ion{Na}{i} 8183~\AA, \ion{Si}{i} 7415~\AA, \ion{Ca}{i} 4585~\AA, \ion{Ti}{ii} 4529~\AA, \ion{Fe}{i} 4494~\AA, \ion{Zr}{ii} 4317~\AA, \ion{Nd}{ii} 4358~\AA, \ion{Eu}{ii} 4129~\AA, \ion{Pb}{i} 4057~\AA, and \ion{Th}{ii} 4019~\AA, in the sample stars. The corrections were computed using the same MARCS models as in LTE\footnote{For a few elements, Mg, Cr, Mn, and Co, we used MAFAGS models instead; however, the difference between the models was tested and shown to be minor, see, e.g. \citet{Hansen2013} and a detailed comparison of the models in \citet{Bergemann2012,Bergemann2019}.}, either with the Detail code \citep{Detail} or MULTI2.3 \citep{Carlsson1986}.

\begin{table} 
 \caption{\label{Tab:nlte} NLTE atomic models used in this study.}
 \centering
 \begin{tabular}{lll}
\hline\hline \noalign{\smallskip}
 Species & Reference & \ion{H}{i} collisions \\
\noalign{\smallskip} \hline \noalign{\smallskip}
 \ion{C}{i}* & Amarsi et al. (2019a,c)  & AK\\
 \ion{O}{i}*    &   \citet{2000AA...359.1085P}, &   \\
               &   \citet{2018AstL...44..411S} & BVM19 \\
\ion{Na}{i}    &   \citet{alexeeva_na}         &  BBD10 \\
\ion{Mg}{i-ii} &   \citet{Bergemann2017}  & BBS12  \\
\ion{Si}{i-ii} &   \citet{Mashonkina2020}  &  BYB14 \\
\ion{Ca}{i-ii} &   \citet{mash_ca,2017AA...605A..53M} &  BVY17 \\
\ion{Sc}{ii}   &   \citet{Zhang2008}  &   SH84 (0.1)\\
\ion{Ti}{i-ii} &   \citet{sitnova_ti}   & SYB20  \\
\ion{Cr}{i-ii} &   \citet{Bergemann2010Cr} & SH84 (0.0)  \\
\ion{Mn}{i-ii} &   \citet{Bergemann2019}  & BV17, BGE19  \\
\ion{Fe}{i-ii} &   \citet{mash_fe,2019AA...631A..43M} & YBK18, \\
               &                                      & YBK19 \\
\ion{Co}{i-ii} &   \citet{Bergemann2010} & SH84 (0.05)  \\
\ion{Zr}{ii}   &   \citet{Velichko2010_zr}    & SH84 (0.1) \\
\ion{Nd}{ii}   &   \citet{2005AA...441..309M} & SH84 (0.1) \\
\ion{Ba}{ii}   &   \citet{Gallagher2020} & BY17, BY18 \\
\ion{Eu}{ii}   &   \citet{mash_eu}             & SH84 (0.1) \\
\ion{Pb}{i}    &   \citet{Mashonkina_pb}       & SH84 (0.1) \\
\ion{Th}{ii}   &   \citet{Mashonkina_pb}       & SH84 (0.1) \\
\noalign{\smallskip}\hline \noalign{\smallskip}
\end{tabular}
\tablefoot{
Collisions with \ion{H}{i} were treated following 
AK \citep{2019A&A...624A.111A,Kaulakys1991},
BVM19 \citep{2019MNRAS.487.5097B},  
BBD10 \citep{barklem2010_na},  
BYB14 \citep{Belyaev2014_Si},
BVY17 \citep{2017ApJ...851...59B}, 
SH84 \citep{Zhang2008}, 
SYB20 \citep{sitnova_ti2019}, 
YBK18 \citep{2018CP....515..369Y}, 
YBK19 \citep{2019MNRAS.483.5105Y},
BGE19 \citep{Bergemann2019}, 
BV17 \citep{BelyaevV2017}, 
BBS12 \citep{BarklemB2012},
BY17 \citep{BelyaevY2017}, 
BY18 \citep{BelyaevY2018},  and
SH84 (0.1) \citep{Steenbock1984} with a scaling factor of \kH\ = 0.1.
SH84 (0.0) \citep{Steenbock1984} and with a scaling factor of \kH\ = 0.0. The * indicates that 3D, NLTE abundances were computed, see Table~\ref{tab:3D}.}
\end{table}

{\bf \ion{O}{i}.} Our NLTE calculations show strengthened \ion{O}{i} 7771, 7774, and 7775\,\AA\ lines and negative NLTE abundance corrections, which is in line with many previous studies \citep[see the pioneering paper of][]{1991A&A...245L...9K}. 
The difference between this study and \citet{2018AstL...44..411S} lies in using recent data on the \ion{O}{i} + \ion{H}{i} collisions from \citet{2019MNRAS.487.5097B} instead of those from \citet{2018A&A...610A..57B}. This update leads to slightly smaller NLTE effects and smaller magnitude of $\Delta_{\rm NLTE}$, for example, by 0.02\,dex in the model atmosphere with 
\Teff\~ = 6000\,K, 
log~g = 3.90, and [Fe/H] = $-2.1$. In the same model atmosphere, we find $\Delta_{\rm NLTE} = -0.15$, $-0.14$, and $-0.13$\,dex for \ion{O}{i} 7771, 7774, and 7775\,\AA, respectively.  
For the sample stars, 
the NLTE abundance corrections change from $-0.3$~dex at [Fe/H] $\simeq -1$ to $-0.1$~dex for [Fe/H] $< -2$. The stronger \ion{O}{i} 7771-5\,\AA, the larger the magnitude of $\Delta_{\rm NLTE}$ is (for 3D NLTE see Sect.~\ref{sec:3DNLTE}.).  

{\bf \ion{Na}{i}.} The NLTE effects are rather large, with $\Delta_{\rm NLTE}$ ranging between $-0.2$ and $-0.5$~dex. As shown in many previous studies \citep[one of the first being][]{1992A&A...265..237B}, \ion{Na}{i} is subject to over-recombination in the atmospheres of late-type stars, resulting in strengthened lines and negative NLTE abundance corrections. For the four stars BD-07\_163, BD-08\_619, BD-12\_106, and HD136343 with both \ion{Na}{i} 8183 and 8194\,\AA\  lines  measured, the NLTE values are close and in good agreement,
so that the dispersion in the single line measurements around the mean, $\sigma = \sqrt{\Sigma(\overline{x}-x_i)^2 / (N_l-1)}$, is smaller than in LTE. Here, $N_l$ is the number of measured lines. For example, for BD-07\_163, $\sigma = 0.11$ and 0.03~dex in LTE and NLTE, respectively. The NLTE effects decrease towards lower metallicity due to weakening of the lines.

{\bf \ion{Mg}{i}.} The NLTE corrections to the Mg lines (4167.3, 4351.9, 4571.1, 4703.0, 5172.7, 5183.6, and 5528.4\,\AA) originate in \citet{Bergemann2017}. The neutral atom is dominated by photoionisation  and hence it is prone to overionisation by a non-local blue radiation field. In turn, this weakens the spectral line and leads to a positive correction (see also Fe-peak elements below for more details). For instance, some lines, such as the high-excitation \ion{Mg}{i} lines, are only affected at the level of $-0.03$ to $+0.08$ dex (similarly for the Mg b triplet in TYC5329-1927-1). In the specific case of BD-08\_619 (with [Fe/H] $\approx -1$), we obtain only modest NLTE corrections of 0.013 on average for \ion{Mg}{i}. These corrections slightly increase with decreasing metallicity  to 0.06 for \ion{Mg}{i} in HE0420+0123a (at [Fe/H] $\sim -2.5$).

{\bf \ion{Si}{i-ii}.} In the stellar parameter range, with which we are concerned, number densities of neutral and singly ionised silicon have comparable values in the line-formation layers. In a competition for enhanced photoionisation of the low-excitation (\Eexc\ $<$ 2\,eV) levels of \ion{Si}{i} with a photon suction caused by bound-bound transitions from many levels close to the ionisation limit down to the lower levels, the latter prevails and increases the populations of the ground state and low-lying (\Eexc $\le5.6$\,eV) levels of \ion{Si}{i}. This results in a strengthened \ion{Si}{i} 7415\,\AA\ line and negative $\Delta_{\rm NLTE}$ of $-0.19$ to $-0.12$\,dex in different stars. In BD-15\_779, \ion{Si}{ii} 6347 and 6371\,\AA\ lines were measured. They are weak, with $\Delta_{\rm NLTE} = -0.03$\,dex.

{\bf \ion{K}{i}.} The NLTE effect of potassium is dictated by the source function and caused by resonance scattering. Similar to the sodium D lines, an overpopulation of the ground states shifts the line formation slightly outwards, which deepens the lines. This means that the effect is governed by the radiation field and rates \citep{Reggiani2019}. Here we interpolate in their grid of NLTE K corrections over all stellar parameters including EW. In doing so, we adopt a multi-D linear interpolation, as the grid is not evenly spaced in abundances and EW. We find a slight offset in  K abundances between our study and \citet{Reggiani2019}, and we furthermore find slightly higher corrections (by $\sim0.12$\,dex) if we interpolate in LTE abundances rather than EW. This difference is, however, negligible at metallicities below $\sim-2$. 
The atomic data (oscillator strength and excitation potential) in this and the analysis of \citet{Reggiani2019} are identical but there might be slight differences in the spectrum synthesis code, model atmospheres, and possible damping treatment. We chose to interpolate in EW, even if we ended up slightly underestimating the K NLTE corrections. Since we relied on EW, we mainly used the 7698\,\AA\ line, as the 7664\,\AA\ line has a silicon blend. However, for two stars (BD-08\_619 and BD-12\_106), we were forced to use the 7664 K line since the telluric A-band is interfering with the, otherwise cleaner, 7698 K line. For BD-08\_619, we find a correction of $-0.15;$ while for the more metal-poor HE0420+0123a, we obtain a correction of $-0.31$\,dex.

{\bf \ion{Ca}{i}.} As shown in previous studies \citep[see, for example,][]{mash_ca}, \ion{Ca}{i} is subject to the ultraviolet (UV) overionisation in the atmospheres of late-type stars. The overionisation tends to weaken the lines. However, in mildly metal-poor stars, another NLTE mechanism is working in the opposite direction. This effect is the lowering of the line source function ($S_\nu$)
below the Planck function ($B_\nu$) in the uppermost atmospheric
layers, where the cores of strong \ion{Ca}{i} lines form, and it tends to make the lines stronger. 
In a given star, $\Delta_{\rm NLTE}$ is positive for weak lines, which form in the layers affected by the overionisation, and it can be negative for strong lines.
The net effect is that the difference between average NLTE and LTE abundances is slightly negative for the [Fe/H] $> -1.5$ stars, but positive for the lower metallicities. 

{\bf \ion{Ti}{i-ii}.} Compared with \cite{sitnova_ti}, the model atom was updated by implementing quantum-mechanical rate coefficients for the \ion{Ti}{i} + \ion{H}{i} and \ion{Ti}{ii} + \ion{H}{i} collisions. 
Calculations of collisional data and their impact on the NLTE results are presented by \cite{sitnova_ti2019}. The NLTE computations lead to weakened lines of \ion{Ti}{i} and positive NLTE abundance corrections, which vary from 0 to 0.23\,dex, depending on the spectral line and stellar parameters. 
For \ion{Ti}{ii}, \dnlte\ is negative (down to $-0.13$\,dex) for strong lines with an equivalent width of EW $ > 80$\,m\AA, but it is positive (up to 0.07\,dex) for weak lines. To calculate average titanium abundances, we employed lines of \ion{Ti}{ii}, as recommended by \citet{2011MNRAS.413.2184B}, \citet{2016AstL...42..734S}, and \citet{sitnova_ti2019}.

{\bf Fe-peak (\ion{Cr}{i}, \ion{Mn}{i}, and \ion{Co}{i}):} The NLTE corrections for the lines of Cr, Mn, and Co were computed based on model atoms of \citet{Bergemann2010Cr}, \citet{Bergemann2019}, and \citet{Bergemann2010}, respectively. All of these elements can be observed in neutral and singly-ionised stage; however, in our spectra, only lines of neutral species  could be used as a diagnostic, owing to the limited wavelength range of the spectra. The neutral atoms of all three elements are photoionisation-dominated ions (see \citealt{Bergemann2014} for a detailed discussion on the physics behind NLTE), which means that they are sensitive to overionisation that is driven by the non-local high-energy (near-ultraviolet to blue) radiation field. This generally leads to weakening of the low-excitation spectral lines of these species, and, therefore, to positive NLTE abundance corrections. In other words, the LTE analysis underestimates the abundances of these elements. However, the amplitude of NLTE corrections differs amongst the spectral lines of the same element. Some spectral lines, such as the Mn I resonance triplet at 4030\,\AA, show NLTE corrections of up to $0.4$\,dex. In addition, the NLTE corrections tend to grow with a decreasing metallicity and surface gravity, and with an increasing T$_{\rm eff}$. For an F-type main-sequence star with [Fe/H] $\approx -1,$ such as BD-08\_619, we obtain, on average, only modest NLTE corrections of 0.07 for \ion{Cr}{i} and 0.19 for \ion{Mn}{i}. The NLTE values for Mn I are comparatively high because only resonance lines are available to us and these are very sensitive to the effects of NLTE and 3D inhomogeneities \citep{Bergemann2019}. On the other hand, the average estimates of NLTE corrections for an RGB star with a metallicity of [Fe/H] $\approx -2.5$ (e.g. HE0420+0123a) are 0.34 for \ion{Cr}{i} and 0.43\,dex for \ion{Mn}{i}. 

{\bf \ion{Fe}{i-ii}.} NLTE mechanisms for \ion{Fe}{i} are very similar to those of \ion{Ca}{i,} resulting, in  most cases, in weakened lines of \ion{Fe}{i} and positive NLTE abundance corrections, which grow towards lower metallicity \citep[see, e.g.][for a discussion]{Bergemann2012}. In one of our most metal-poor stars, BD+24\_1676, $\Delta_{\rm NLTE}$ varies from 0.11 to 0.35\,dex for different \ion{Fe}{i} lines. In mildly metal-poor stars, $\Delta_{\rm NLTE}$ can be slightly negative (of $-0.05$~dex) for some strong \ion{Fe}{i} lines. 
We note that \ion{Fe}{ii} is a majority species in the atmospheres of our sample stars, and the NLTE effects are, in general, minor for lines of \ion{Fe}{ii}: $\Delta_{\rm NLTE}$ does not exceed 0.02\,dex, in absolute value, for 4233, 6238, and 6247\,\AA, and it reaches a maximal value of $-0.10$\,dex for \ion{Fe}{ii} 4923 and 5018\,\AA\ in TYC5329-1927-1 (for 3D corrections see Sect.~\ref{sec:3DNLTE}).

{\bf \ion{Zr}{ii}, \ion{Nd}{ii}, \ion{Eu}{ii}, \ion{and Th}{ii}.} The ionised species are the majority ones for these corresponding chemical elements and they are subject to similar NLTE mechanisms. The investigated spectral lines arise in the transitions from the low-excitation (\Eexc\ $<$ 1~eV) levels, which keep the thermodynamic equilibrium populations, while the UV pumping transitions produce enhanced excitation of the upper levels. Therefore, NLTE leads to weakened lines of these species. The magnitude of the NLTE effects is small. In a given star and for a given species, the difference in $\Delta_{\rm NLTE}$ between different lines amounts to 0.02 to 0.10~dex. The strongest line exhibits the maximal value for $\Delta_{\rm NLTE}$, however, it does not exceed 0.16, 0.16, 0.08, and 0.08~dex for \ion{Zr}{ii} 4208\,\AA, \ion{Nd}{ii} 4061\,\AA,  \ion{Eu}{ii} 4205\,\AA, and \ion{Th}{ii} 4019\,\AA, respectively.

{\bf \ion{Ba}{ii}.} We computed new 1D NLTE corrections for the stars in our sample using the Ba model atom described in \citet{Gallagher2020}. Deviations from LTE are caused by strong-line scattering and radiative pumping, causing a level population of Ba that is far from LTE. 
On average, we obtain small NLTE corrections ($-0.07$\,dex) for the 4554\,\AA\ line in our sample and slightly higher corrections for the 6496\,\AA\ line.

{\bf \ion{Pb}{i}.} Lead is strongly ionised in the atmospheres of the sample stars, and the UV overionisation is the main NLTE mechanism for \ion{Pb}{i}, resulting in depleted absorption in the \ion{Pb}{i} 4057\,\AA\ line. This line was measured only in the star, BD-07\_163, and its NLTE abundance correction amounts to 0.45\,dex.

\subsection{3D NLTE abundances}\label{sec:3DNLTE}
\begin{figure}[htp]
    \begin{center}
        \includegraphics[scale=0.33]{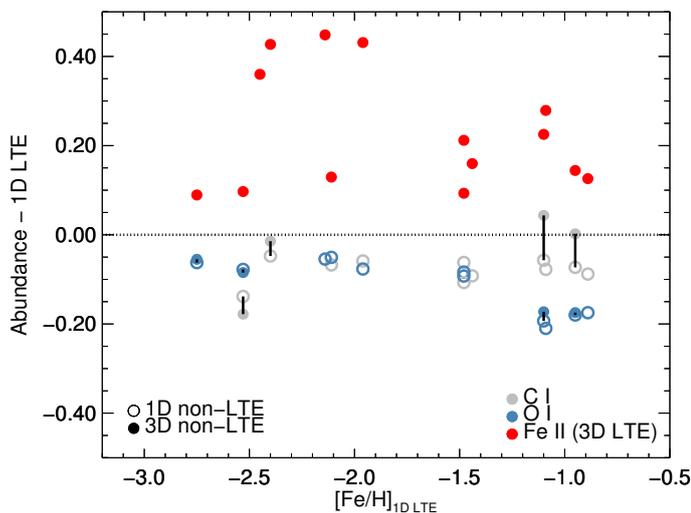}
        \caption{Line-averaged
            \ion{C}{I}, \ion{O}{I}, and 
            \ion{Fe}{II} abundance corrections (1D non-LTE $-$ 1D LTE, and
            3D non-LTE $-$ 1D LTE, abundance differences) for the sample
            of stars where available. The \ion{Fe}{II} results
            presented here strictly assume zero departures from LTE.
            Vertical lines between two data points indicate that they
            correspond to the same star.}
        \label{fig:3dnlte} 
    \end{center}
\end{figure}

The atmospheres, and thus the emergent spectra,
of late-type stars are susceptible to the effects of convection
occurring just below the visible surface.
These effects can be accounted for through the use of 
ab initio 3D hydrodynamic stellar atmosphere 
simulations \citep[e.g.][]{2007A&A...469..687C,Freytag2012}.  
So-called 3D NLTE methods arguably allow for the 
most realistic spectral synthesis and thus most reliable abundance
analysis. However, these methods are extremely computationally expensive; additionally, for many chemical species,
the lack of accurate atomic data makes pursuing this approach even more challenging. Hence, to
date, 3D NLTE abundance analyses remain sparse.

Nevertheless, 
while the majority of the analysis presented in this paper is based 
on 1D NLTE methods (Sect.~\ref{sec:NLTE}), 
3D NLTE abundance results are available
for a handful of important chemical species.  
\citet{2019A&A...622L...4A,2019A&A...630A.104A} recently presented
3D NLTE abundance corrections across the \textsc{stagger}-grid
of 3D model stellar atmospheres
\citep{2013A&A...557A..26M} for the chemical species
\ion{C}{I} and \ion{O}{I} as well as 3D LTE abundance
corrections for \ion{Fe}{II}.

In Fig.~\ref{fig:3dnlte} we show the line-averaged 
3D NLTE abundance corrections
(3D NLTE $-$ 1D LTE abundance differences)
for \ion{C}{I} and \ion{O}{I},
based on the grids from \citet{2019A&A...630A.104A}, interpolated
onto the sample of stars. Unfortunately, these abundance corrections are only
available for the dwarfs and subgiants in the sample
(only four out of the $14$~stars).
For comparison, for \ion{C}{I} and \ion{O}{I,} we also show
1D NLTE abundance corrections that were calculated using
the same atomic models for the entire sample.
We also show 3D LTE abundance corrections for \ion{Fe}{II}
for which abundance corrections are available for the entire sample.

The 3D NLTE abundance corrections for \ion{C}{I}
are slightly, but significantly dissimilar from the 1D NLTE abundance corrections (from the spectrum modelling perspective),
at least according to these particular models and 
for the particular four dwarfs and subgiant shown in Fig.~\ref{fig:3dnlte}.
The mean absolute difference between the 3D NLTE and 1D NLTE abundances
is $0.06\,\mathrm{dex}$, and the largest discrepancy is
$0.10\,\mathrm{dex}$~for BD-08\_619. 
These differences are much smaller for \ion{O}{I}:
The mean absolute difference between the 3D NLTE and 1D NLTE abundances
is $0.01\,\mathrm{dex}$, and the largest discrepancy is
$0.02\,\mathrm{dex}$, again for BD-08\_619. 
It is unknown how large the 3D NLTE effects are for these species,
for the giants in the sample.
\begin{table}[]
    \centering
       \caption{Absolute, 3D LTE or 3D NLTE abundances for Fe, C, and O (and their standard deviation).}
    \label{tab:3D}
    \begin{tabular}{lccc}
    \hline\hline
    Star ID & A(Fe)$_{3\rm{D,LTE}}$ &A(C)$_{3\rm{D,NLTE}}$ & A(O)$_{3\rm{D,NLTE}}$\\ 
    \hline
      BD-08\_619   & 6.74 (0.10) & 8.21 (0.10) & 8.50 (0.03) \\
      BD+09\_2190   & 4.96 (0.10)& --- & 6.70 (0.03) \\
      BD+24\_1676   & 5.17 (0.10) & 6.08 (0.11) &  7.05 (0.01)\\
      HD17492   & 6.88 (0.10) & 7.86 (0.16) & 8.44 (0.01)\\
      2MASS J0023 & 6.05 (0.10) & 6.78 (0.10) &  --- \\   
      \hline
    \end{tabular}
\end{table}

The 1D NLTE abundance corrections for \ion{C}{I} and \ion{O}{I}
are negative for these stars: The 1D NLTE abundances
are smaller than the 1D LTE abundances.  For \ion{O}{I}, the 
3D NLTE abundance corrections (with respect to 1D LTE) are also all
negative: The 3D effect and the NLTE effect go in the same direction
\citep{2016MNRAS.455.3735A}.
For \ion{C}{I,} however, the 3D NLTE abundance corrections
can be positive: The 3D effect and the NLTE effect can 
go in opposite directions \citep{2019A&A...624A.111A}.

The 3D LTE abundance corrections for \ion{Fe}{II} shown in Fig.~\ref{fig:3dnlte} can be larger than the 'typical' upper limit of $+0.15$ dex reported in \citet{2019A&A...630A.104A}. For most stars in our sample, applying these
corrections to our 1D LTE abundances from \ion{Fe}{II} would typically worsen the agreement with \ion{Fe}{I} (i.e. the ionisation balance).
There are several possible reasons for this.  For one, this may reflect neglected 3D NLTE effects in \ion{Fe}{I} lines \citep{2016MNRAS.463.1518A}, keeping in mind that LTE is expected to be a fairly good approximation for \ion{Fe}{II} \citep{2012MNRAS.427...50L}. Another possibility is that it reflects issues with the adopted microturbulence relation. We note that the \ion{Fe}{II} lines used in this study are rather strong, with reduced EWs (log(EW/$\lambda$)) of greater than $-5$\,dex, and the 3D abundance corrections are very sensitive to the adopted
microturbulence parameter in this regime.

In summary, we provide 
3D NLTE abundances for \ion{C}{I} and \ion{O}{I } in this work. 
We use 1D NLTE abundances for the remaining species (including Fe) discussed in
Sect.~\ref{sec:NLTE}, and 1D LTE abundances otherwise (see Figs.\ref{fig:1dnlte}--\ref{fig:3dnlte}, and \ref{fig:all_patterns}). Hence, the 3D NLTE pattern shows two elements, the 1D NLTE up to 17 elements, and finally the 1D LTE pattern includes 32 elements (34 if limits are counted as well). The various methods are never mixed within a pattern.

\section{Results}\label{sec:results}
\begin{figure*}
    \centering
    \includegraphics[scale=0.95]{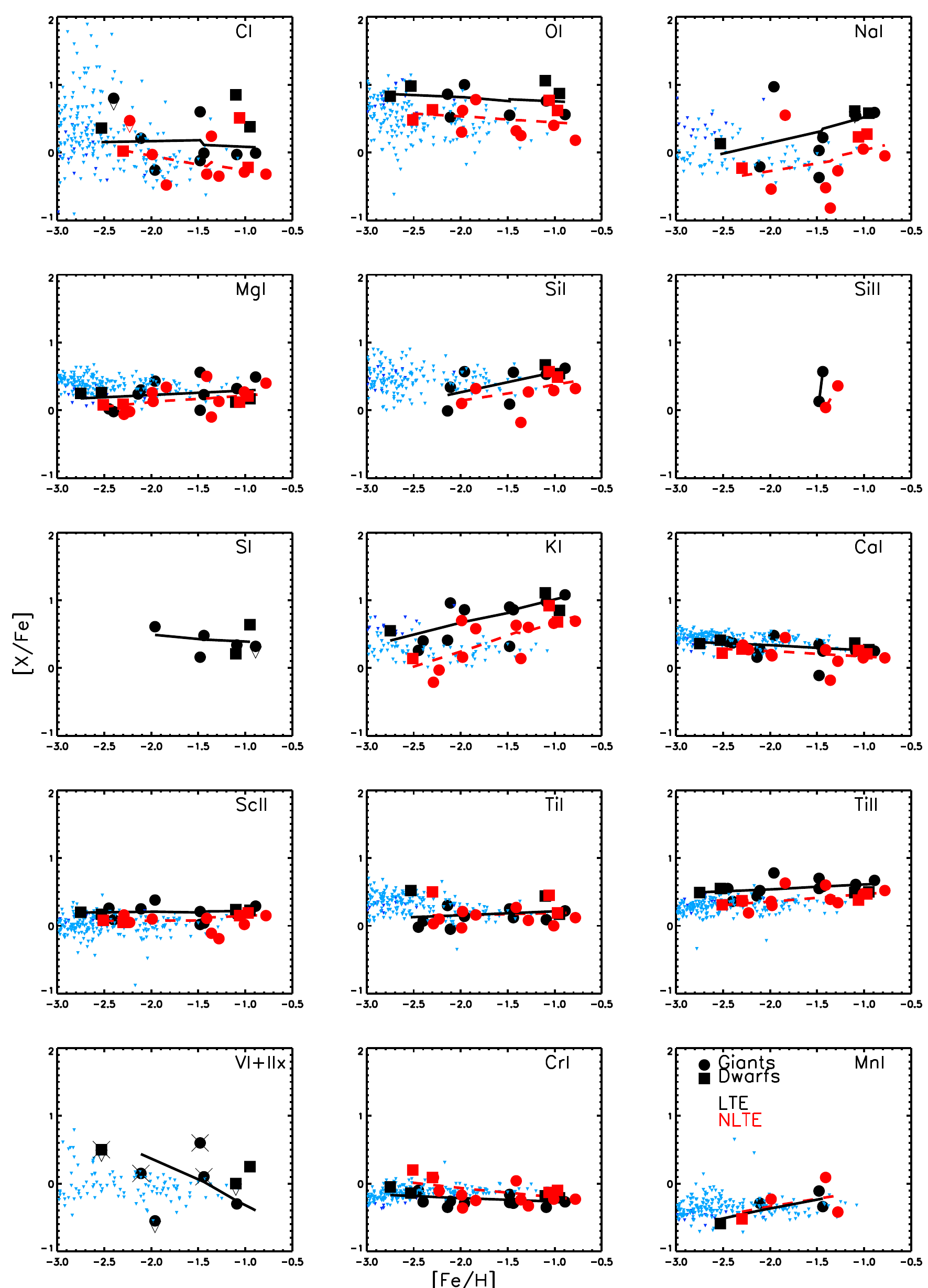}
    \caption{Relative abundances [X/Fe] versus [Fe/H] of element (X) up to Mn. LTE abundances are illustrated by black filled symbols, while NLTE abundances are shown as red symbols, with giants depicted as circles and dwarfs as squares. The 'x' in the vanadium panel indicates stars for which we measured VII rather than VI. LOWESS trends are indicated for LTE (black solid line) and NLTE (red dashed line). In blue, we show the abundances from \citet{Cayrel2004} and \citet{Roederer2014}.}
    \label{fig:light}
\end{figure*}

\begin{figure*}
    \centering
    \includegraphics[scale=0.95]{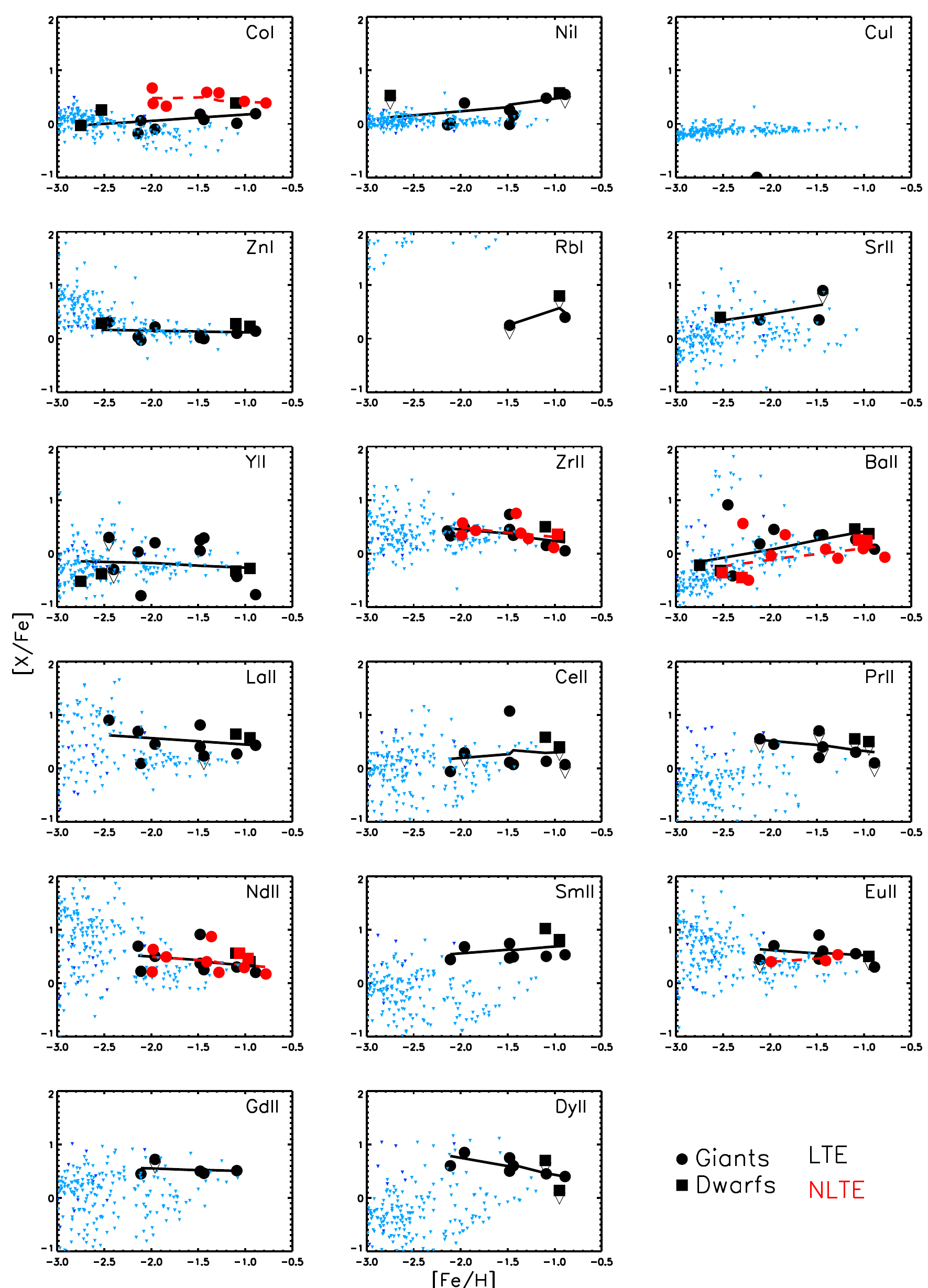}
    \caption{Relative abundances [X/Fe] vs. [Fe/H] of elements X from Co up to Dy. LTE abundances are illustrated in black filled symbols, while NLTE abundances are shown as red open symbols, with giants depicted as circles and dwarfs as squares. Locally weighted scatter-plot smoother trends are indicated for LTE (black solid line) and NLTE (red dashed). In blue, we plot the data from \citet{Cayrel2004}, \citet{Francois2007}, and  \citet{Roederer2014}.}
    \label{fig:heavy}
\end{figure*}

The sample spans a broad range of metallicities, and we therefore briefly discuss the trends in the grand scheme of GCE. For each element in Fig.~\ref{fig:light} and \ref{fig:heavy}, we compare the element to literature samples \citep{Cayrel2004,Francois2007,Roederer2014}; furthermore, to illustrate the difference in LTE and NLTE trends, we overplotted a locally weighted scatter-plot smoothed (LOWESS) line.

Starting with our lightest element Li, we find a value of A(Li)~=~1.92\,dex for BD+24\_1676, which is slightly below  the trend seen for metal-poor dwarfs \citep[][see their Fig. 10]{Sbordone2010}. Due to the temperature of $\sim6300$\,K in BD+24\_1676, there is, however, a good agreement (within 0.1\,dex) with their Li trend in their Fig. 12.

\begin{figure}
    \centering
    \includegraphics[scale=0.5]{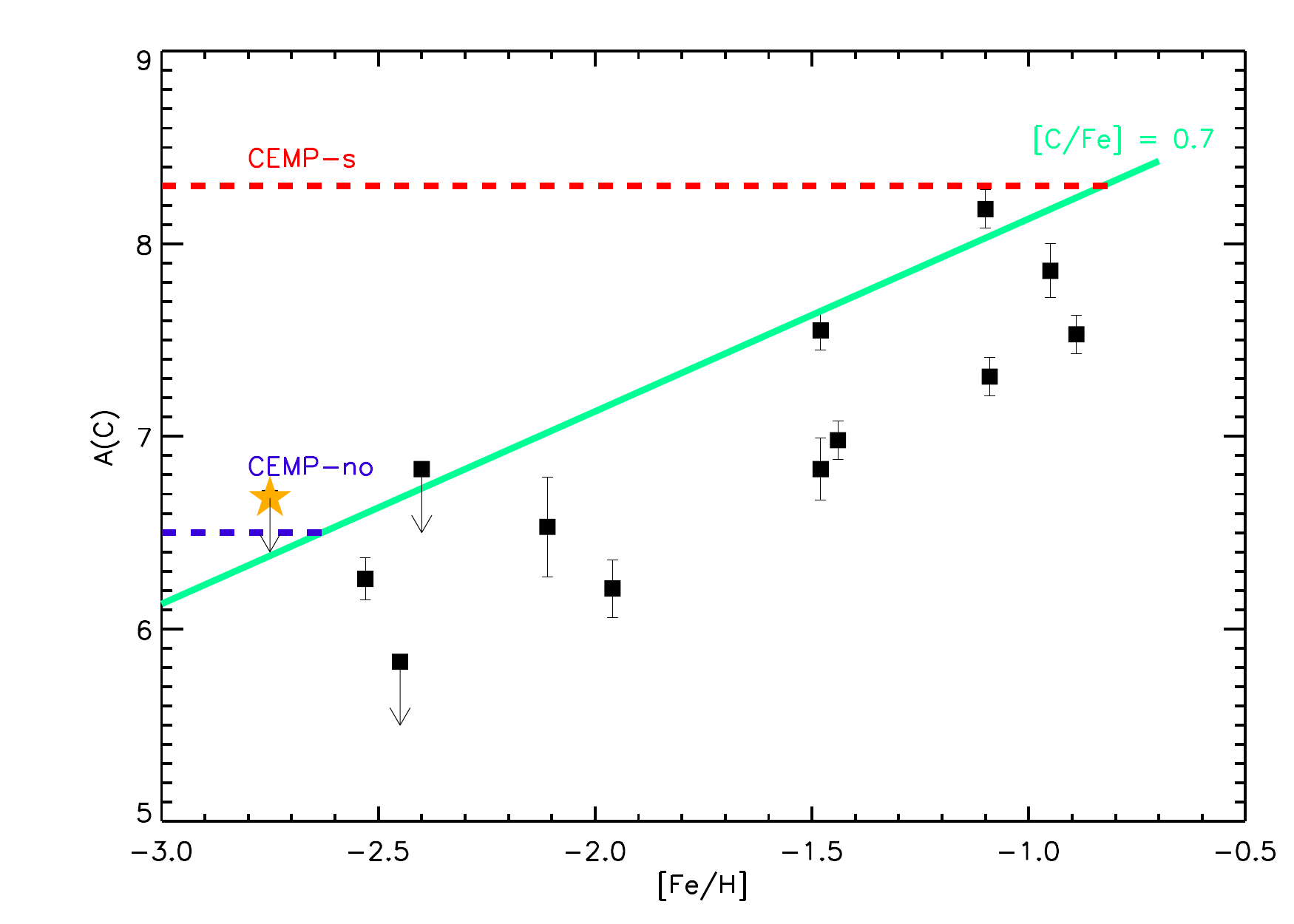}
    \caption{Absolute C abundances versus [Fe/H] (LTE). Indications of CEMP-no (blue line - low A(C)), CEMP-s (red line - high A(C)), and C-normal metal-poor stars (separated by the green line [C/Fe]=0.7) have been indicated. BD+90\_2190 is shown by a star.}
    \label{fig:AC}
\end{figure}

The next element in increasing atomic number is C. The carbon abundances show a large spread, with a C-enriched  metal-poor star (\object{2MASS J0023}), which can be classified as a CEMP star with its [C/Fe]~$\lesssim0.8$ (see Fig.~\ref{fig:AC}). Also TYC5481 with its high [C/Fe] almost classifies as a CEMP star, but it would more likely be a CH star owing to its [Fe/H]. TYC5329 seems to show enhancements in the heavy elements, typical of a CEMP-s or -r/s star, but we could not derive C from this star due to the wavelength coverage. In all dwarfs and subgiants, the 1D LTE abundances are larger than the corrected 1D NLTE values, while the 3D NLTE corrected abundances in three out of four stars almost bring the C-abundances from the atomic red lines ($>9000$\,\AA) back to the 1D LTE level. We note that the C corrections are within our uncertainties. The trend of atomic C in evolved giant stars remains unknown and is needed before the 3D NLTE GCE nature of C can be reliably traced and understood. We also compare our 1D NLTE C abundances to a larger NLTE study from \citet{Zhao2016}. The trend is similar, but we have a larger star-to-star scatter due to the inclusion of CEMP and C-rich stars in our study.

In addition to the atomic C abundances, we also derived C from CH using the G-band. Based on the upper limit we placed for BD+09\_2190 ([C/Fe]~$ \leq 1.0$), it most likely qualifies as a CEMP-no star (using [C/Fe]~$>0.7$ as a criterion for CEMP stars -- see \citealt{Beers2005} and \citealt{Hansen2019} for further classification details). In all cases, the molecular 1D LTE C abundances are lower than what we derived from the near-infrared atomic C lines. Except from one extreme case, BD-08\_619, the average difference between the atomic and molecular C abundances are $\sim0.45$, with the molecular values always being lower. A part of the discrepancy could be due to modelling issues or slightly uncertain stellar parameters, which propagate through and affect the atomic and molecular lines differently. We note that the NLTE corrected \ion{C}{I} agrees with 1D LTE CH within $\sim\pm0.1$\,dex in a couple of cases. According to \citet{Dobrovolskas2013}, the 3D corrections for CH can amount to $0$ to $-0.2$\,dex for giants, while \citet{Gallagher2016} studied the 3D nature in dwarfs and found considerably larger corrections. In the latter study, they found clear dependencies on the stellar parameters, such as [Fe/H], but also on the absolute C abundance (A(C)), where lower A(C) stars are more affected. The 3D A(C) corrections to CH range from 0 to $-0.9$\,dex, where the more metal-poor stars are most affected. However, in order to get our atomic A(C) in 3D NLTE to agree with the future 3D NLTE molecular values, the future NLTE corrections to CH would need to be large and positive ($0.5$ to $1.0$\,dex) so as to compensate for the negative 3D corrections and the offset between molecular and atomic C abundances. 

Oxygen behaves like a typically enhanced $\alpha-$element and follows the well-known trends as shown in \citet{Cayrel2004}, for example. The NLTE corrections clearly lower the average value for O, as seen in Figs. \ref{fig:1dnlte}--\ref{fig:3dnlte}, and \ref{fig:all_patterns}. In the latter figure (\ref{fig:all_patterns}), there is a difference in the 1D NLTE abundances of O, which likely arises due to the use of different codes and treatment of collision rates. As shown for Mn in Fig. 10 of \citet{Bergemann2019}, the use of Kaulakis collisions can reduce or enhance the NLTE correction by $\sim \pm 0.05$\,dex. The overall NLTE trend is in agreement with that of \citet{Zhao2016}, and we see a hint of a knee, but we also have a few stars with low NLTE O values around [Fe/H] $\sim-1.5$. These are discussed further in Sect.~\ref{sec:kinem}.

Sodium shows a gradually increasing trend with metallicity, except for BD-10\_3742, which is very enhanced in Na by a factor   of 10  compared to the Sun in LTE which is reduced to [Na/Fe] = 0.5 in NLTE. This high level of Na is atypical for halo field stars, but it is quite common in second generation globular cluster stars \citep[e.g.][]{Carretta2009}. However, the low C/O ratio could also indicate a very evolved giant that has undergone severe stellar mixing events \citep[see, e.g.][]{Placco2014b}. Generally, the star-to-star scatter is fairly large and the NLTE corrections clearly reduce the average Na value. The large scatter was also shown in \citet{Zhao2016}.

Both Mg and Si (neutral and ionised) show typical enhanced levels in metal-poor stars. A few peculiar cases show very high [Mg/Fe]$_{\rm LTE}\sim0.5$\,dex (BD-15\_779) and  very low ($\sim0$) values have been derived for TYC5481-00786-1, 2MASS J0023, and HE0420+0123a (see Sect.~\ref{sec:kinem} for details). In BD+09\_2190, we find a high [Mg/Fe] of $\sim0.6$\,dex from the 4571\,\AA\ line in the PEPSI spectra, while high-resolution, high S/N archival UVES  spectra yield a considerably lower Mg abundance by $0.3-0.5$\,dex for 5172, 5183, and 5528\,\AA. Here, we provide an average value for all four lines, both in LTE and NLTE. Both NLTE averages and trends of Mg and Si are in agreement with the findings of \citet{Zhao2016}; however, due to our smaller sample size and inclusion of chemically peculiar stars, we find some outliers and generally detect less of a knee in our abundances.
As is seen in the large sample of sulphur by \citet{Duffau2017}, our trend agrees well with theirs and spans broadly from [S/Fe]~$\sim0.1-0.6$\,dex.

Potassium in our sample is fairly high and, despite all efforts to avoid blends, a few cases might suffer from their effect, returning values slightly above literature values \citep{Cayrel2004,Reggiani2019}. However, below [Fe/H] = $-2$, our values drop ([K/Fe] $< 0.5$) and they are in good agreement with the listed literature studies. In comparison to \citet{Zhao2016}, we find a larger spread and a higher NLTE abundance average for K. This could indicate that unnoticed blends might interfere more at higher metallicities. The GCE behaviour is clearly different in LTE and NLTE (see Fig.~\ref{fig:light}).

Calcium shows a clean trend with little scatter as is generally seen in the literature. We note that one star has a particularly low [Ca/Fe] (see Sect.~\ref{sec:discussion}--Sect.~\ref{sec:kinem}). Similarly to Ca, Sc shows a flat trend with low star-to-star scatter, which is also the case for neutral and ionised Ti. This is in agreement with the NLTE results from \citet{Zhao2016}; however, we do not detect an upturn in Sc around [Fe/H] $=-1$ and our average is $\sim0.1$\,dex higher than what they find for \ion{Sc}{ii} and \ion{Ti}{ii,}  which is well within the uncertainty. We note that the average \ion{Ti}{i} and \ion{Ti}{ii} differ in most cases, as we did not enforce ionisation equilibrium, unless the  gravities determined via parallaxes happen to satisfy this.

Vanadium was derived in five stars and an additional three show upper limits. The trend is very spread with metallicity. This was also noted in \citet{Roederer2014} who studied \ion{V}{i} and \ion{}{ii} as a function of temperature and [Fe/H]. Similar to them, we find higher \ion{V}{ii} than \ion{V}{i} (see 'x' in Fig.~\ref{fig:light}). They also note the lack of atomic data for hyperfine splitting (hfs) for \ion{V}{ii}; however, if we could implement the missing, hyperfine split oscillator strength, the abundances of \ion{V}{ii} would likely decrease. The magnitude of the effect also depends on the strength of the lines, which in some stars amount to $\sim100$\,m\AA\ and may cause significant changes in the abundances if hfs could be included. In both cases, a part of the explanation could be that three out of the four stars with high \ion{V}{ii} do not show ionisation equilibrium. We note that \ion{Ti}{i} and \ion{}{ii} also exhibit a similar trend with higher \ion{Ti}{ii} values. The difference in \ion{Ti}{i} and  \ion{}{ii} can, in part, be explained by the \ion{Ti}{i} value being driven by one or two lines.

In LTE, both Cr and Mn show subsolar trends ([X/Fe]~$<0$), which increase with metallicity that is in agreement with \citet{Cayrel2004} and \citet{Bonifacio2009}. The NLTE corrected [Cr/Fe] and [Mn/Fe] are higher. The 3D NLTE abundances would be even higher than the abundances derived using the 1D NLTE approach  \citep[as also seen in][]{Eitner2020, Bergemann2019}.

Unlike the incomplete Si-burning elements, Co, which is produced in the complete Si burning, shows slightly enhanced abundances ($\sim0.1$\,dex in LTE c.f. Fig.~\ref{fig:heavy}). Nickel seems to show two levels, one around Solar and one enhanced at higher metallicities. For copper, we only present an upper limit, which is about 0.4\,dex lower than the metal-poor stars in \citet{Zhao2016}. Zinc on the other hand, stays flat around 0 and is in good agreement with the analysis of \citet{Duffau2017}.

The first heavy element ($Z>30$) we analyse is Rb. The abundances are based on the 7800\,\AA\ line, which is hard to analyse accurately as it suffers from a strong Si-blend and possible telluric contamination as well. Few stellar abundances have been published and we add one detection and two upper limits to the literature, all of which are slightly enhanced.

The three elements Sr, Y, and Zr show the typical trends with Sr and Zr abundances that are higher than those of Y. The GCE trends are in good agreement with \citet{Francois2007} and \citet{Hansen2012,Hansen2013}. Owing to the lack of observations in the blue-most PEPSI CD1 setting, only four stars have wavelength coverage of the Sr lines. For Zr, we find an excellent agreement with the NLTE Zr abundances from \citet{Zhao2016}.

The heavy elements between Ba and Sm show a large star-to-star scatter \citep{Francois2007,Hansen2012}. We highlight that Pr and Sm show particularly flat trends with lower scatter; however, this could be due to our limited samples size.
Similarly, Gd--Dy seem to yield a flat trend at a slightly enhanced level. Generally, the heavy elements are derived from the majority species and are hence less biased by the LTE approximation. Therefore, the LTE and NLTE abundances of the four heavy elements NLTE corrected here (Zr, Ba, Nd, and Eu) show similar GCE behaviours. The NLTE study by \citet{Zhao2016} shows that Ba was widely spread around zero, which is similar to what we find. Also our Eu NLTE abundances are in good agreement with their trend.

Finally, we report upper limits of Pb ([Pb/Fe]$_{\rm LTE} \lesssim0.8$) and Th ([Th/Fe]$_{\rm LTE} \lesssim0.5$) in BD-07\_163.\ Additionally, we note that these values are very uncertain and we refrain from using them further.

\section{Discussion}\label{sec:discussion}
Here, we discuss  stars with peculiar abundance ratios  and focus on their heavy element patterns. To explore the origin of the neutron-capture elements, we compare their abundances to AGB yield predictions \citep{Cristallo2011,Cristallo2015} and assess s-process contamination at higher metallicities. We explore the r-process using stars with stellar parameters that are representative of our sample with a well-studied r-process pattern. For this purpose we use HD20, a metal-poor giant ([Fe/H] $= -1.60$) in which 48 elements (58 species) including 28 neutron-capture elements have been studied, making it an excellent benchmark star for the study of heavy elements. Moreover, this star has extremely accurate stellar parameters \citep[e.g. asteroseismic gravity, see][for details]{Hanke2020}, slight r-process enhancement, and the r-process contribution to each element has subsequently been filtered at an absolute level, log $\epsilon(X)$ + log $\epsilon (X_r$), where the r-fraction is taken from Table 5 in \citet{Hanke2020}. This enables a purer comparison to the r-process at a lower metallicity than the Sun. Hence, we use the pattern of HD20 as a 'metal-poor Sun' with a more pure r-process representation, owing to its lower metallicity and r-pattern instead of comparing it to a biased solar scaled r-process pattern.

\begin{figure}
    \centering
    \includegraphics[scale=0.35]{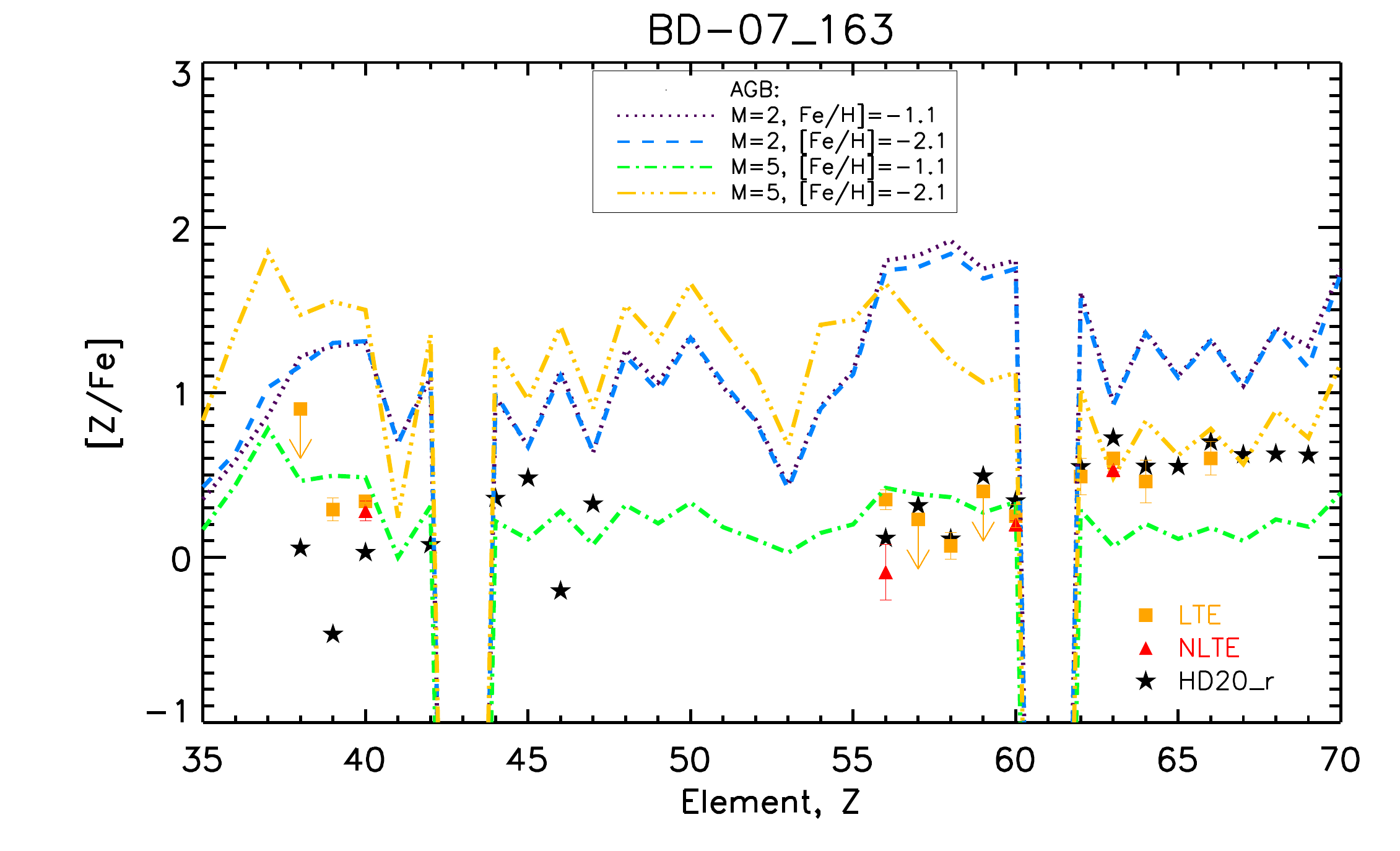}
    \includegraphics[scale=0.35]{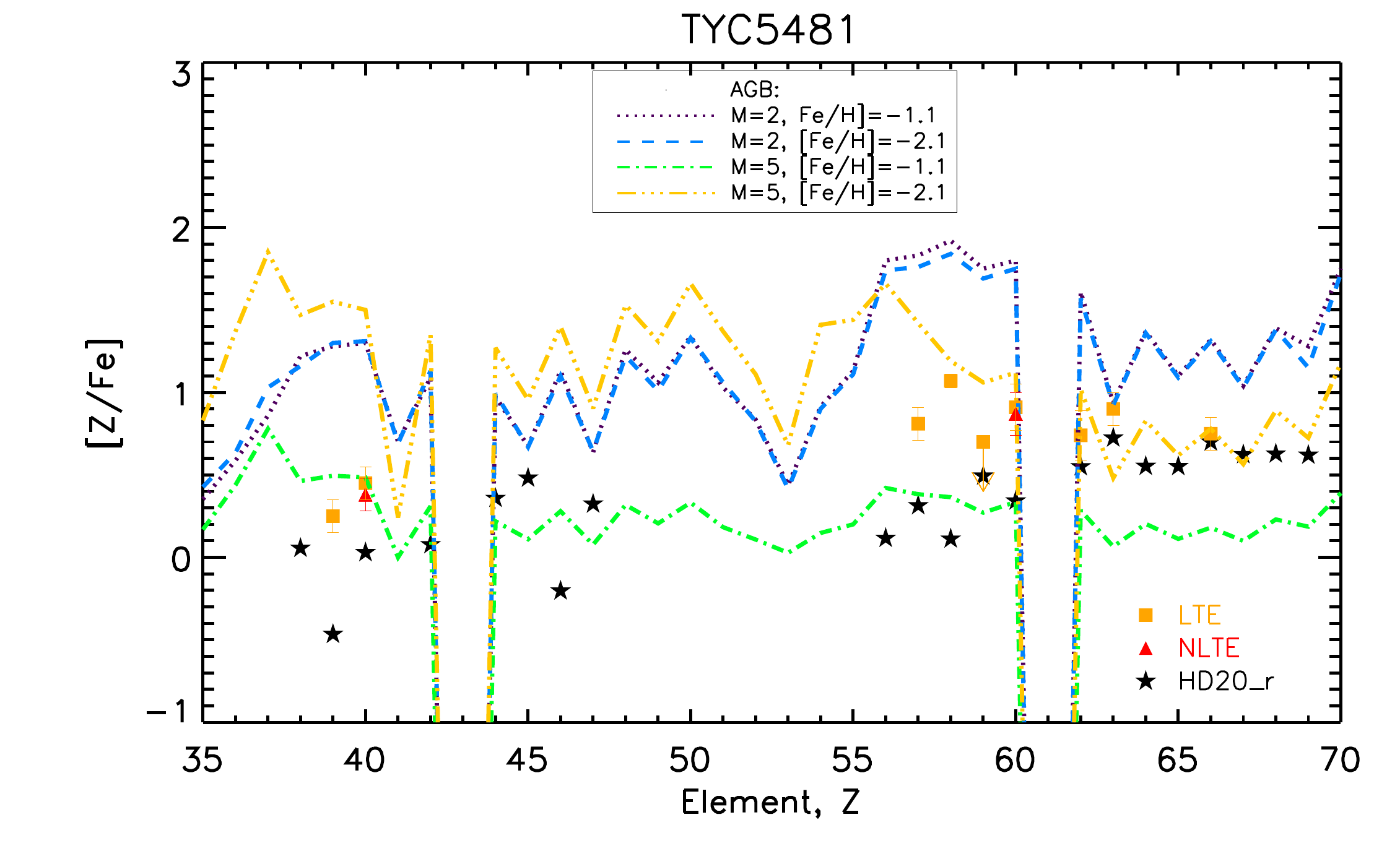}
    \includegraphics[scale=0.35]{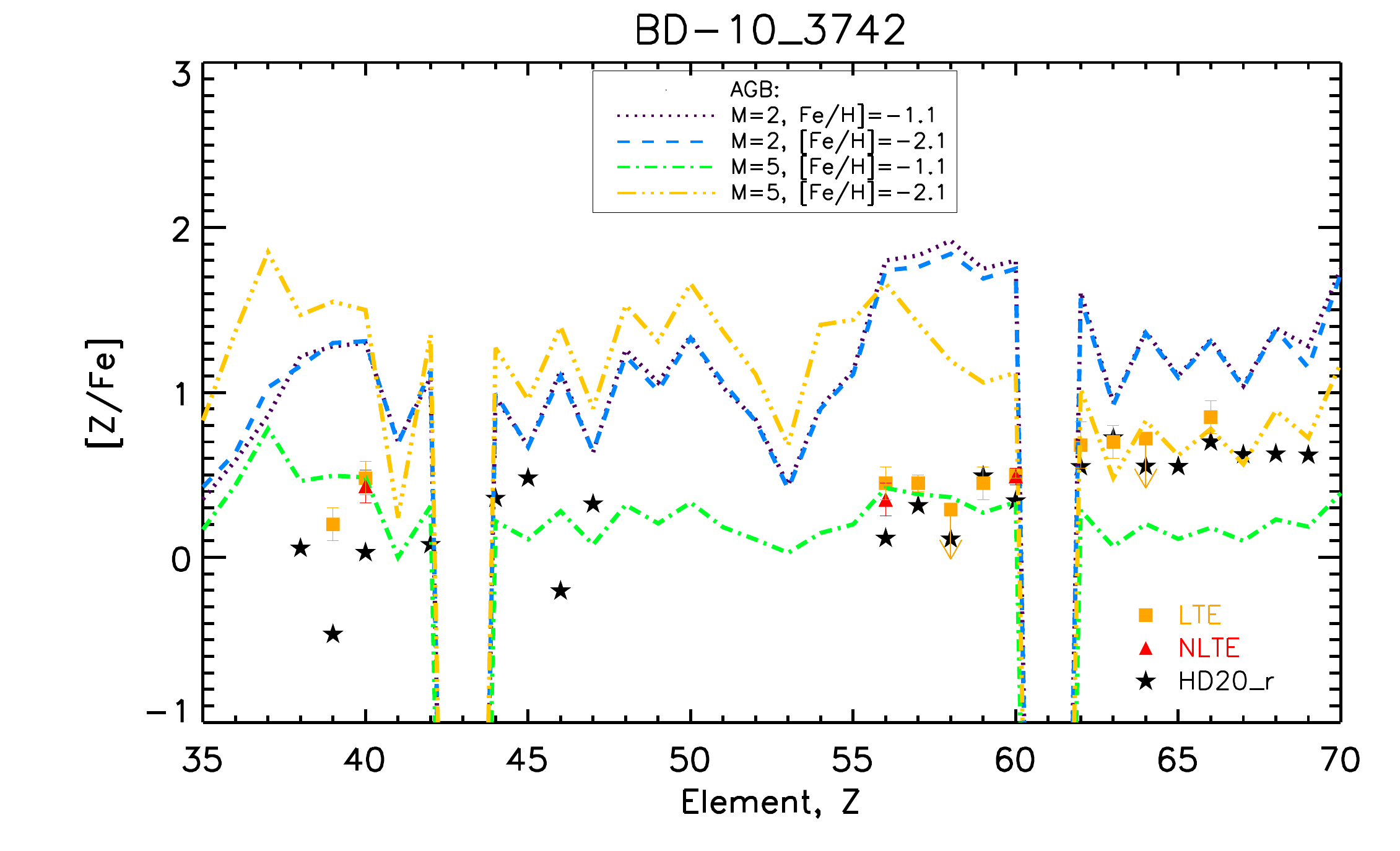}
    \caption{Abundance patterns in 1D+LTE, 1D+NLTE compared to the pure r-process representation of HD20 (LTE), and the s-process yields from the rotating AGB stars with 2 and 5\,M$_\odot$ and [Fe/H] = $-1.1$, and $-2.1$, respectively. }
   \label{fig:yields}
\end{figure}

The most metal-rich stars in our sample, [Fe/H] $>-1$, are clearly enriched from a highly mixed gas by numerous different events (see Sect.\ref{sec:mono}).
Hence, we only attempted to trace AGB and r-process contributions in stars with [Fe/H] $\sim-1.4$ and below. In order to do so, we computed the $\chi^2$ between the observations and the yields, using the line-to-line scatter listed in the online Table as uncertainty. 
From Fig.~\ref{fig:yields} and the computed $\chi^2$, we find that BD-07\_163 and especially TYC5481 are strongly polluted by the s-process. None of the models that are contrasted here provide a good fit to all of the derived abundances. In general, the heavier elements tend to show an upward trend in Fig.~\ref{fig:yields}. This is somewhat driven by Dy, which from Fig.~\ref{fig:heavy} is also seen to be high, so we cannot exclude that we overestimated our Dy abundances a bit. However, based on spectrum synthesis, we estimate that this is on the $0.1$\,dex level. From the yield comparison, we see that the lighter elements
could have been formed by a massive (5M$_{\odot}$), fast rotating (V$_r~=~30$\,km/s), metal-rich ([Fe/H] $=-1.1$) AGB star. Both BD-07\_163 and TYC5481 show heavy element contamination that is better fitted by the more metal-poor AGB star, so the metallicity of BD-07\_163 may prevent a direct trace of the s-process enrichment and it may be multi-enriched. However, TYC5481 with an A(C)~=~7.55, [Fe/H] $\sim-2$ and [La/Eu] $=-0.1$, using La as a proxy for Ba and in taking into account that La is an odd element which typically yields lower abundances than the even elements (Ba), this star is likely a CEMP-r/s star.\ This also explains why neither the pure s- nor r-patterns provide a satisfactory fit on their own to the observations of TYC5481.  
Meanwhile, the metal-poor giant BD-10\_3742 shows a pattern that strongly resembles the r-fraction of HD20, and it likely has predominantly been enriched in the heavy elements by the r-process.
Our metal-poor dwarf star, BD+24\_1676, shows a mixed r- and s-process pattern
 despite the low metallicity (see also Sect.\ref{sec:mono}).

\subsection{Chemically peculiar stars}
\begin{figure}
    \centering
    \includegraphics[scale=0.4]{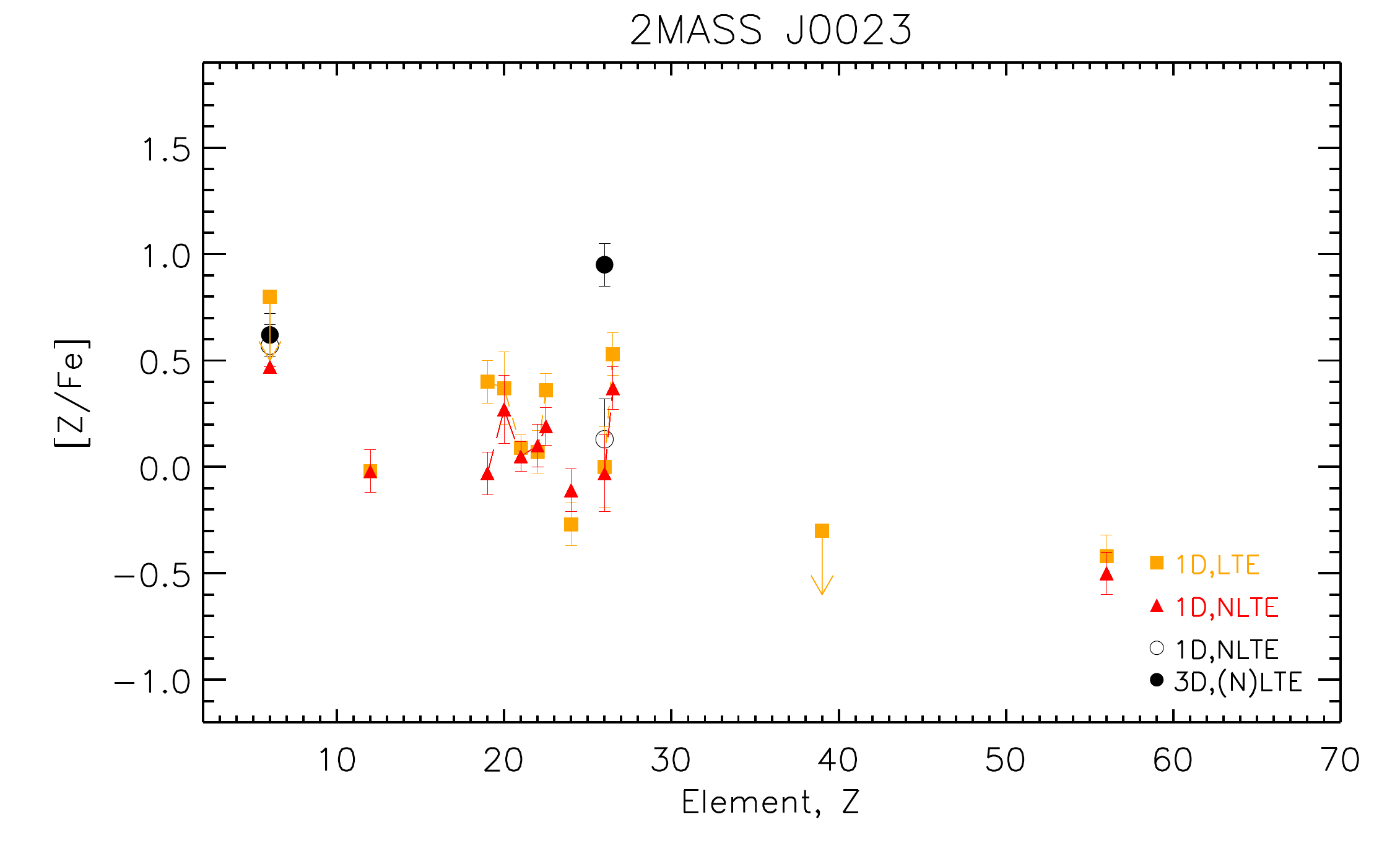}
    \caption{Abundance patterns in 1D+LTE, 1D+NLTE, and 3D+NLTE of 2MASS J0023. }
   \label{fig:2mass_pattern}
\end{figure}

Unfortunately, neither Sr nor Eu could be derived from the spectra of 2MASS J0023, so classifications adopted in \citet{Beers2005} or \citet{Hansen2019} cannot be applied directly. However, with both [Y/Fe]~$<-0.3$ and [Ba/Fe]~$=-0.42$ being subsolar, 2MASS J0023 is very likely a CEMP-no star, as indicated by the limit on A(C) (Fig.~\ref{fig:AC}). This is in line with the lack of heavy element detections. The pattern is shown in Fig.~\ref{fig:2mass_pattern}, while the remaining patterns in 1D LTE and NLTE as well as 3D NLTE are shown in the online Fig.~\ref{fig:all_patterns}.

Moreover, the high [C/Fe] (in LTE) and heavy element content of TYC5481 likely mean that it is a CEMP-r/s star; however, the subclassification is more uncertain since we are missing key elements (Ba) for this purpose. TYC5329 could be a CEMP-s star due to the s-process enhancement in this star; however, we could not derive C here. 
Also BD-08\_619 is C rich ([C/Fe] = 0.85) if we trust the atomic C lines; however, the G-band indicates a subsolar [C/Fe], and further C corrections are needed to fully understand this star. However, with the higher metallicity, it more likely belongs to the CH group than the CEMP stars.

Finally, we note a broad spread in [$\alpha$/Fe] with several low-$\alpha$ stars and a few above 0.5. In Sect.~\ref{sec:kinem}, we explore their kinematics in order to probe if the stars could be accreted into the Milky Way \citep[see e.g.][]{Nissen2010, Bergemann2017b,Hansen2019}. Low-$\alpha$ halo stars have likely been accreted from dwarf galaxies, which, compared to the MW, 
offer less gas and in turn form a larger population of lower-mass supernovae, thereby explaining the lower $\alpha-$abundances \citep[see e.g.][for different views]{McWilliam2013,Reichert2020}.

\subsection{Kinematics of chemically peculiar stars}\label{sec:kinem}
For a sub-sample of the chemically most intriguing stars with high or low $\alpha-$content and for the (candidate) CEMP stars 
(BD+09\_2190, BD-15\_779, HE0420+0123a, 2MASS J0023, and TYC5481), we performed an orbital analysis
using proper motions and parallaxes from the second data release (DR2) of Gaia 
\citep{GaiaDR2} as well as their radial velocities. For the remaining sample, we confirm that these are `normal' disc and inner halo stars.
As in \citet{Hansen2019}, stellar orbits were integrated backwards for 12 Gyr in a simple 
Galactic potential that consists of a 
logarithmic halo and spherical bulge \citep{Fellhauer2008} and a disc model by \citet{Dehnen1998}.
For comparison purposes, in Fig.\ 10, we show a Toomre diagram including 
the CEMP-no and C-normal stars from \citet{Hansen2019}, where a negative rotational velocity 
component $V$ indicates retrograde orbits and T denotes the combined 
vertical and radial components relative to the local standard of rest (LSR). 
As three of the stars show deviations
from the LSR of more than 210 km\,s$^{-1}$, we consider them to be typical halo objects
\citep{Koppelman2018}. Two of the stars, BD-15\_779 and TYC5481, at intermediate LSR velocities between $\sim$100 -- 200 km\,s$^{-1}$ could be 
assigned to the metal-poor tail 
of the thick disc \citep{Kordopatis2013} with their [Fe/H] $\sim-1.5$; while at $Z_{\rm max}\la1.7$ kpc,
an overlap with an inner halo cannot be excluded.

Our CEMP-no candidate 2MASS J0023 is of particular interest, where
\citet{Hansen2019} asserted that (79\%, 45\%, and 64\%) of their CEMP-no stars were characterised by 
($e~>$ 0.5, $R_{\rm apo}<$ 13 kpc, and $R_{\rm peri}>$ 3 kpc); all of the criteria are fulfilled 
by this star. This points to it being a typical contender for its class (CEMP-no). 
\begin{figure}
    \centering
    \includegraphics[scale=0.45]{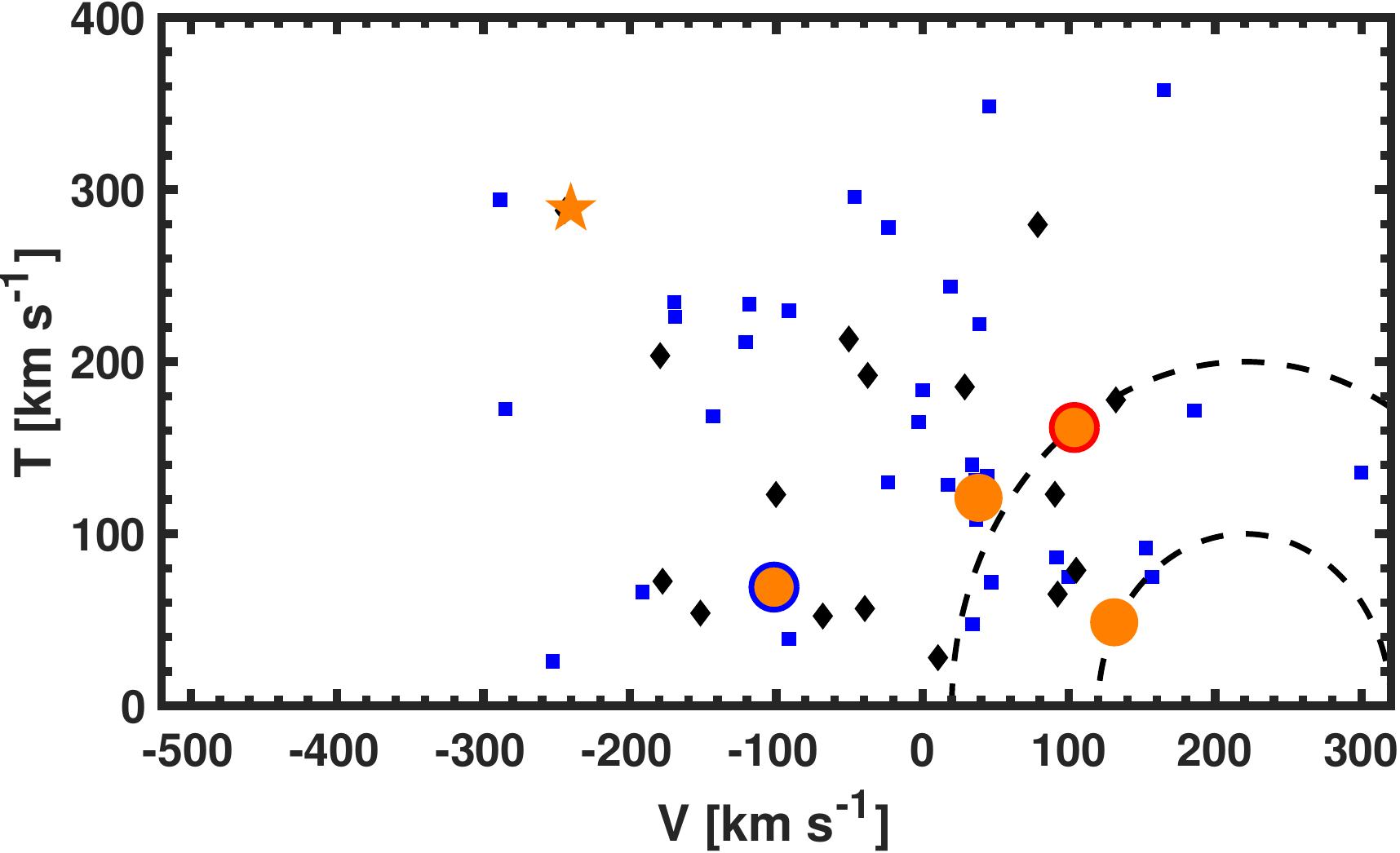}
    \caption{Toomre diagram for a subset of our  sample (orange circles)
in comparison with the sample of CEMP-no (blue squares)
and C-normal metal-poor halo stars (black diamonds)
from \citet{TTHansen2015,TTHansen2016no,Hansen2016}.
Only stars BD+09\_2190 (star), BD-15\_779, HE0420+0123a, 2MASS J0023, and TYC5481 are shown for their chemical peculiarities. 
The CEMP-no candidate 2MASS J0023 is singled out by a large blue circle and TYC5481, a CEMP-s candidate, is shown by a red circle.
The dashed circles correspond to a space velocity relative 
to the local standard of rest of 100 and 200 km\,s$^{-1}$, 
centred on $V_{\rm LSR}$ = 232 km\,s$^{-1}$.
}
    \label{fig:Toomre}
\end{figure}

\begin{figure}
    \centering
    \includegraphics[scale=0.45]{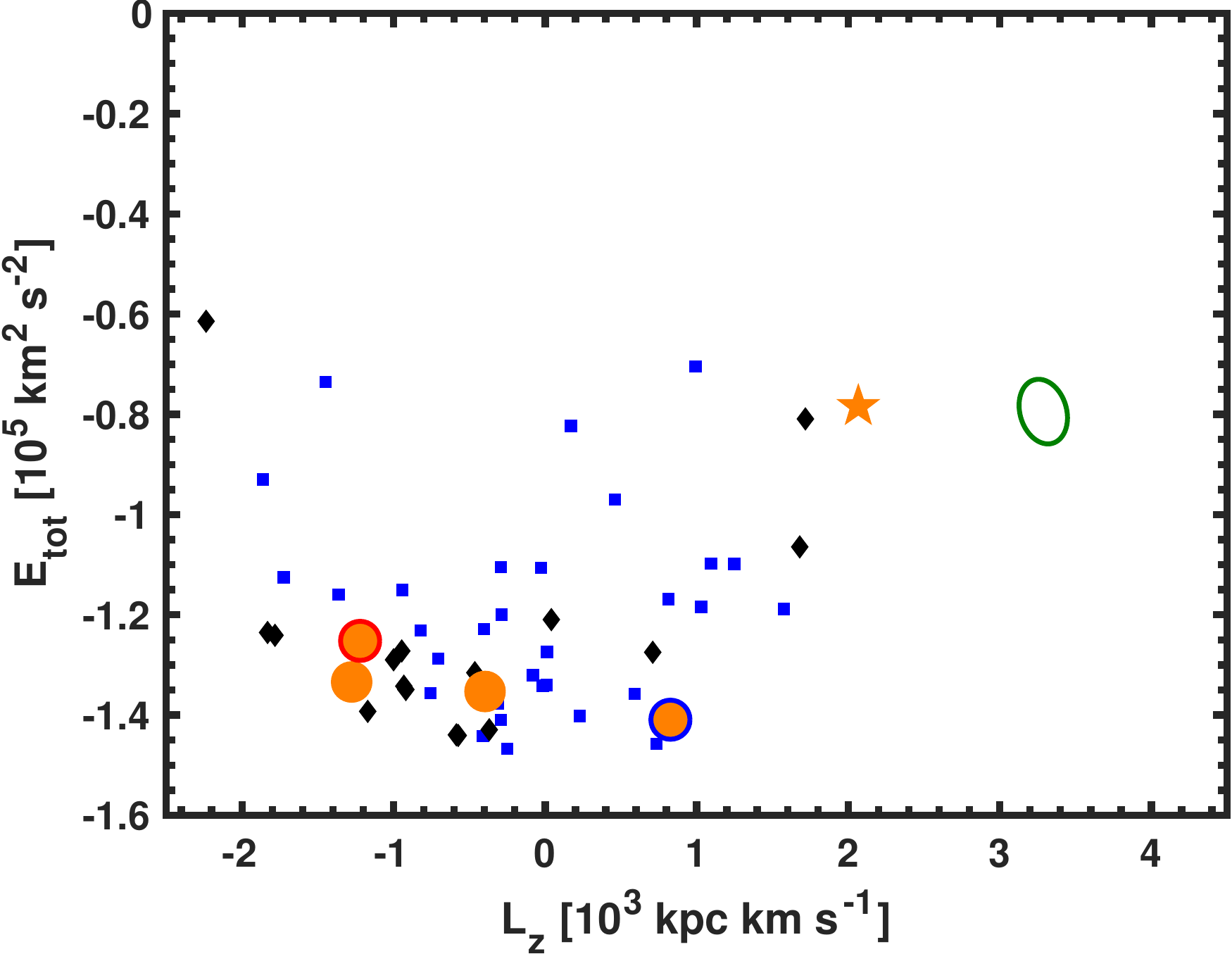}
    \caption{Lindblad diagram for the same stars as in Fig.~\ref{fig:Toomre}.
The green ellipse [at ($L_z$,$E_{\rm tot})\sim (3200$ kpc\,km\,s$^{-1},-0.8\times 10^5)$ km$^2$\,s$^{-2}$] 
illustrates the location of the high-energy, counter-rotating Sequoia merger candidates \citep{Myeong2019}. The same symbols are used as in Fig.~\ref{fig:Toomre}.
}
    \label{fig:Lindblad}
\end{figure}

Secondly, star BD+09\_2190 deserves special mentioning, in particular, in light of its peculiar
enrichment patterns and low $\alpha-$content hinting at a mono-enrichment event (see Sect.~\ref{sec:mono}-\ref{sec:constr}).
At an apocenter distance of 62 kpc and an orbital ellipticity of 0.82, it is an excellent candidate
for an outer halo object. With the Lindblad diagram in Fig.~\ref{fig:Lindblad}, we attempt to 
identify a possible progenitor. 
Here, we show the total specific orbital energy and the 
specific angular momentum in terms of the azimuthal action $L_z=-J_{\phi}$, both of 
which are integrals of motion in stationary, axisymmetric potentials and thus 
suited to identify coherent groups of accreted systems \citep[e.g.][]{Gomez2010,Roederer2018}.
Recently, \citet{Myeong2019} identified a major accretion event in phase space, dubbed `Sequoia', 
which brought in a large number of stars and globular clusters on high-energy, retrograde orbits
(see also \citealt{Massari2019,Koch2019}). Its location in action space is indicated in Fig.~\ref{fig:Lindblad}.
While the star BD+09\_2190 (at (2,-0.8) in Fig.~\ref{fig:Lindblad}) is rotating more slowly than the average Sequoia-component, its energy is 
comparable. However, and most importantly, its [Mg/Fe]$_{\rm LTE}\sim 0.3$ and [Fe/H]$_{\rm LTE}\sim -2.75$ are compatible with a metal-poor 
extension of Sequoia's metallicity and abundance distribution as shown in \citet{Myeong2019} using LTE abundances. 
It is worth noting, however, that our NLTE abundances of this star would rather put this view into question. 
While a unique identification of this star's origin in this particular event is therefore pending, 
an accretion of BD+09\_2190 is irrefutable through its chemo-dynamics. 

\subsection{Mono-enrichment}\label{sec:mono}
\begin{table*}[!ht]
    \centering
        \caption{Constrained and unconstrained best fitting models to the two most metal-poor stars. The reported value is the median, the limits are the 2.3 and the 97.7th percentile corresponding to 2$\sigma$ lower and upper limits. The limits have been left empty where they are identical to the median value, which indicates that the limit is not well-resolved in the posterior distribution. 
    }
    \label{tab:starfit}
    \begin{tabular}{l cccc | cccc}
    \hline \hline
 &\multicolumn{4}{c}{BD+09\_2190 ([Fe/H]~=~$-2.75$)}&\multicolumn{4}{c}{BD+24\_1676 ([Fe/H]~=~$-2.53$)}\\
 \hline
Abund.$/$Model  & \multicolumn{2}{c}{Mass [M$_{\odot}$]} & \multicolumn{2}{c}{Energy [Foe]} & \multicolumn{2}{c}{Mass [M$_{\odot}$]} & \multicolumn{2}{c}{Energy [Foe]}  \\
log$\epsilon$ & Unconst. & constr.  & Unconstr. & constr.  & Unconstr. & constr.  & Unconstr. & constr. \\
\hline
LTE   &$25.5$   &  $25.5^{26.5}$   &  $5.0$ & $3.0_{2.4}$ & $25.5^{26.5}$   &  $25.5^{26.5}$  & $3.0^{5.0}_{2.4}$ & $1.5$\\
LTEred& $25.5$   &  $25.5^{26.5}$   &  $10.0_{5.0}$ & $3.0_{2.4}$&  $25.5^{26.5}$ & $26.5_{25.5}$  & $3.0_{2.4}^{5.0}$  & $1.5$\\
NLTE  & $11.2_{10.6}^{26.5}$   &  $19.2_{18.1}^{20.5}$   & $0.6_{0.3}^{10.0}$  & $1.5^{1.8}$ & $27.0_{23}^{28.5}$   &  $45.0_{42.0}$   &$10 $ & $5.0_{3.0}$\\
\hline
    \end{tabular}
\end{table*}
Here, we explore the question as to whether the most metal-poor stars could have been enriched by a single supernova event, making them mono-enriched, true second generation stars, or if they are multi-enriched by several different events.
As suggested by \citet{Hartwig2018}, this can, at first glance, be observationally assessed by inspecting the [Mg/C] ratio as a function of [Fe/H]. This test allows one to asses whether a star, provided it is a second-generation star, is likely enriched by one or several SNe. A mixture of random SNe is far less likely to result in stars that pass this test than a random individual SN. This is, however, no guarantee that the star actually is a second generation star, and we therefore conducted a constrained comparison to a large grid of supernova model predictions using a Bayesian approach \citep[for details see][or Section \ref{sec:constr}]{Magg2020}.
\begin{figure}
    \centering
    \includegraphics[scale=0.62]{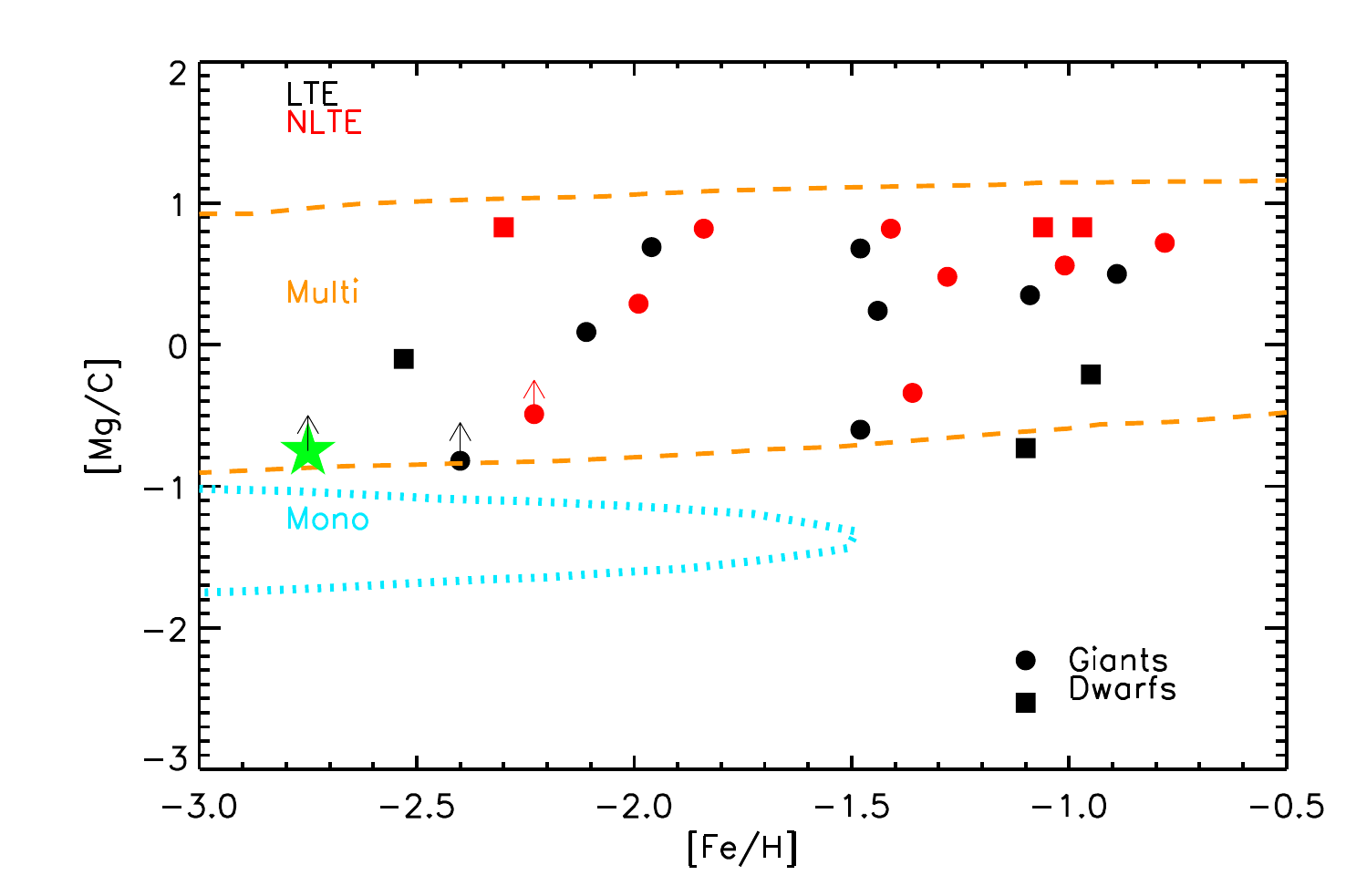}
    \caption{[Mg/C(I)] vs. [Fe/H] in LTE (black) and NLTE (red). The symbols show dwarfs and giants, while the enclosed dashed and dotted area indicates if the stars are likely to be multi- or mono-enriched (true second generation stars), respectively. The green star represents BD+09\_2190 (LTE).}
    \label{fig:MgC}
\end{figure}

Figure~\ref{fig:MgC} shows that there are no obvious candidates, but our two CEMP-no stars (2MASS J0023 and BD+09\_2190) are right at the limit of multi-enrichment, and they are furthermore sufficiently metal-poor to be potential mono-enriched stars. However, the observations require that both C and Mg be measured, which excludes the two most metal-poor dwarfs with [Fe/H]~$<-2.5$ from Fig.~\ref{fig:MgC}, but we retain 
them in the following detailed enrichment analysis.  
For this purpose, we used the 1D LTE and 1D NLTE abundance patterns to contrast the SN yields, and we note that the 3D NLTE abundances for C and O fall within the NLTE {error bars}. Moreover, to fit the models, we used the statistical errors on the observationally derived abundances.

A simple comparison to the grid of supernova models from \citet{Heger2010} could indicate, at first glance, that the metal-poor stars BD+24\_1676 and BD+09\_2190 (in LTE) would be enriched by a massive supernova of 25.5\,M$_{\odot}$. For BD+24\_1676, the inferred mass from the best fit SN model with the minimum $\chi^2$ and the $2\sigma$ limits agree within 1\,M$_{\odot}$ and they yield consistent progenitor masses and a slightly broader range of explosion energies. This consistency could misleadingly indicate a good match to the model, which is not the case (see Sect.~\ref{sec:constr}). Similarly, for BD+09\_2190, the mass varies very little but the energy varies by a factor of two, pushing it into the energetic hypernovae regime. This is the outcome of simply using different models and adopting different mixing, dilution, and ejected masses.
We note that since not all elements can be NLTE corrected, the NLTE pattern always, or so far, consists of fewer element abundances than the more complete, but biased, LTE abundance pattern. Hence, we used three sets of patterns, namely a complete set of LTE abundances ('LTE'), a set of LTE abundances reduced in number of elements to match the NLTE case ('LTEred'), and finally the NLTE pattern ('NLTE').

\begin{figure*}
    \centering
    \includegraphics[scale=0.8]{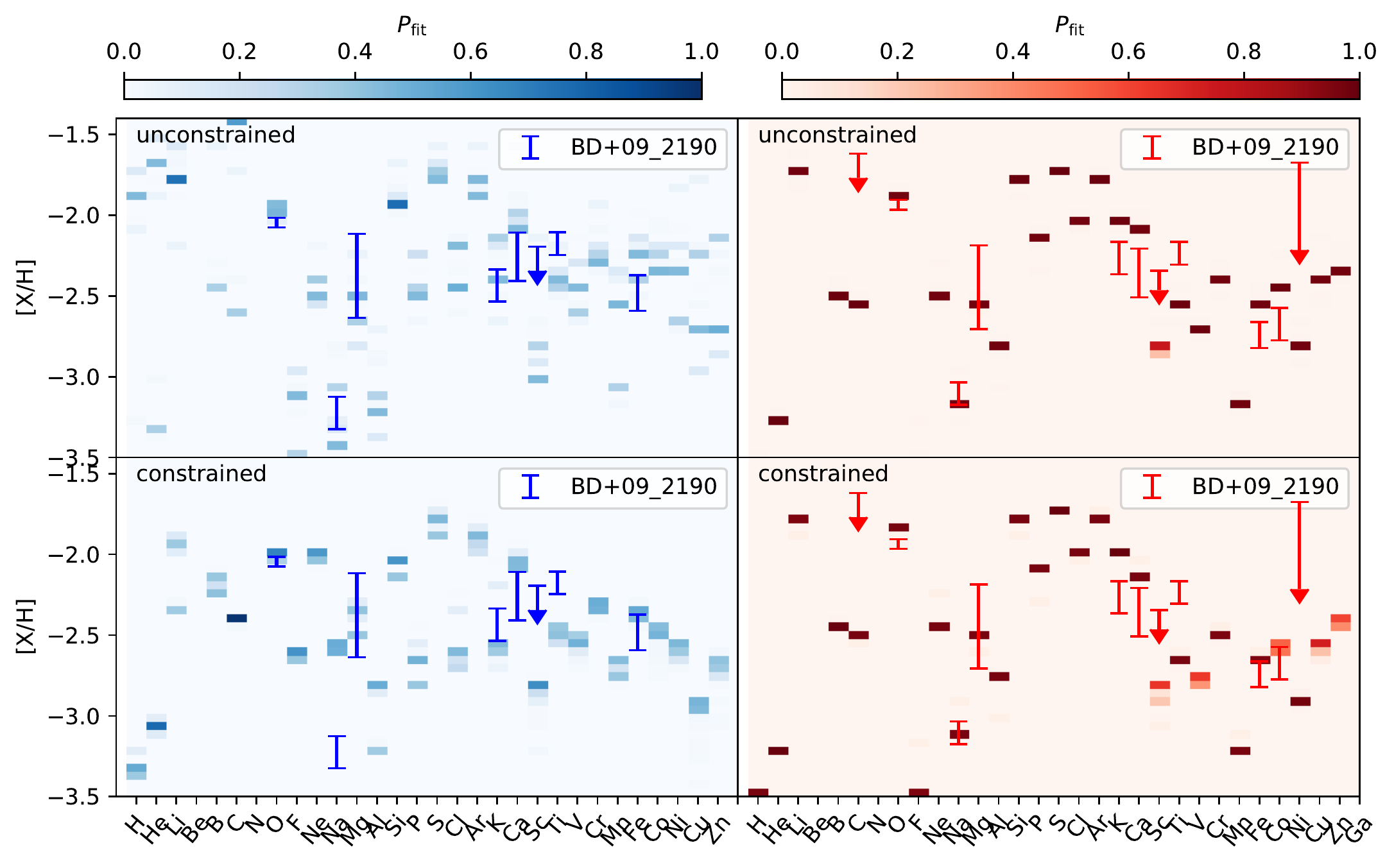}
    \caption{Bayesian inference of best fit models to all elements in the range of H to Zn. The column of dashes above each element represents the  posterior distribution of abundances in that element as obtained from fitting the abundance patterns to the \citet{Heger2010} SN II yields. Top panels show the unconstrained approach, while the two bottom panels show the constrained counter part. Blue indicates NLTE, while red shows LTE. The normalised probability of any model fitting is indicated by the colour bar.  Observations are shown as symmetric error bars around the abundance. The limit on Sc applies to the models, not the observations. We note that even with the constrained dilution, almost every abundance of any individual element could be reproduced. Therefore, the white does not indicate a lack of models in the abundance space, but merely a lack of well-fitting ones.} 
    \label{fig:Bd09all}
\end{figure*}

Following, we tested the impact of using LTE (red) versus NLTE abundances (see also, e.g. Fig.~\ref{fig:Bd09all}). By blindly adopting the best fit model, the LTE versus NLTE case could result in progenitor masses differing by up to 14--20\,M$_{\odot}$ and energies spread by a factor of $\sim$10.

\subsection{Constrained, diluted fits to supernova yields}\label{sec:constr}

Here, we use a Bayesian template fitting approach described in \citet{Magg2020}. In this method, the observed abundance patterns are compared to SNe models from \citet{Heger2010}. We use this set of yields as it includes around 18000 models with a wide range of stellar masses and explosion energies, offering the possibility to fit a wide range of abundance patterns. While the fitting model could also be used, for example, with the yields from \citet{Ishigaki2018}, its applicability is more limited in this case because these SNe models are designed to be aspherical. This issue is discussed further in \citet{Magg2020}. These yields are based on a grid of 1D simulations of non-rotating Pop~III core-collapse SNe, spanning a mass range of $9.6\,\Ms \le M_*\le 100\,\Ms$ and explosion energies (in units of $10^{51}\,\mathrm{erg}$, abbreviated foe) of $0.3 \le E_{51} \le 10$. As only elements up to Zn are included in these yields, we disregard heavier elements in the fits. Each model is assigned a likelihood based on how well its predictions agree with the observationally derived abundances. Upper limits on detections are treated as a strict upper limit with a Gaussian error, such that the resulting likelihood is shaped as a Gaussian error-function. 
To obtain posterior probabilities from the likelihoods, the likelihoods are normalised, such that the sum over all likelihoods equals one. The treatment of errors and limits is similar to the StarFit tool as used in \citet{Fraser2017}.
In such fitting routines, usually only ratios of abundances are fitted, which is commonly realised by an arbitrary dilution factor that is chosen to optimise the fit likelihood. This dilution factor describes how much metal-free gas the SN ejecta mix with before forming a new star and, therefore, change the overall metallicity of the star. \citet{Magg2020} show that these dilution masses are commonly picked too low and are incompatible with the expected expansion of SN remnants. Fits that we mark here as 'constrained' are fits that were performed while enforcing the minimum dilution mass based on the analytic limit on dilution from \citet{Magg2020}, whereas fits labelled as 'unconstrained' were performed with arbitrary dilution factors. The latter are merely shown for reference purposes and they should not be considered to be viable progenitor scenarios for the origin of the observed abundance patterns. We note that our results rely on the SN yields from \citet{Heger2010}, and different results may be obtained if a different set of yields, for example from \citet{Ishigaki2018}, is used.
Below, we discuss individual, potential mono-enriched, metal-poor stars.

In Table~\ref{tab:starfit} we show our comparisons to SN models, indicating the likelihood distributions and limits. We show two cases, a simple 'unconstrained' case with freely varied dilution and mixing, and our second comparison which considers dilution \citep[]['constrained']{Magg2020}. Both use a Bayesian inference with priors, however, the latter  'constrained' case ensures a lower limit of reasonable dilution and we consider it to be better and more physical.

As seen from Table~\ref{tab:starfit}, there is little difference in the best fit model as well as the model ranges between the LTE and the reduced LTE (LTEred). This is likely because there is no difference in the abundances listed for each individual element. However, there are fewer elements in the LTEred case, which means that there might be a minimum number of elements per pattern needed to provide a decent fit (for BD+24\_1676, LTE has 14 elements with $Z<30,$ while in LTEred there are only ten). For BD+09\_2190, the reduced LTE pattern has nine elemental abundances (in this case, the same as LTE). Hence, a critical number seems to be $\sim10$ elements per star. Furthermore, the combination of elements also matters (see Sect.~\ref{sec:pop3}), which could explain why BD+24\_1676 does not provide a range of best fitting models, but it might be stuck with one 
best fit model, which still may not yield a satisfactory fit. 
Below, we explore if some elements or combinations thereof have better predictive powers in tracing the supernova progenitor.

Finally, we note that for BD+24\_1676, we see a strong increase in mass and energy when moving from LTE to NLTE (see Fig.~\ref{fig:diagramBD24} in the appendix). This seems to hinge on the NLTE corrected $\alpha$, Cr, and Mn abundances. We also note that this is where we see the largest difference between the constrained and unconstrained fits. This is discussed below.

\subsection{Mono-enriched candidates}
\begin{figure}
    \centering
    \includegraphics[scale=0.55]{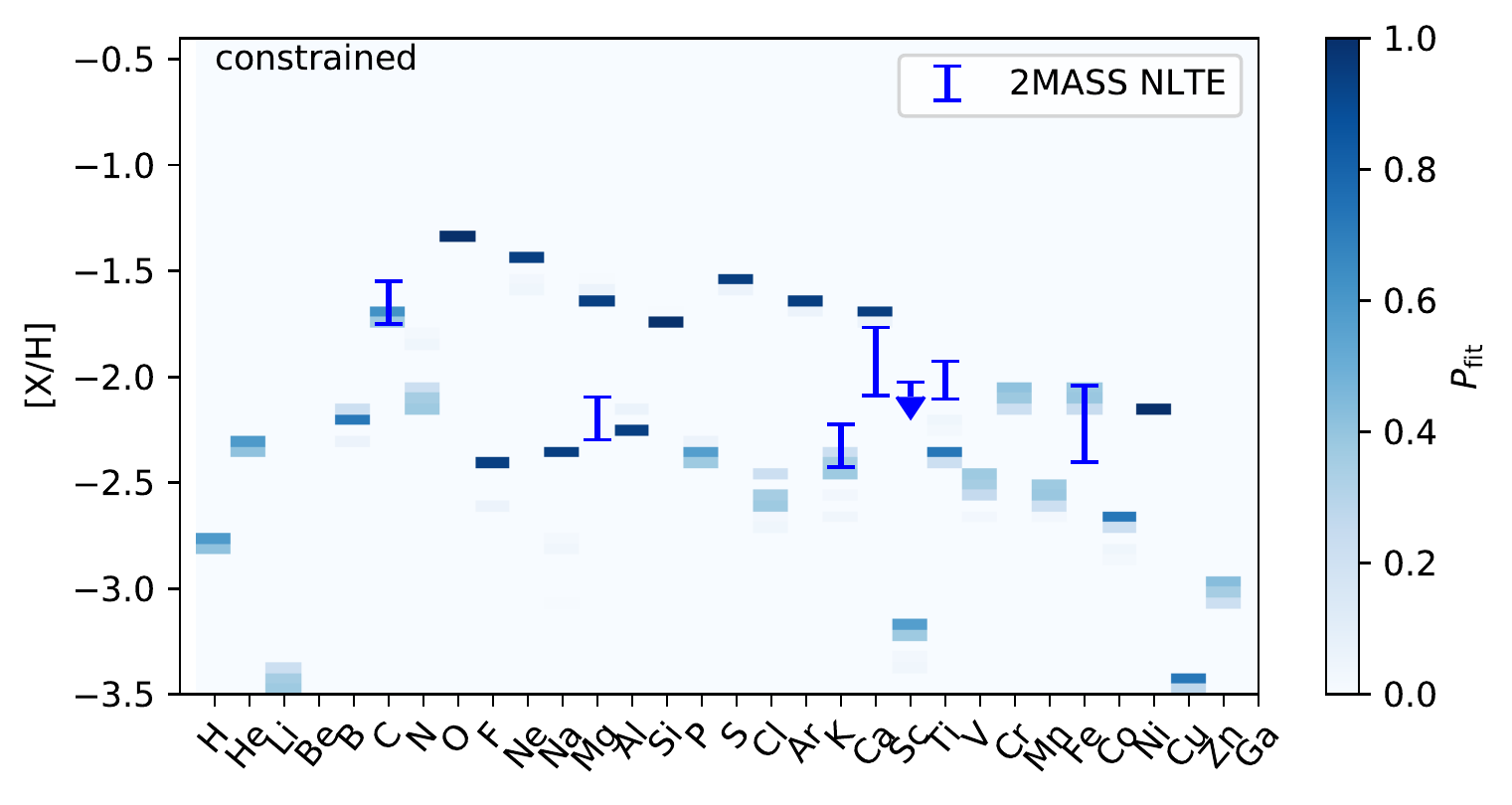}
    \caption{Constrained model comparison to NLTE abundances of our Mg/C-predicted mono-enriched metal-poor star 2MASS J0023. As in Fig.~\ref{fig:Bd09all}, the normalised likelihood is shown by the colour bar.}
    \label{fig:SNconstrained}
\end{figure}

The CEMP-no candidate 2MASS J0023 poses an interesting case in this comparison, as CEMP-no stars have been proposed to be bona fide second generation stars. Also Fig.~\ref{fig:MgC} indicates, based on the [Mg/C] ratio, that it is right at the limit, and we seek to find the nature of the progenitor as it could provide important clues as to the formation scenario of CEMP-no stars. By inspecting Fig.~\ref{fig:SNconstrained}, only one model stands out (the one model in dark blue without likelihood distributions). This means that there is only one or no well-fitting model in the grid of models from \citet{Heger2010}. However, the fit is poor; therefore, we can exclude that 2MASS J0023 is mono-enriched, despite its Mg/C-ratio and what simple unconstrained fits would lead to. Considering the criteria laid out by \citet{Hartwig2018} that this star is unlikely to form from a combination of multiple SNe, this may indicate that 2MASS J0023 is not a true second-generation star altogether. However, using a different set of SN models, including more exotic kinds of SN, could challenge this conclusion.

\begin{figure*}
    \centering
    \includegraphics[scale=0.8]{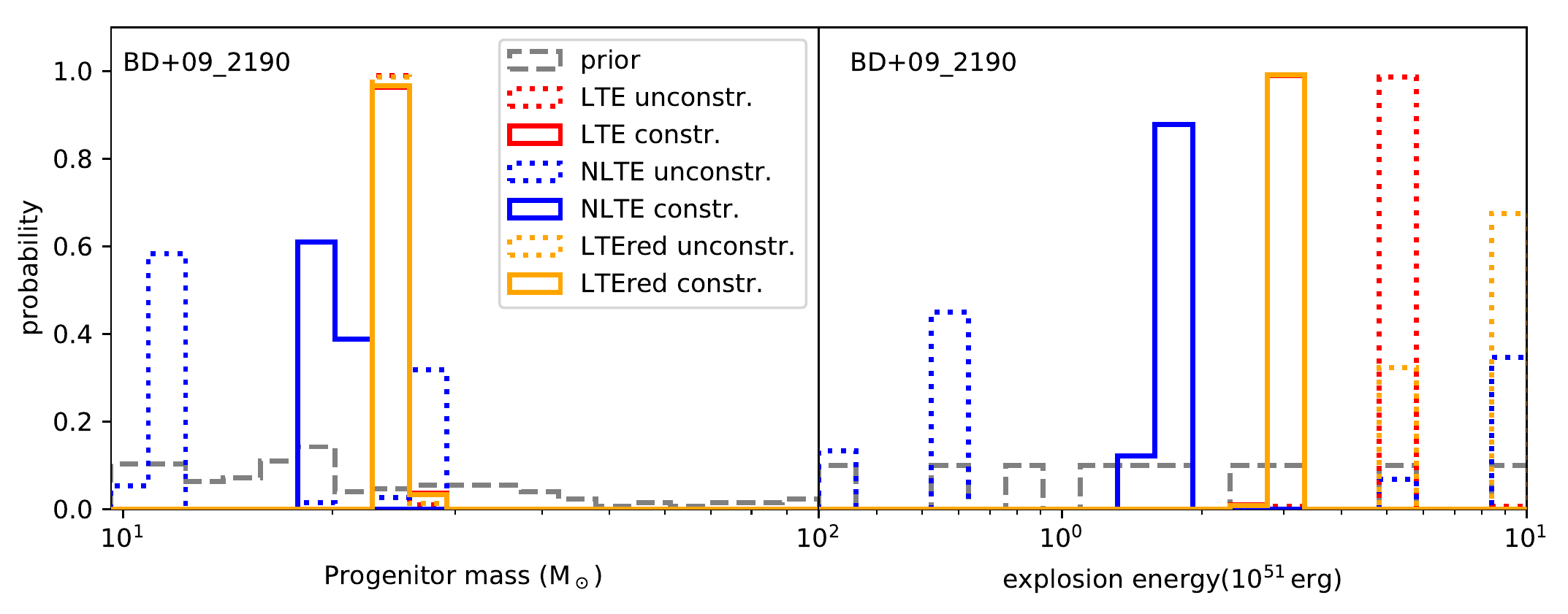}
    \caption{Probability from unconstrained and constrained model comparisons, showing preferred mass and energy in the following three cases: LTE (red), LTEred (yellow), and NLTE (blue) for BD+09\_2190. }
    \label{fig:diagram}
\end{figure*}

We now turn to our most metal-poor star. Unfortunately, we could not derive C but only place an upper limit and hence we have a star that is right at the verge of the multi- and mono-enriched area in the [Mg/C] diagram.
However, the constrained fit including all observed elements (in NLTE) shows a perfect case with statistical likelihood distributions for each element (see Fig.~\ref{fig:Bd09all}). For BD+09\_2190, we could not derive Na from the PEPSI spectra due to the lack of wavelength coverage, and we therefore derived Na from UVES spectra \citep[from][]{Hansen2012}. Despite only having  nine NLTE abundances (11 LTE abundances), this star is the most promising mono-enriched candidate we have.
The unconstrained comparison to SN yields indicates that the most likely Pop III progenitor (Z~=~0) would be 25.5\,M$_{\odot}$ with an explosion energy of 5\,foe (10\,foe in the reduced LTE case); nevertheless, these values change in NLTE and turn out to be lower (11.2\,M$_{\odot}$ and 0.6\,foe). In the constrained case, the mass and energy inferred from NLTE and LTE are closer. In LTE, we obtain 25.5\,M$_{\odot}$ and 3.0\,foe, while NLTE yields 19.2\,M$_{\odot}$ and 1.5\,foe, respectively.  However, NLTE yields higher values than LTE for BD+24\_1676, so this is not a general trend that NLTE provides lower progenitor mass and energy than LTE (see Table~\ref{tab:starfit} and Fig.~\ref{fig:diagram}). 
This manifests the importance in comparing accurate stellar abundances with carefully treated models and yield predictions, since simple $\chi^2$ matching in large samples could skew the inferred masses and in turn the initial mass function (IMF) at the lowest metallicity of mono-enriched stars in the Galaxy.

Similar to the metal-poor stars in Fig.~\ref{fig:SNconstrained} and \ref{fig:diagram}, we compared all other stars in our sample to the `constrained models'. The obtained fits were poor, their [Mg/C] high, and we can therefore exclude that all the more metal-rich stars in our sample with [Fe/H] $>-2.4$ are mono-enriched. The nature of their composition is the result of multiple enrichment events, where several supernovae and AGB stars have contributed to the resulting patterns we derived (see Sect.~\ref{sec:discussion} where we discuss the dominant contribution to the stellar photospheric composition).

\subsection{Elements with strong predictive powers on the Pop III progenitor - in LTE and NLTE}\label{sec:pop3}
We now explore if some elements provide stronger constraints than others on the supernova progenitor when comparing observationally derived stellar abundances to yield predictions. 
Figure~\ref{fig:Bd09all} shows the unconstrained (top panels) and the dilution constrained
(bottom panels) predictions compared to the LTE and NLTE abundances of BD+09\_2190. As not all possible nucleosynthetic sources of Sc are accounted for, modelled Sc abundances are only lower limits \citep{Heger2010}. Therefore, we treat the observed Sc abundances as upper limits, despite them being detections, which is equivalent to the modelled abundances being (observational) lower limits.
In addition, we note that other elements, such as Ti, may also be difficult to model accurately as nucleosynthesis theory underproduces Ti \citep[see e.g.][]{Kobayashi2020}.

\begin{figure*}
    \centering
    \includegraphics[scale=0.8]{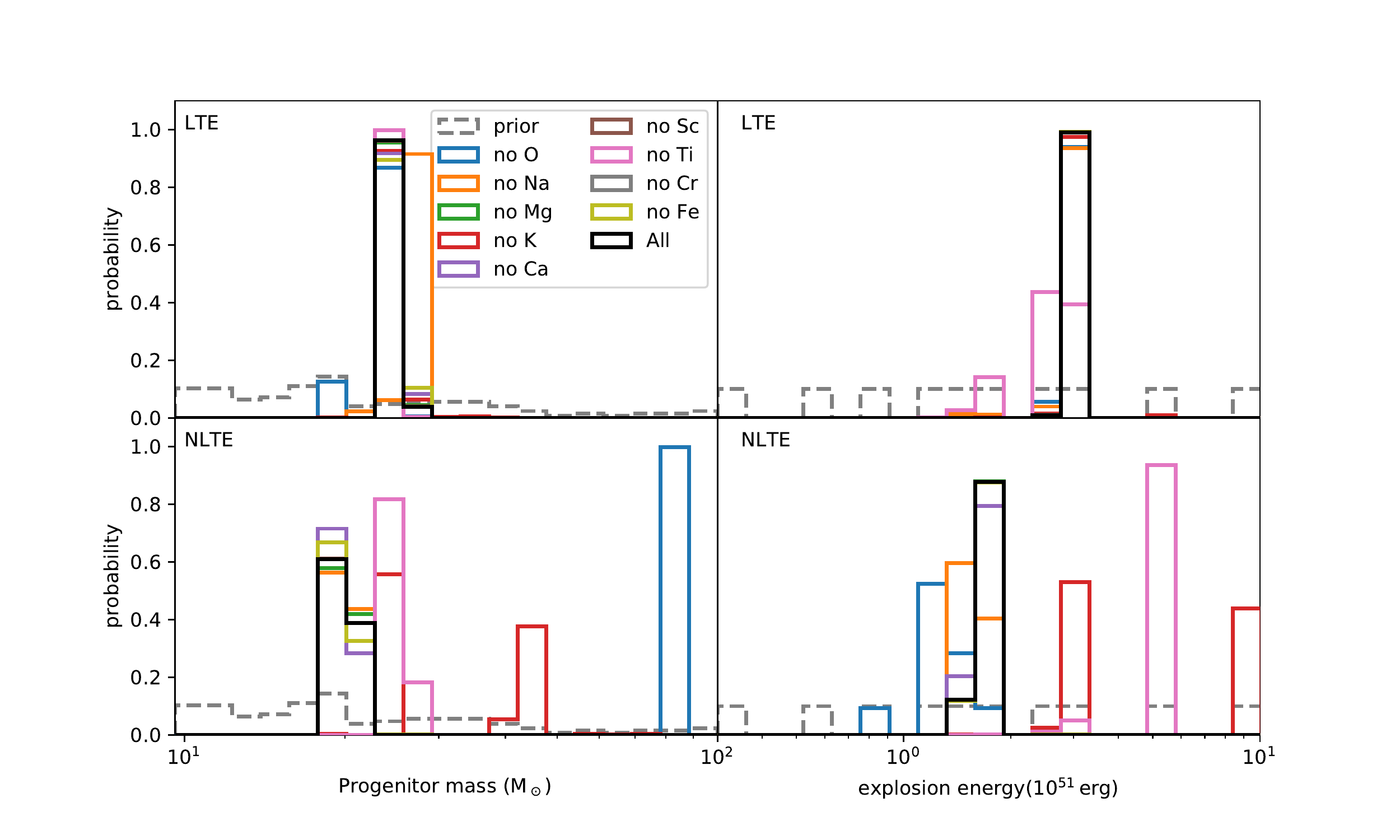}
    \caption{Constrained fits for the star BD+09\_2190 performed without the element listed (see legend) indicating the impact on mass and energy (c.f. top panel of Fig.~\ref{fig:diagram}).}
    \label{fig:elem}
\end{figure*}

The unconstrained (direct) comparison shows clear distributions with peaking likelihoods for all elements, except for O (see Figs.~\ref{fig:Bd09all}, \ref{fig:diagram}, and \ref{fig:elem}). These resolved posterior distributions indicate that the method is able to recover a range of SN models that provide a good fit to the observed abundance patterns.
In some cases (e.g. for BD+24\_1676; see Fig.~\ref{fig:diagramBD24} and the appendix for more details), some elements show a double peaked distribution. Here, the observational constraints are particularly important as they help to find the better solution. In this regard, limits are also useful. For comparison, the constrained case shows that some elements reach a maximum likelihood and then their sensitivity drops. Based on this, elements such as O, (F), Na, Mg, Al, and (Si), but also K, Ca, Ti, and Mn, offer strong constraints when combined. This could be explained in part by their different sensitivity to the SN progenitor mass and explosion energy through their odd-even distribution. We emphasise that the different Na abundance in LTE and NLTE play an important role in constraining the best fit model, and we note that Na$_{\rm NLTE}$ does not present a good fit in either of the unconstrained or constrained cases.

However, to test the element-to-element sensitivity, the  pattern is fit several times by removing one element at a time. Figure~\ref{fig:elem} shows such fits to our best mono-enriched candidate, BD+09\_2190, in LTE and NLTE. In LTE, there are no strong constraints on the energy, except from Ti; while in NLTE, the combination of Ti, K, and Mn in other stars, provide important traces of the odd-Z and early Fe-peak elements which are energy sensitive. In addition, lighter elements, such as O, Na, and Mg, should be included, otherwise the distribution becomes multi peaked and no strong conclusions can be drawn. To constrain the mass using LTE abundances, we see that oxygen is important in combination with Na and other $\alpha-$elements. However, in NLTE, this becomes even more pronounced. Here, K is also seen to play a vital role, and lacking any of these elements could skew the mass prediction by a factor of $\sim4-5$. Again we see that O, Na, Mg, K, and Mn are important tracers. They represent the odd-even pattern, even in sparsely populated patterns (eight elements), yet they contain Fe-peak elements and $\alpha-$elements that are mass and energy sensitive. The lack of N abundances in this sense prevents a light odd-even pattern around C-N-O, which would have supported the SN traces and might therefore impose a slight bias on our results. However, the current spectra do not allow us to derive N. Future blue spectra in line with what, for example, the CUBES spectrograph (\citealt{CUBES} and Ernandes et al. 2020 in prep.) will provide, will greatly help us solidify the true origin of stars, such as BD+90\_2190, by adding molecular N and O to the patterns. 

We note that due to the sparsely populated 3D NLTE patterns, we did not attempt find a best fitting (constrained) Pop~III model. Nevertheless, we emphasise that it is important to expand 3D NLTE patterns and conduct constrained fits to understand the nature of the first population of stars and, in particular, the mass spectrum and energetics
of the explosion.

\section{Conclusion}\label{sec:concl}
Stellar archaeology is a powerful tool for exploring and understanding the chemical evolution of the Galaxy. Despite the limited sample size of our study, the broad range of stellar parameters permits one to explore the chemical evolution and the richness of the derived abundance patterns, allowing us to contrast competing models in a meaningful way. 

Our high-resolution PEPSI analysis is an attempt to accurately explore the Milky Ways true chemical evolution and the nature of Pop III supernovae. The GCE trends of the 32 studied elements (34 when including limits) are in agreement with the previous literature samples and only potassium appears slightly higher than anticipated. For the lighter elements, such as C, O, Na, K, Mn, and Co, NLTE corrections play an important role, to the point where the GCE of the MW appears different in LTE versus NLTE.
For Si, Ti, V, and Fe, we derived both neutral and ionised abundances; furthermore, in most cases, there is a difference or a scatter due to the lack of ionisation equilibrium in the stellar sample.
For the heavy elements, where the abundances are generally derived from the majority species, the NLTE corrections are smaller, and the GCE-trends appear consistent in both LTE and NLTE (see e.g. Zr, Ba, Nd, and Eu).

Typically, 1D LTE abundances are used in most studies as their analysis is faster and more straightforward. However, as shown in this study (and highlighted by Fig.~\ref{fig:all_patterns}), the overall abundance pattern can look quite 
different in LTE and NLTE, even if 1D hydrostatic models atmospheres are used. Yet, this 
is not the complete picture as we show by adding 3D (and 3D NLTE) corrected abundances for a small 
number of elements (C, O, and Fe). The optimal and most physical answer to understanding the 
chemical evolution of the Galaxy or to 
pinning down the nature of the Pop~III supernovae can 
best be answered with a pattern of 3D NLTE corrected elements in a statistical 
sufficient sample using a dilution constrained method, including key elements, such as C and O. 

The comparison to the yield predictions returned three important results, namely, that the most robust conclusions can be inferred when the stellar pattern consists of $\sim 8-10$ or more elements. The conclusions differ depending on the physical basis of the radiative transfer models (LTE, NLTE, 1D hydrostatic equilibrium, and 3D hydrodynamics), but also from the yield or SN model side. Here the unconstrained versus the dilution constrained methods provide different results. These results are so different that they can skew and bias the inferred IMF of the Galaxy if not properly dealt with. In the case of BD+24\_1676, the inferred progenitor mass shows a discrepancy of $\sim20$\,M$_{\odot}$ between NLTE and LTE and unconstrained and constrained methods, respectively. If this would be the outcome of the inferred mass from numerous mono-enriched second generation stars, the mass distribution of the Pop III stars could look very different. Finally, when comparing observations and theory, it is not only important to have a sufficiently rich, well-sampled pattern, but also the combination of odd-even elements with $Z<30$ (such as C, O, Na, Mg, K, Ca, Ti, and Mn) need to be present to best constrain the mass and explosion energy of the Pop~III progenitor. We highlight that the results also depend on the set of SN model predictions that are included in the comparison.

Based on $\alpha-$abundances and kinematics, we traced the chemo-dynamical origin of several chemically peculiar stars and find strong indications of two fast halo stars (BD+09\_2190 and 2MASS J0023) where BD+09\_2190 was likely accreted into the MW long ago, while all other stars move on common disc-like orbits. Relying on the atomic C-abundances, we find strong evidence for a CEMP-no (2MASS J0023) and a CEMP-$s$ star (TYC5481). We tested if the CEMP-no star was a bona fide second generation star, as many CEMP-no stars are; however, in following both constrained and unconstrained dilution mass model comparisons, we could reject the mono-enrichment hypothesis for this specific star with the yields we adopted here. The CEMP-$s$ star was likely enriched by an intermediate-mass and -metallicity AGB star.

We placed constraints on the possible (dominant) AGB donor at high metallicities, seeing how they add s-process to metal-poor stars, such as TYC5481, while the heavy element content of BD-10\_3742 shows a perfect r-process pattern as represented by the `metal-poor Sun' HD20 that serves as a neutron-capture benchmark star with its almost pure r-process pattern. Finally, we find that BD+09\_2190 is likely a true second generation star, which could be mono-enriched by a $\sim25.5$ or $19.2$\,M$_{\odot}$ supernova with a normal explosion energy (3 or 1.5\,foe) in LTE or NLTE, respectively. This shows the importance of carefully assessing the stellar abundances as well as conducting a physical meaningful (constrained) comparison to supernova yields in order to properly understand the nature of the first stars. 

\begin{acknowledgements} 
CJH acknowledges support from the Max Planck Society and from the ChETEC COST Action (CA16117), supported by COST (European Cooperation in Science and Technology). CJH is grateful to A. Amarsi for providing 3D and NLTE computations for C, O, and Fe, and also to S. Cristallo and H. Reggiani for fruitful dialogues and helpful input.  
A.K., M.B., and R.S.K. gratefully acknowledge funding by the Deutsche Forschungsgemeinschaft (DFG, German Research Foundation) -- Project-ID 138713538 -- SFB 881 ("The Milky Way System"), subprojects A03, A05, A10, A11, as well as subprojects B01, B02, and B08. 
EC gratefully acknowledge support from the French National Research Agency (ANR) funded project ``Pristine'' (ANR-18-CE31-0017).
H.W.Z. acknowledges the National Natural Science Foundation of China No. 11973001 and National Key R\&D Program of China No. 2019YFA0405504.
MM was supported by the Max-Planck-Gesellschaft via the fellowship of the International Max Planck Research School for Astronomy and Cosmic Physics at the University of Heidelberg (IMPRS-HD).  
R.S.K. also acknowledges support from Germany's Excellence Strategy in framework of the Heidelberg Cluster of Excellence STRUCTURES (grant EXC-2181/1 - 390900948), and he thanks for funding from the European Research Council via the ERC Synergy Grant ECOGAL (grant 855130). M.M. and R.S.K. further more acknowledge access to the data storage service SDS\@hd and computing services bwHPC supported by the Ministry of Science, Research and the Arts Baden-W\"urttemberg (MWK) and the German Research Foundation (DFG) through grants INST 35/1314-1 FUGG as well as INST 35/1134-1 FUGG.
LBT Corporation partners are the University of Arizona
on behalf of the Arizona university system; Istituto Nazionale di Astrofisica, Italy; LBT Beteiligungsgesellschaft, Germany, representing the Max-Planck Society, the Leibniz-Institute for Astrophysics Potsdam (AIP), and Heidelberg University; the Ohio State University; and the Research Corporation, on behalf
of the University of Notre Dame, University of Minnesota and University of Virginia. It is a pleasure to thank the German Federal Ministry (BMBF) for the year-long support for the construction of PEPSI through their Verbundforschung
grants 05AL2BA1/3 and 05A08BAC as well as the State of Brandenburg for the continuing support  (see https://pepsi.aip.de/).
This work has made use of data from the European Space Agency (ESA) mission
{\it Gaia} (\url{https://www.cosmos.esa.int/gaia}), processed by the {\it Gaia}
Data Processing and Analysis Consortium (DPAC,
\url{https://www.cosmos.esa.int/web/gaia/dpac/consortium}). Funding for the DPAC
has been provided by national institutions, in particular the institutions
participating in the {\it Gaia} Multilateral Agreement. Finally, we would like to thank the anonymous referee for useful comments.
\end{acknowledgements}

\bibliographystyle{aa}
\bibliography{38805corr_arxiv.bbl}

\newpage
\onecolumn
\begin{appendix}
\section{Online material}
Online material available on CDS and in the published paper (open access).
\begin{figure}
    \centering
    \includegraphics[scale=0.35]{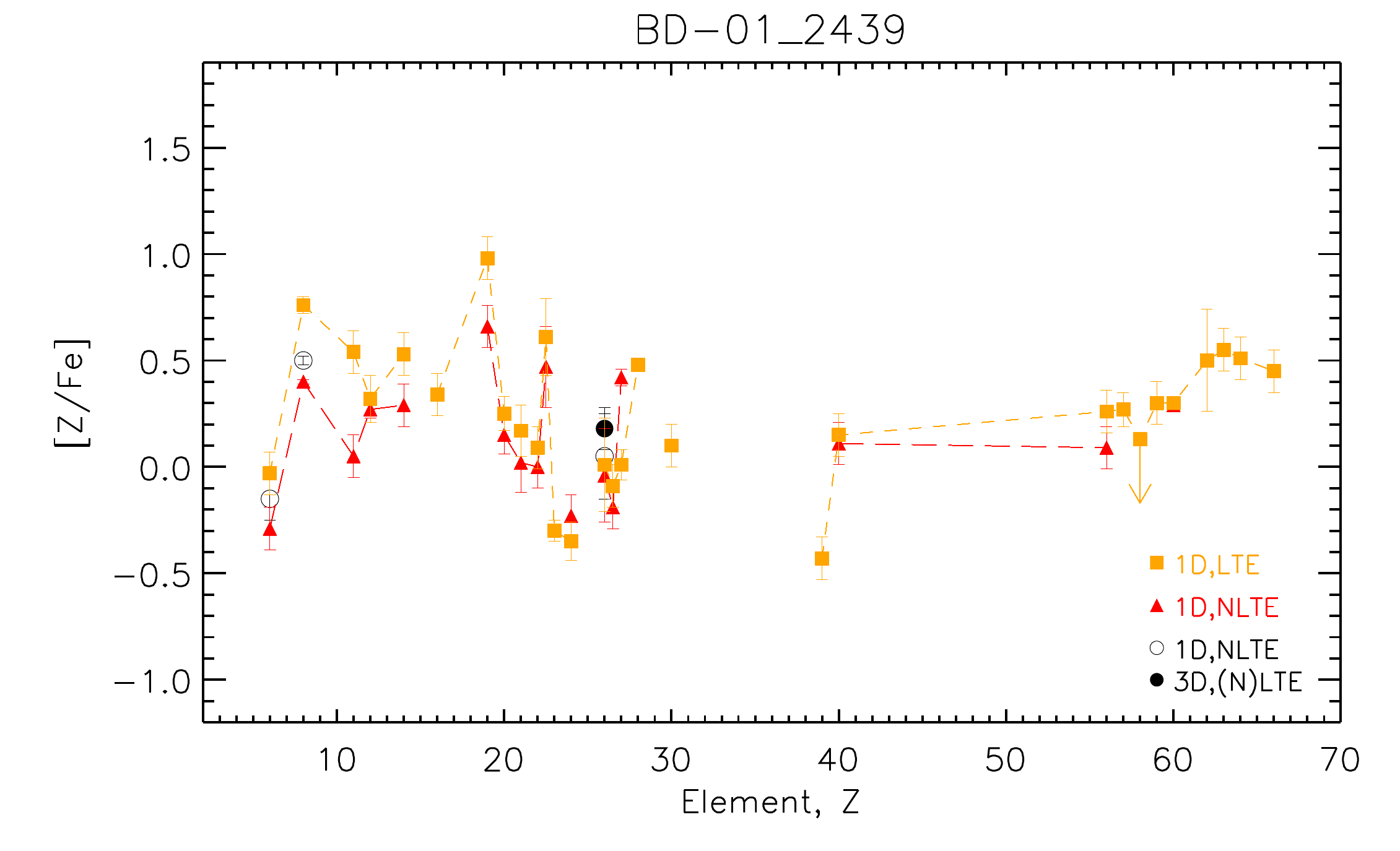}
    \includegraphics[scale=0.35]{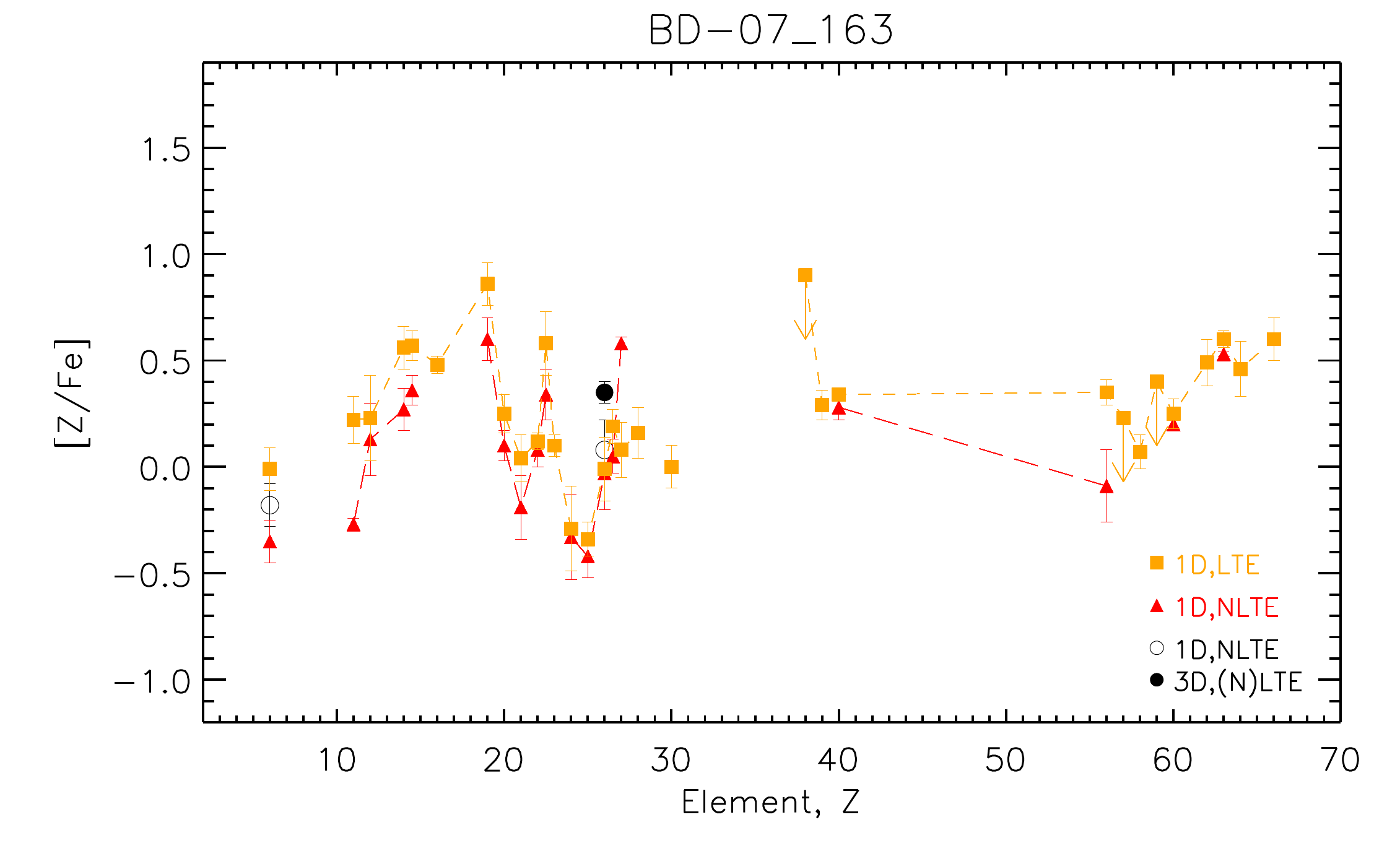}
    \includegraphics[scale=0.35]{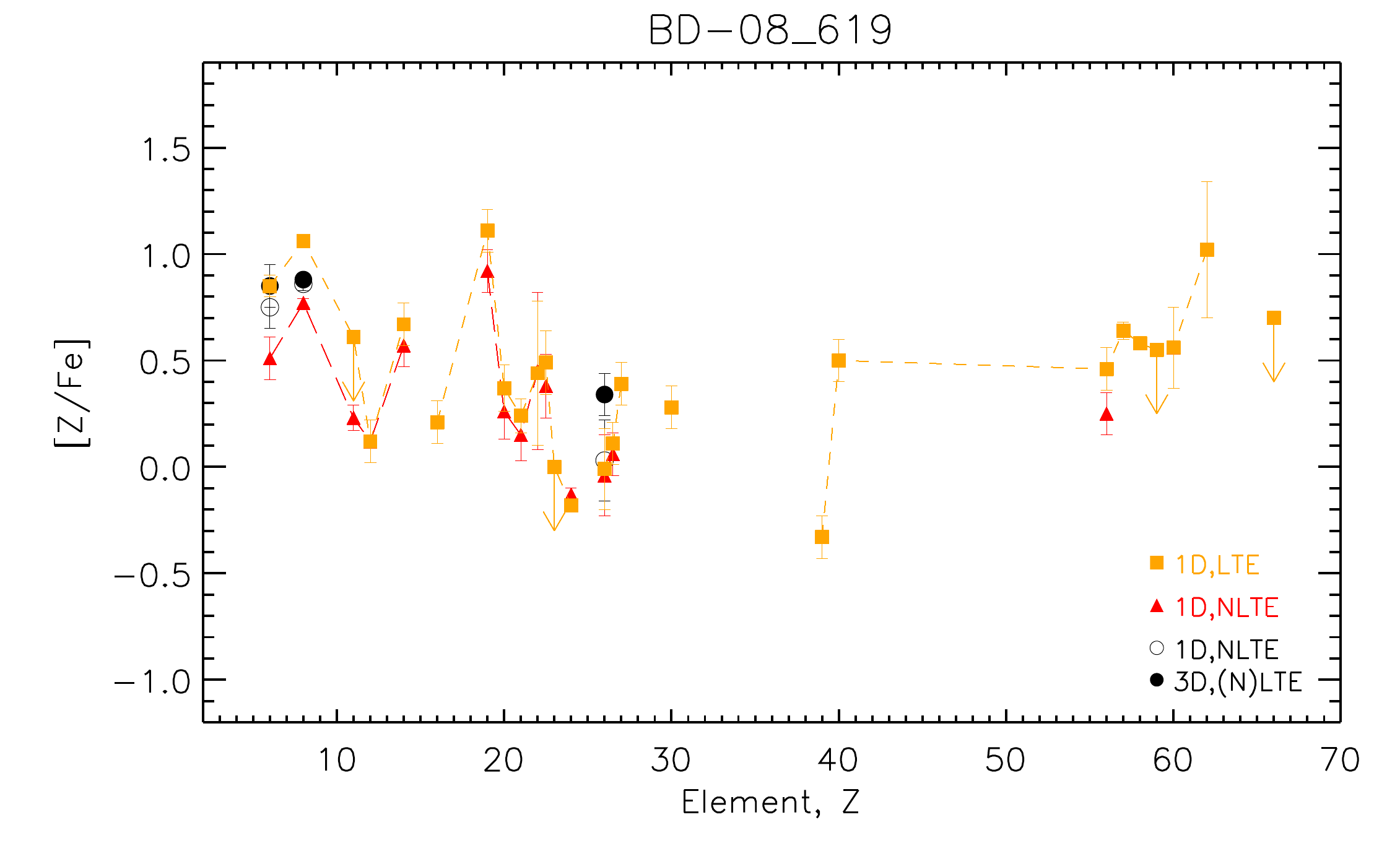}
    \includegraphics[scale=0.35]{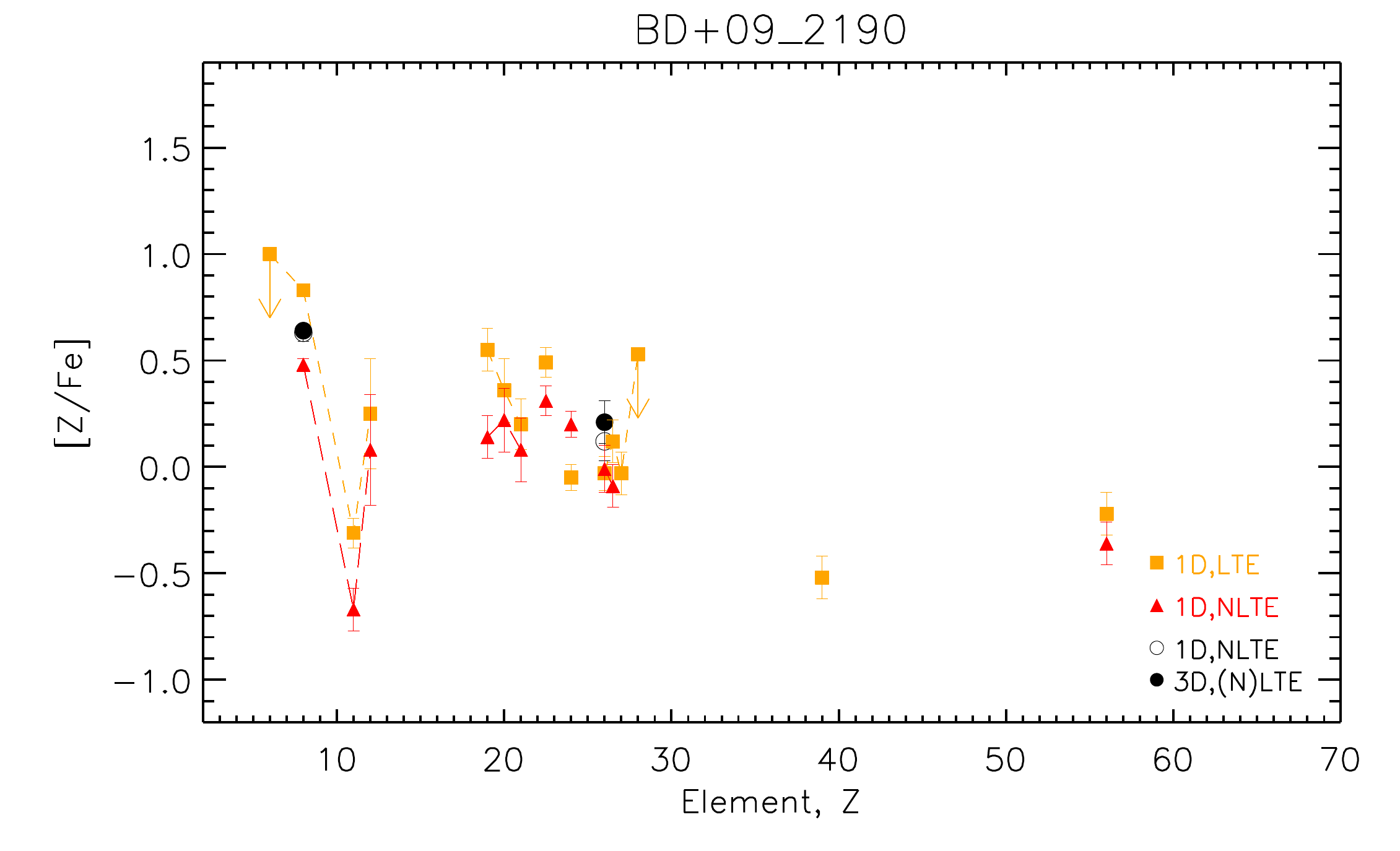}
    \includegraphics[scale=0.35]{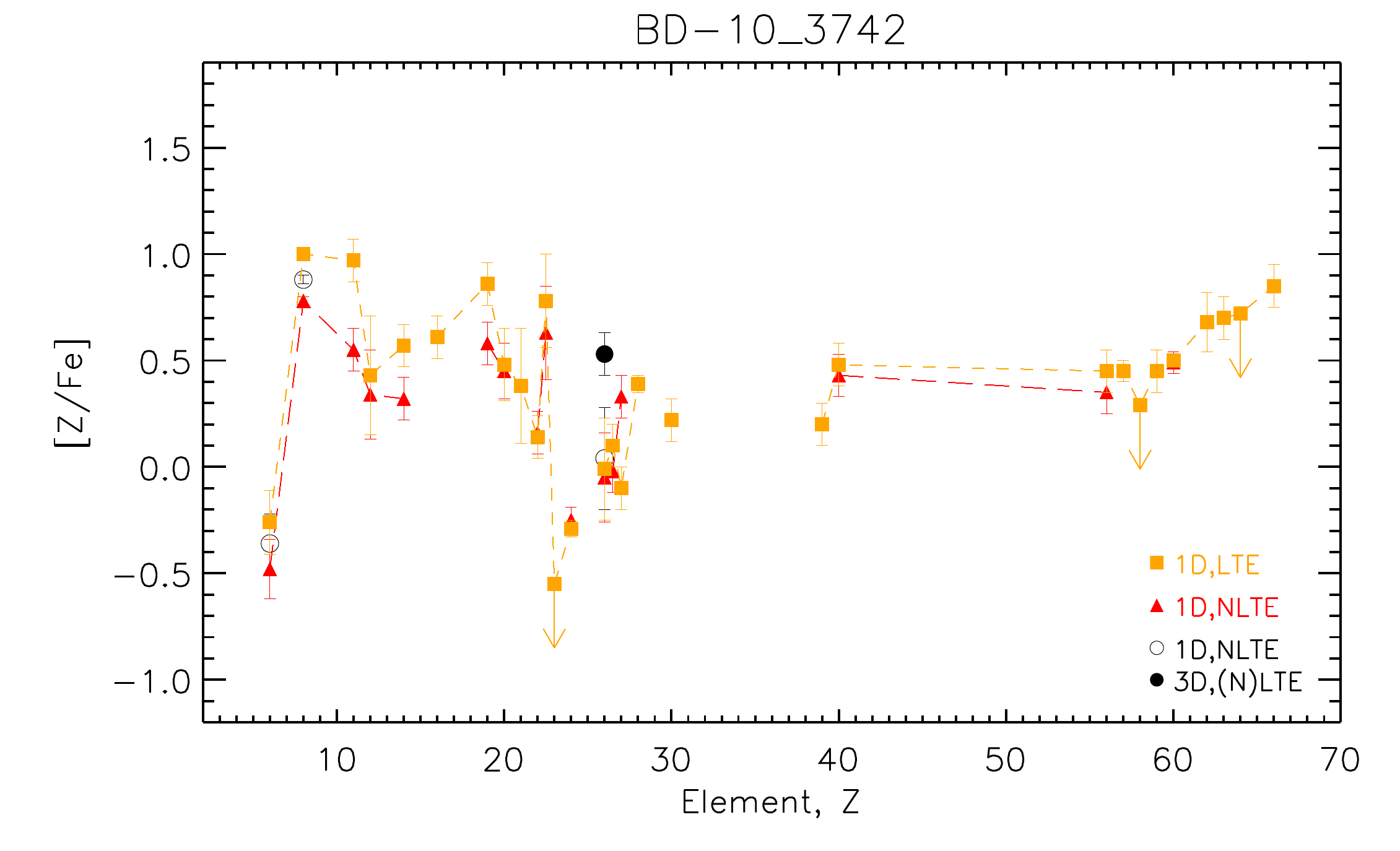}
    \includegraphics[scale=0.35]{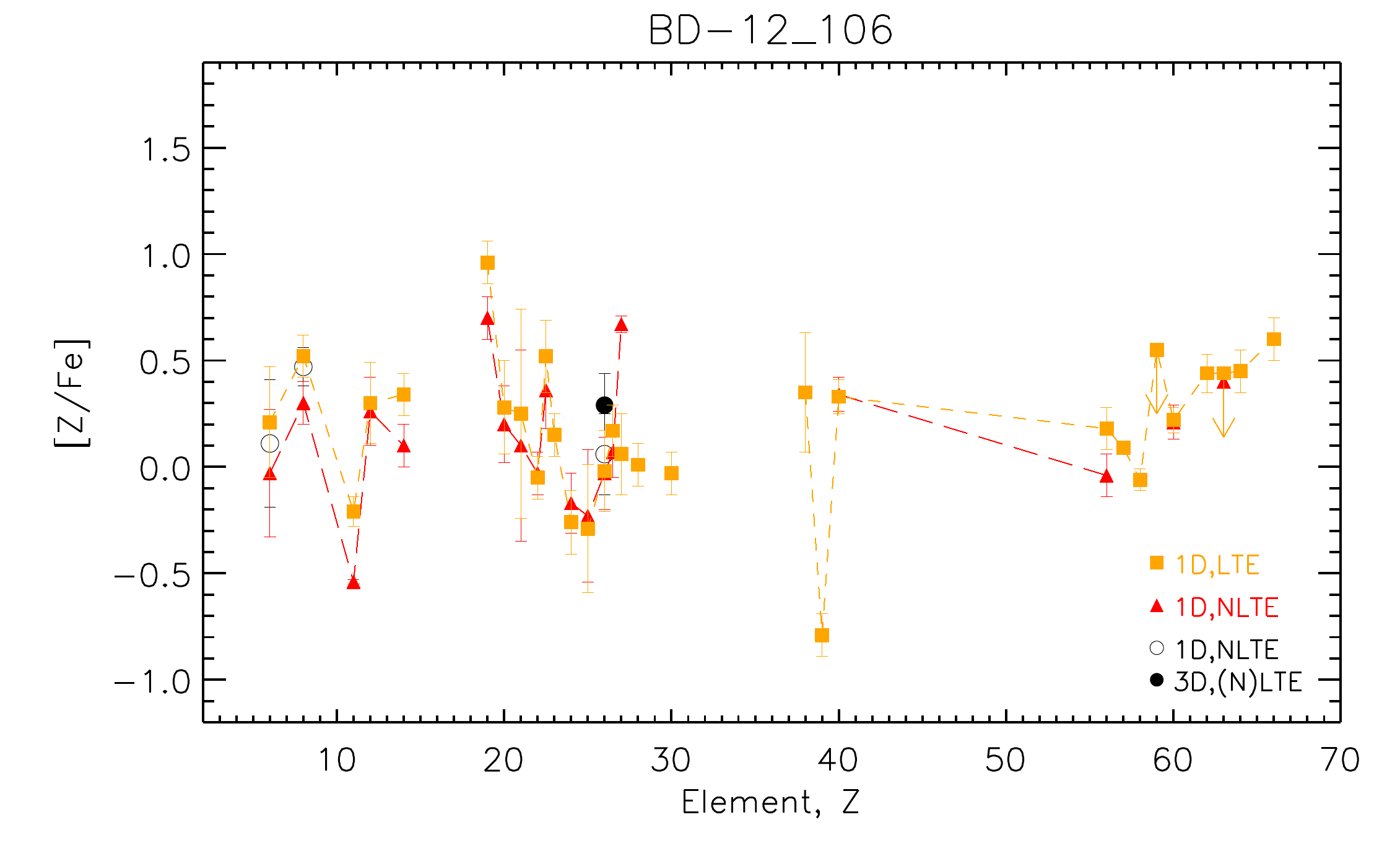}
    \includegraphics[scale=0.35]{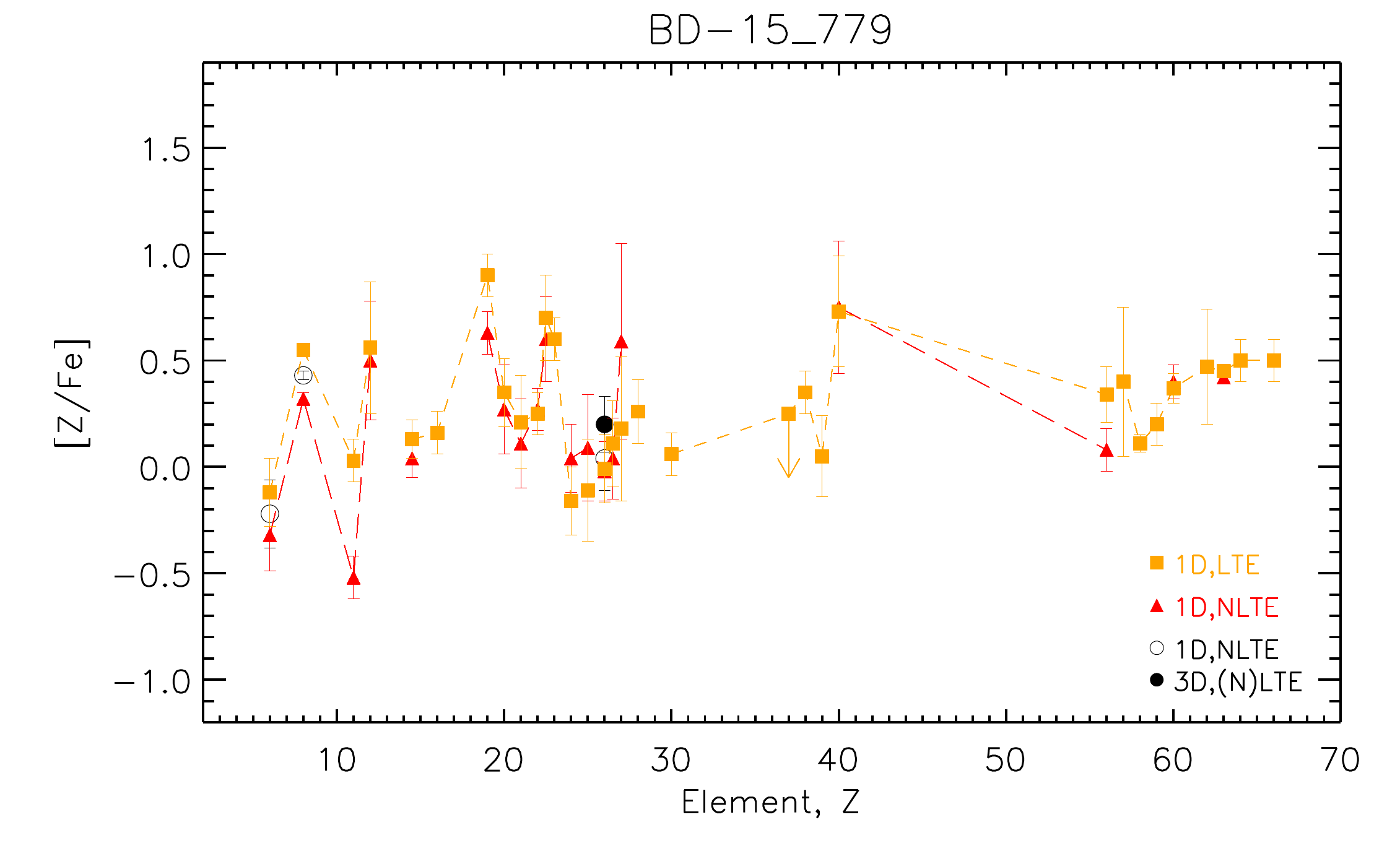}
    \includegraphics[scale=0.35]{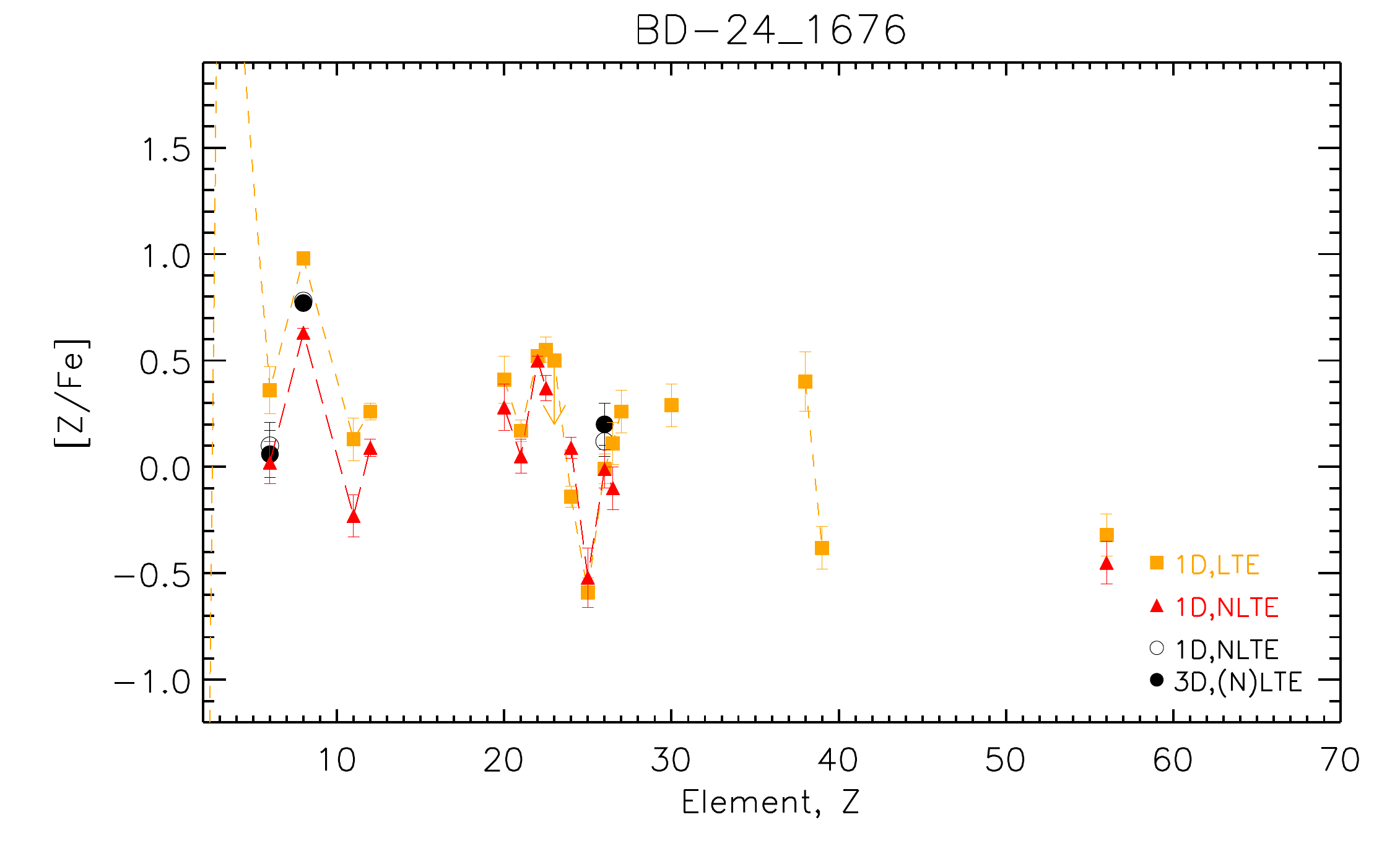}
    \includegraphics[scale=0.35]{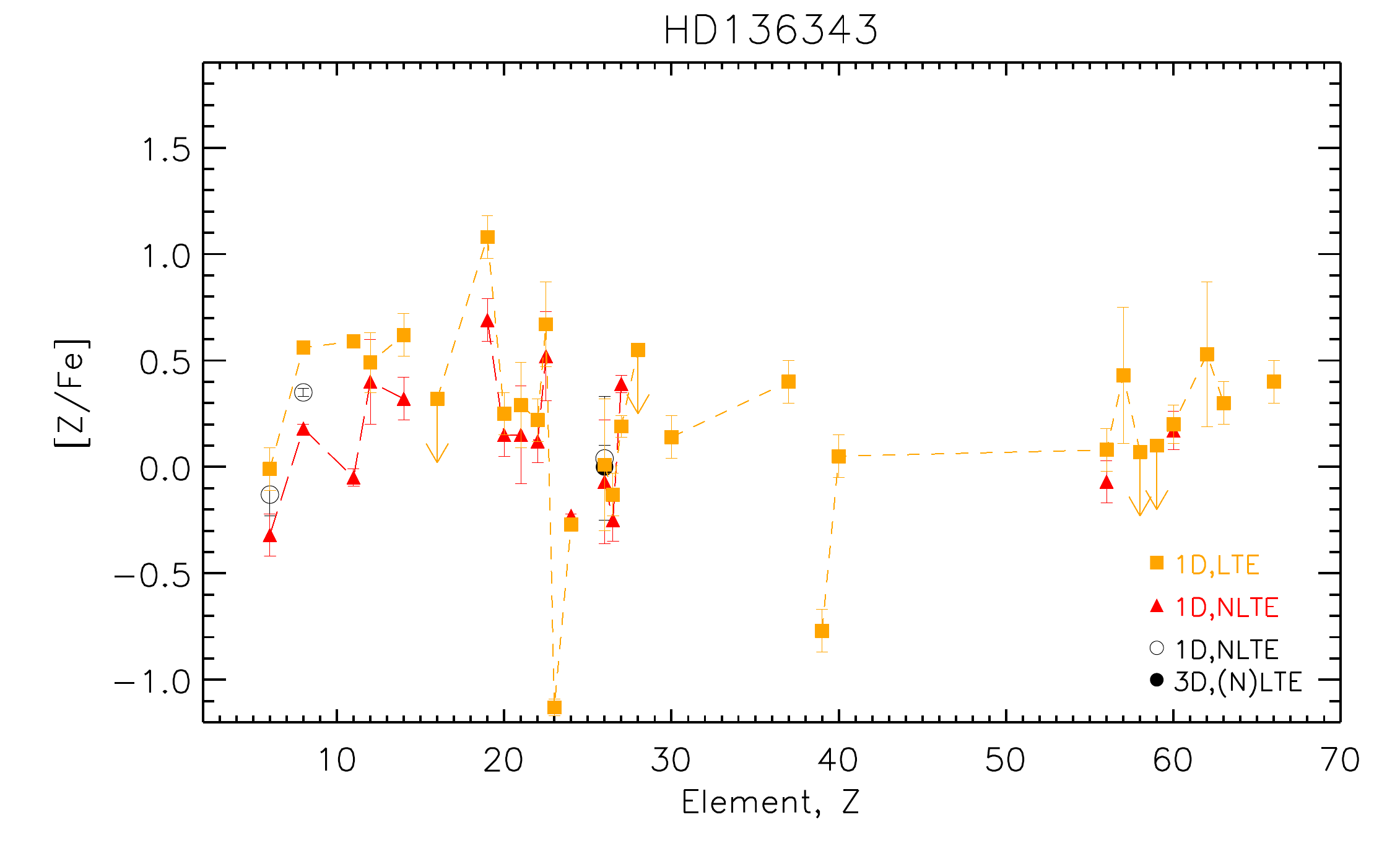}
    \includegraphics[scale=0.35]{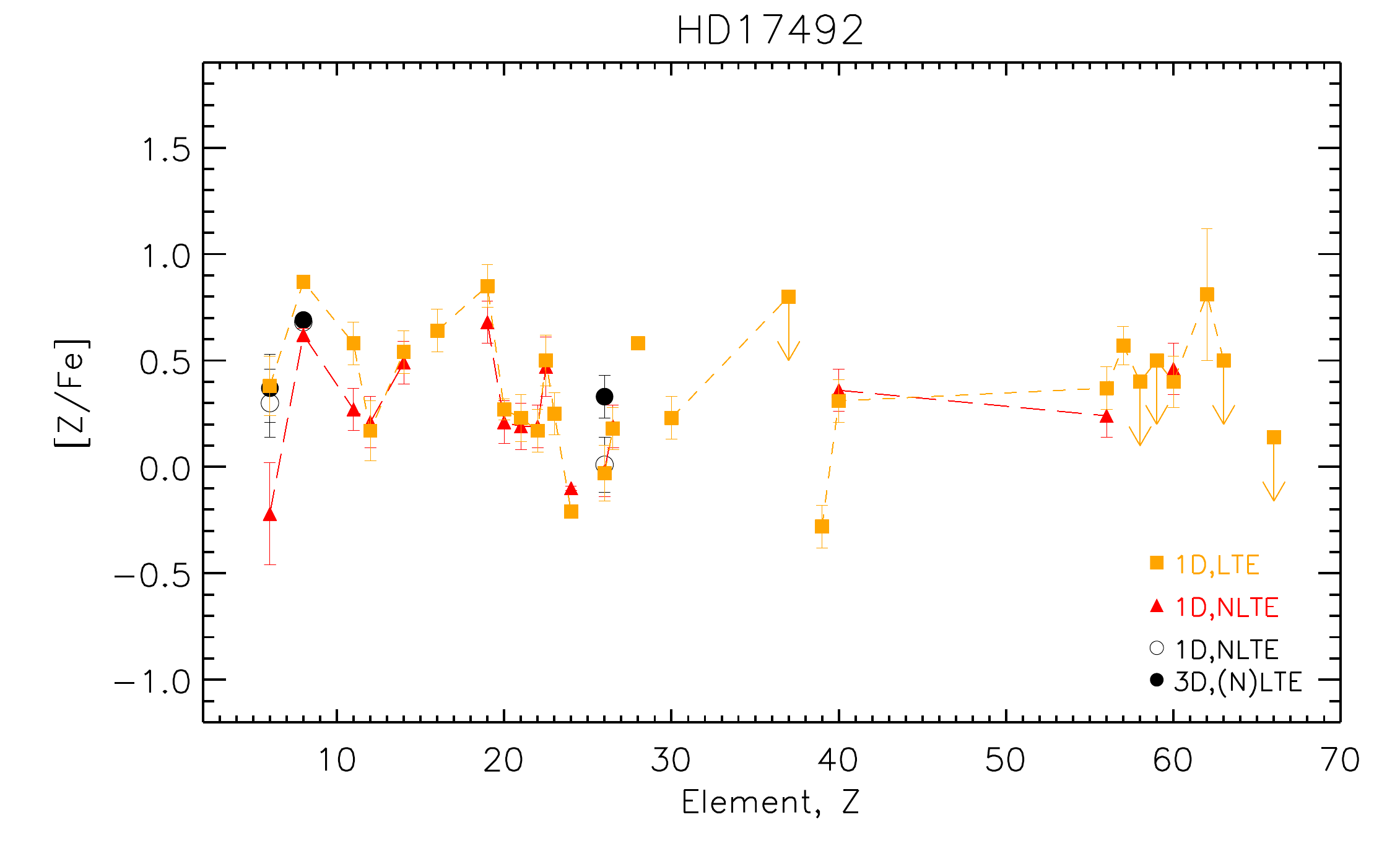}
     \label{fig:all_patterns}
\end{figure}
\begin{figure}
    \centering
    \includegraphics[scale=0.35]{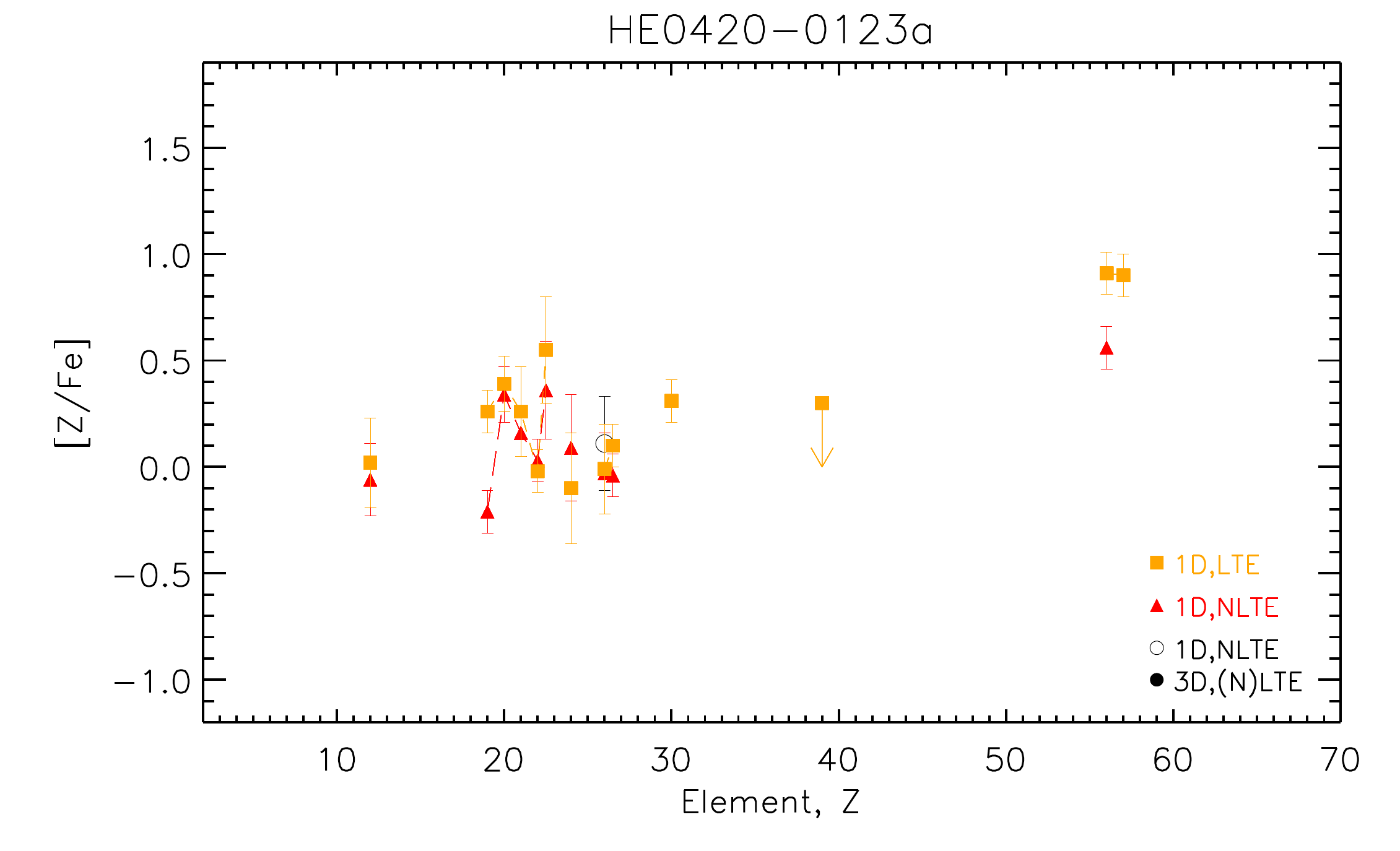}
    \includegraphics[scale=0.35]{M2l_pattern_err.pdf}
    \includegraphics[scale=0.35]{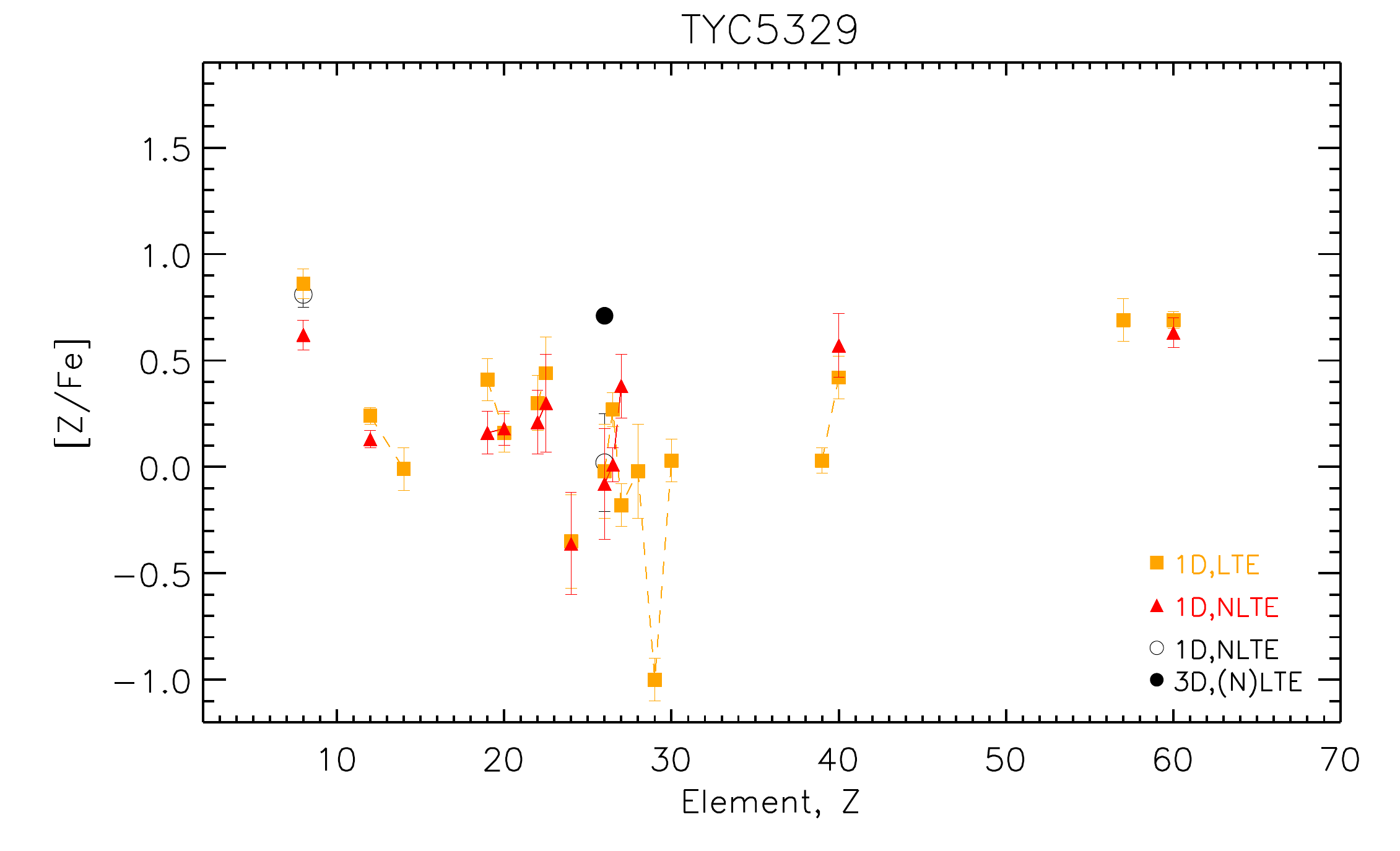}
    \includegraphics[scale=0.35]{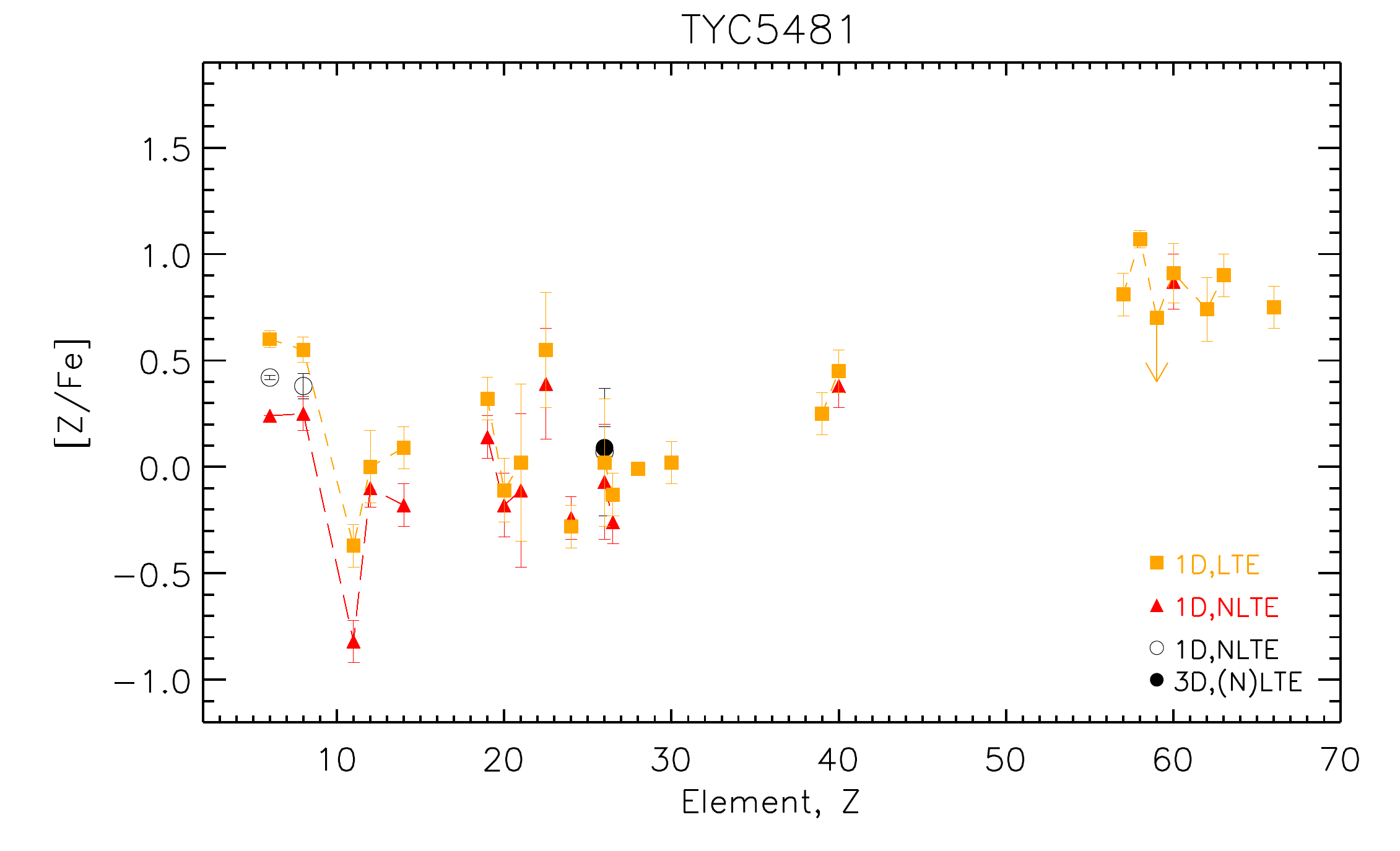}
    \caption{Abundance patterns in 1D LTE (yellow squares), 1D NLTE (red triangles with corrections as listed in Table~\ref{Tab:nlte} with the exception of C, which in the red triangles are based on \citet{Alexeeva2015C}), and 1D NLTE and 3D NLTE for C and O (in black), while 3D LTE is shown for Fe \citep[corrections from ][as described in Sect.~\ref{sec:3DNLTE}]{2019A&A...622L...4A,2019A&A...630A.104A}.}
   \label{fig:all_patterns}
\end{figure}

\newpage
\begin{figure*}[!ht]
    \centering
    \includegraphics[scale=0.8]{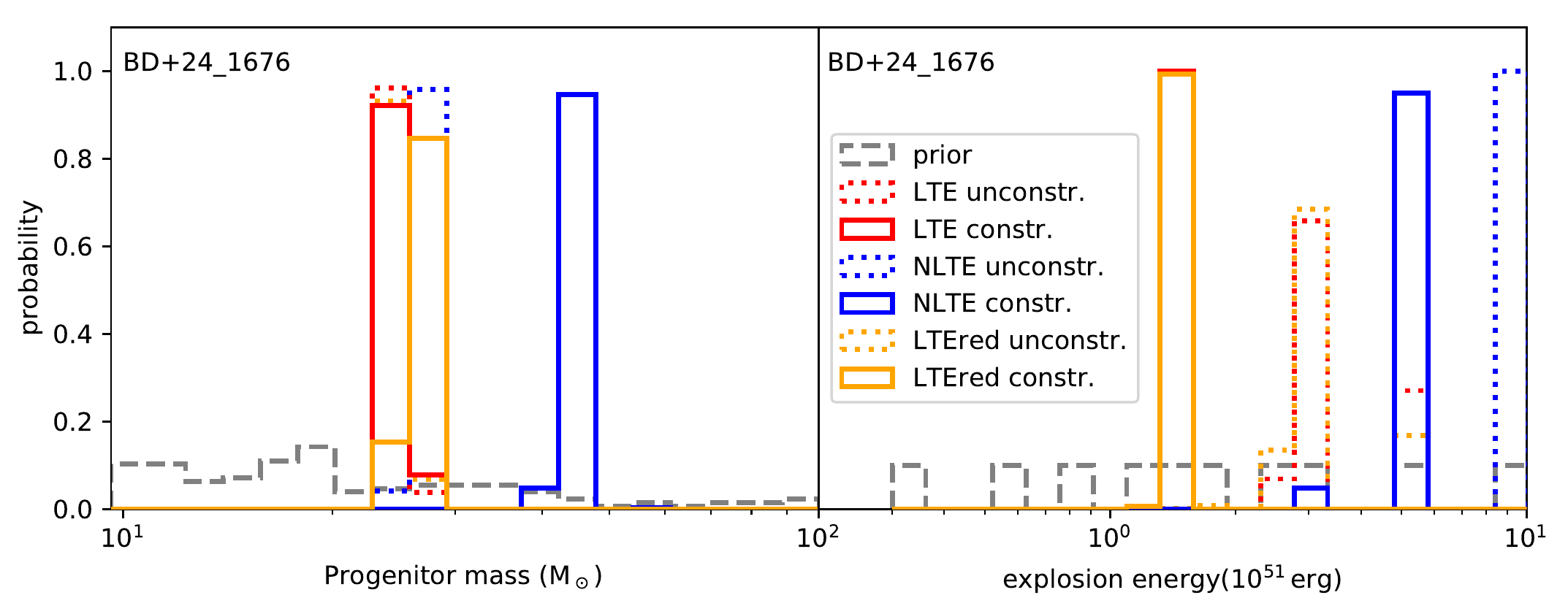}
    \caption{Probability from unconstrained and constrained model comparisons, showing preferred mass and energy in the following three cases: LTE (red), LTEred (yellow), and NLTE (blue) for BD+24\_1676. }
    \label{fig:diagramBD24}
\end{figure*}
\subsection{Testing mono versus multi enrichment at low-metallicity}
 For our second most metal-poor dwarf, BD+24\_1676, we compared it to both constrained and unconstrained models. We note that Na has a strong impact on the inferred Pop III parameters, and we explore the impact of individual elements in Sect.~\ref{sec:pop3}. Based on the bottom panels in Fig.~\ref{fig:diagramBD24}, the model fit is better than for 2MASS J0023; however, the strongly peaked posterior distributions indicate that there is only one or no well-fitting model in the SN library that was used. 
As we do not find a range of well-fitting models, we will not attempt to further investigate the posterior distributions.  This can either be due to a fundamental mismatch between the observed and modelled abundances or due to an insufficiently sampled parameter space. We, therefore, conclude that this star is likely multi-enriched (or not a true second-generation star), which is in good agreement with the r+s mixed heavy element pattern.

\end{appendix}
\end{document}